\documentclass[fleqn,usenatbib]{mnras}

\usepackage{newtxtext,newtxmath}

\usepackage[T1]{fontenc}
\usepackage[english]{babel}
\usepackage[autostyle]{csquotes}

\DeclareRobustCommand{\VAN}[3]{#2}
\let\VANthebibliography\thebibliography
\def\thebibliography{\DeclareRobustCommand{\VAN}[3]{##3}\VANthebibliography}

\usepackage{graphicx}	
\usepackage{amsmath}	
\usepackage{threeparttable}
\usepackage{booktabs, makecell, multirow}
\usepackage{url}
\usepackage{caption}
\usepackage[position=bottom]{subcaption}
\usepackage{hyperref}
\usepackage{mathtools}
\usepackage{dirtytalk}
\usepackage{capt-of}
\usepackage{tabularx}
\usepackage{rotating}
\usepackage{adjustbox}
\usepackage{mwe}
\newlength{\tempheight}
\newlength{\tempwidth}

\newcommand{\rowname}[1]
{\rotatebox{90}{\makebox[\tempheight][c]{{#1}}}}
\newcommand{\columnname}[1]
{\makebox[\tempwidth][c]{\textbf{#1}}}

\renewcommand{\thesubfigure}{\alph{subfigure}}
\newcommand{\mycaption}[1]
{\refstepcounter{subfigure}\textbf{(\thesubfigure) }{\ignorespaces #1}}
\title[Stacked broadband radio SEDs of SFGs]{Radio continuum spectra of SFGs in the XMM-LSS Field below-threshold}

\author[Ocran et al.]{
E.F. Ocran$^{1,3,4}$\thanks{E-mail: ocran62@gmail.com}, A.R. Taylor$^{2,3,4}$,  J.M. Stil$^{4,5}$, M. Vaccari$^{2,3,4,8}$, S. Sekhar$^{2,4,7}$, C.H. Ishwara-Chandra$^{2,6}$,\newauthor Jae-Woo Kim$^{1}$
\\
$^1$ Korea Astronomy and Space Science Institute, 776 Daedeokdae-ro, Daejeon 305-348, Korea\\
$^2$ Department of Astronomy, University of Cape Town, Private Bag X3, Rondebosch 7701, South Africa \\
$^3$ Department of Physics and Astronomy, University of the Western Cape, Private Bag X17, Bellville 7535, South Africa \\
$^4$ Inter-University Institute for Data Intensive Astronomy, South Africa \\
$^5$ Department of Physics and Astronomy, University of Calgary, 2500 University Drive NW, Calgary AB, T2N 1N4, Canada\\
$^6$ National Centre for Radio Astrophysics, Tata Institute of Fundamental Research, Pune 411007, India\\
$^7$ National Radio Astronomy Observatory, 1003 Lopezville Road, Socorro, NM 87801, USA
$^8$ INAF - Istituto di Radioastronomia, via Gobetti 101, 40129 Bologna, Italy 
}

\date{Accepted 2025 May 6. Received 2025 May 2; in original form 2024 April 2}

\pubyear{2025}

\begin{document}
\label{firstpage}
\pagerange{\pageref{firstpage}--\pageref{lastpage}}
\maketitle
\begin{abstract}
This study investigates the radio spectral properties of \textit{K}$_{S}$-selected star-forming galaxies (SFGs) in the XMM-LSS field using extensive multiwavelength data. By employing various diagnostics, SFGs are distinguished from quiescent galaxies and AGN across seven redshift bins ($\rm{0.1\leq\,\textit{z}\,\leq\,3.0}$).
The broadband radio frequency spectral energy distribution is analysed at observer-frame frequencies from 144 to 1500 MHz using median stacking techniques correcting for median flux boosting. We investigate the relationship between the radio spectral index, $\alpha$ (where $S\propto\nu^{\alpha}$) and redshift ($z$). Our analysis reveals no significant inverse correlation between $\alpha$ and $z$, indicating that the radio spectrum remains independent with varying redshift.
We fit the stacked median radio SEDs with a power law (\textit{PL}), curved power law (\textit{CPL}) and double power law (\textit{DPL}) models. For the \textit{DPL} and \textit{CPL} models, we observe a consistent steepening of the low-frequency spectral index across all redshift bins. For the \textit{CPL} model, the curvature term $q$ is greater than zero in all redshift bins.
Model comparisons indicate that spectra are generally well fitted by all the models considered. At 1500 MHz, SFGs display both a steep synchrotron component and a flat free-free emission component, with a thermal fraction consistently around 11$\%$ to 18$\%$. 
Further deep radio observations, with higher resolution to better deal with source blending and confusion noise and wider frequency coverage to better separate non-thermal and thermal radio emission, are required to reveal the detailed physical processes, thus clarifying the nature of radio sources.
\end{abstract}



\section{Introduction}
The main mechanism of radio continuum emission in both Active Galactic Nucleus (AGN) and Star-forming Galaxy (SFG) is the synchrotron emission (e.g. see \citealt{1992ARA&A..30..575C,2011ApJ...740...20P,2013MNRAS.436.3759B}). However, various physical mechanisms contribute to shaping the emission spectrum. Unlike
AGN, where the ultimate bulk of the emission is from the central supermassive black hole, in SFGs, synchrotron emission resulting from relativistic plasma accelerated in supernova remnants associated with massive stars $\rm{(\textit{M}_{\star}\, \gtrsim\,8 \textit{M}_{\odot})}$ dominates \citep[see][for a review]{1992ARA&A..30..575C,Padovani2016}. The radio spectra of star-forming galaxies is typically
characterised as a power law ($\rm{\textit{S}_{\nu}\,\propto\,\nu^{\alpha}}$), which encodes information
about the thermal and non-thermal energetic processes that
power them.

Several studies have shown that SFGs exhibit complex radio spectral energy distributions but are generally considered to have a mean spectral index between $\rm{-0.8}$ and $\rm{-0.7}$ with a relatively small dispersion, 0.24 (e.g. see \citealt{1992ARA&A..30..575C,2001MNRAS.327..907J,2009MNRAS.397..281I}). The radio spectral energy distribution (SED) provides valuable information that can be used to differentiate between source types according to their dominant emission mechanisms (e.g. \citealt {2014MNRAS.439.1556M}). \citet{2007A&A...463..519B} and \citet{2008MNRAS38375G} combined 610 MHz and 1.4 GHz data, and
found evidence for flatter spectral indices  at sub-mJy radio fluxes, suggesting that core-dominated
radio-quiet AGN are playing a key role in the sub-mJy radio population. 

In recent decades, numerous studies have focused on characterising the radio  continuum spectra of SFGs to explore
the relative contributions of synchrotron and free-free emission, components \citep[see,][]{1988A&A...190...41K,1998AJ....116.1039R,2000ApJ...544..641H,2015AJ....149...32M,2017ApJ...836..185T,2018A&A...611A..55K,2019ApJ...875...80G}. Studies by \citet{2018MNRAS.474..779G} using radio continuum data between 70 MHz and 48 GHz found  SFGs  to have steeper spectral indices in the high-frequency range than in previous studies. Also, models
by \citet{2010ApJ...717..196L} predict steeper synchrotron
spectra in high redshift 'puffy' submillimeter galaxies, SMGs, than in compact starbursts.
\cite{2021MNRAS.507.2643A} used the early science data from the MeerKAT \citep{2016mks..confE...1J} International GHz Tiered Extragalactic Exploration (MIGHTEE) Survey  \citep{2016mks..confE...6J} as well as the rich ancillary data in the COSMOS field to select SFGs up to $\textit{z}\,\sim3$ and study their radio spectral properties. They found that on average, the radio spectrum of SFGs flattens at low frequency with median spectral indices of $\alpha(0.33\,-\,1.3\,GHz)\,=\,-0.59$ and $\alpha(1.3\,-\,3.0\,GHz)\,=\,-0.74$. 
By combining high-sensitivity  LOw Frequency ARray (LOFAR) 150 MHz, upgraded Giant Meterwave Radio Telescope (\textit{u}GMRT) 400 MHz and 1,250 MHz, GMRT 610 MHz, and the Karl G. Jansky Very Large Array (VLA) 5 GHz data in the ELAIS-N1 field, \cite{2024MNRAS.528.5346A} studied the radio spectral properties of radio-detected SFGs at observer-frame
frequencies of 150 – 5000 MHz. They reported that on average, the radio spectrum of SFGs is flatter at low frequency than at high frequency.
 
\citet{2019A&A...621A.139T} constrained the shape of the average radio SED of a 1.4 GHz-selected sample $\rm{(\textit{SFR} > 100\,\textit{M}_{\odot}yr^{-1})}$ of highly star-forming galaxies (HSFGs) in the 0.4 - 10 GHz frequency range within the COSMOS field. Using survival analysis and fitting a double power law SED, \citet{2019A&A...621A.139T} found that the steepens from a spectral index of -0.51 below 4.5 GHz to -0.98 above 4.5 GHz.

Deep low-frequency ( < 1 GHz) radio surveys are at least one order of magnitude more sensitive for objects with typical synchrotron spectra than previous wide-area high-frequency surveys ($\rm{>\,1.0\,GHz}$) \citep[see,][]{2022A&A...663A.153R}.
Henceforth, deep radio continuum surveys are essential for
our understanding of physical processes at work in SFGs from the local to the distant Universe \citep[see,][]{2008MNRAS38375G,2009MNRAS.397..281I,2010MNRAS.402.2403M,2017A&A...602A...1S,2020MNRAS.491.1127O,2021A&A...648A...5M,2023MNRAS.520.2668H}. A known tool to average together data for a given set of objects is stacking. By employing a 'stacking' methodology to
recover sufficient signal-to-noise ratios on faint objects, one can understand what lies beneath the survey sensitivity limit.

\citet{2023MNRAS.524.5229O} performed a stacking analysis on deep and wide 610MHz data from the Giant Metrewave Radio Telescope (GMRT) in the ELAIS N1 field to measure the average radio luminosity and derive average SFRs of galaxies as a function of stellar mass and redshift ($\rm{\textit{z}<\sim 1.5}$). 
New radio telescopes are in the process of conducting major continuum surveys with equivalent depths of less than a few tens of $\mu$Jy/beam at 1.4 GHz over large regions of the sky. These surveys will be dominated by SFGs and a key requirement in extracting the maximum scientific return will be identifying and separating these objects from other classes such as radio-quiet and radio-loud AGN.
The X-ray Multi-Mirror Mission (XMM) Large-Scale Structure (XMM-LSS) field is one of the best-studied extragalactic deep fields with enhanced  multi-wavelength data products publicly available for use by the community. 
The field is also one of the Deep-Drilling Fields (DDFs)  \citep[DDFs;][]{2018arXiv181106542B,2018arXiv181200516S} selected as targets for the Vera C. Rubin Observatory Legacy Survey of Space and Time (LSST). Taking advantage of our broad spectral coverage of surveys targeting the XMM-LSS, we identify and study the broadband radio spectra of median stacked SFGs. We use new low to mid-frequency observations provided by \citep{2019A&A...622A...4H} at 144 MHz, and the \textit{super}MIGHTEE Pilot Project Data (300 - 950 MHz). In addition to that, we also use 1284 MHz and 1500 MHz data from \citet{2022MNRAS.509.2150H} and 
\citet{2020MNRAS.496.3469H} respectively.  Figure~\ref{stack_sens_res.fig} shows the sensitivity and resolution of these XMM-LSS radio surveys.

A well-determined radio spectrum for SFGs is critically important for studies that are based on the rest-frame radio power, especially those at high redshift, which are most sensitive to
the assumed \textit{k}-corrections. The study of the radio spectral index has a clear bias introduced
by the detection threshold in each map. In order to reduce the effect of this bias, many deep multi-frequency radio imaging restrict their statistical analysis to those sources with flux density ($S$) above a certain flux  density limit \citep[see][and references therein]{2009MNRAS.397..281I,2016MNRAS.463.2997M,2020MNRAS.491.1127O,2022ApJ...938..152D}.
It is the goal of this paper to investigate the shape of the radio continuum spectrum of SFGs below the survey noise threshold by binning our sample into different redshift bins. In this study, we selected SFGs from the SPLASH-SXDF Multi-Wavelength Photometric Catalogue \citep{2018ApJS..235...36M}. Deep radio
observations in this field, as well as multi-wavelength observations such the \citet{2018ApJS..235...36M} catalogue will enhance our knowledge of these radio
galaxies, and our understanding of galaxy evolution.
By studying the spectral shape of the total continuum radiation, between 144 MHz and 1500 MHz using median stacking techniques up to $\rm{\textit{z}\simeq3}$, we can comprehend the mechanisms that characterise the radio spectra at low and high frequencies, and their correlation with emission in other wavebands.  
The aim is to reach much lower noise levels, providing a statistical detection of complete samples that includes sources that are individually undetected in the original images.  

Is there an evolution in the radio spectral indices with redshift? This is the principal question we address in this work.

This paper is arranged as follows. 
In Section~\ref{data.sec}, we
present the radio and other ancillary data used for this work. Section~\ref{method.sec},
discusses the compilation of our radio galaxy sample, the classification of \textbf{SFG} candidates. In Section~\ref{results.sec}, we describe modelling the  stacked radio SEDs of our classified SFGs. In Section~\ref{res_disc.sec}, we discuss the possible physical mechanisms that determine the radio spectra over the  frequency range of this study. Finally, our results are concluded in Section~\ref{conclusion.sec}. 
We adopt the convention that the flux density \textit{S} is proportional to the observing frequency $\rm{\nu}$ according to: $\rm{\textit{S}_{\nu}\,\propto\,\nu^{\alpha}}$.
We assume a flat cold dark matter ($\rm{\Lambda}$CDM) cosmology with  $\rm{\Omega_{\Lambda} \ = \ 0.7}$, $\rm{\Omega_{m} \ = \ 0.3}$ and $\rm{\textit{H}_{o} \ = \ 70 \ km\,s^{-1} \ Mpc^{-1}}$ for calculation of intrinsic source properties.

\section{DATA}\label{data.sec}

To evaluate the radio SED using stacking techniques, we combined radio images in the XMM-LSS field obtained at different frequencies; 144 MHz with the LOFAR \citep{2019A&A...622A...4H}, \textit{super}MIGTHEE Band-3 (300 - 500 MHz) and Band-4 (550 - 950 MHz) with the  \textit{u}GMRT, 1284 MHz  with MIGHTEE \citep{2022MNRAS.509.2150H} and 1500 MHz obtained with the  VLA \citep{2020MNRAS.496.3469H}. 

\begin{figure*}
\centering
\centerline{\includegraphics[width=0.95\textwidth]{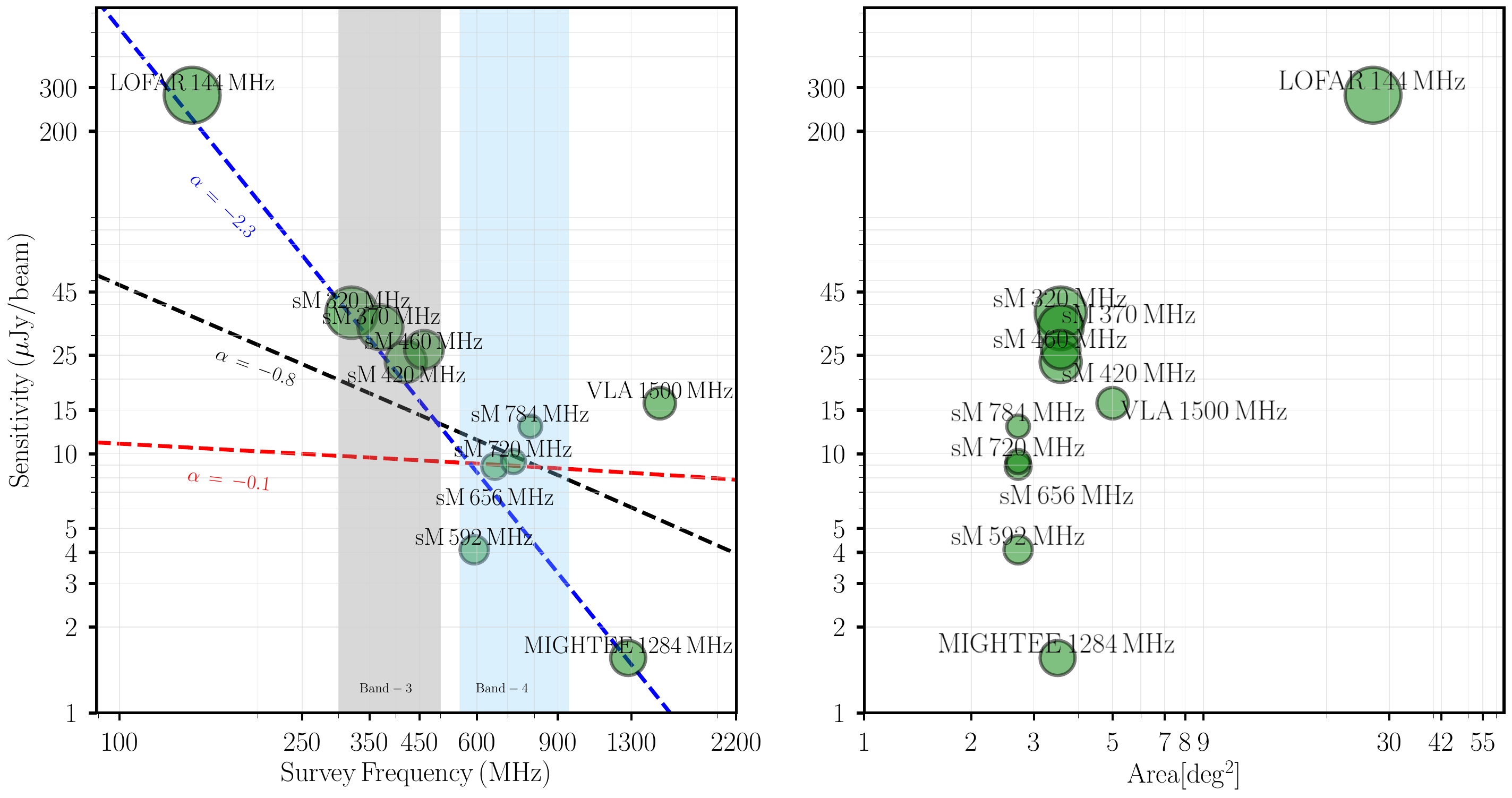}}
    \caption{Left: Comparison of sensitivities and survey frequencies of the XMM-LSS radio surveys used in this work. The sizes of the green circles are proportional to the survey resolution as shown in Table~\ref{tab_data}. The dashed black, red and blue lines represent  fiducial sources with a spectral index $\rm{\alpha\,=\, -0.8}$ (synchrotron), $\rm{\alpha\,=\, -0.1}$ (free-free emission), and $\rm{\alpha\,=\, -2.3}$ (ultra-steep), respectively.  The grey and lightblue shaded areas represent the \textit{super}MIGHTEE Band-3 and Band-4 frequency range respectively. Right: Sky area versus sensitivity for the selected radio surveys  we use in this study.}
\label{stack_sens_res.fig}
\end{figure*}

A summary of the observational parameters for these radio surveys is presented in Table~\ref{tab_data}.
The left panel of Figure~\ref{stack_sens_res.fig} is a pictorial representation of the sensitivities and frequencies of the XMM-LSS radio surveys used in this work. The surveys also have
comparable angular resolution and similar rms noise sensitivity for
sources with $\rm{\alpha\,=\,-2.3}$ with the exception of VLA data at 1.5 GHz. The right panel shows the sky area versus sensitivity ($\rm{\sigma_{rms}}$) for the selected radio surveys  we use in this study. The XMM-LSS field is also covered by legacy data sets at X-ray, ultraviolet, and infrared wavelengths. Multi-wavelength data is used for identifying and classifying the sources.

\begin{table*}
 \centering
 \caption{Table summarising the observational parameters for the radio surveys in the XMM-LSS field.}
 \scalebox{0.95}{
\begin{tabular}{c cccccc}
\toprule
Frequency, ($\rm{\nu}$) & Min RMS & Median RMS & Area&RA&DEC&Beam size\\ 

MHz&$\rm{\mu Jy\,bm^{-1}}$&$\rm{\mu Jy\,bm^{-1}}$&$\rm{sq\,deg}$&&&$(arsec^{2})$\\
\midrule
$\rm{144}$& 280 &400 &$\sim$27& 02h20m00.00s& -04$^{\circ}$30$^{\prime}$00$^{\prime\prime}{.}$0&$\rm{7.5\times8.5}$ 
\\
$\rm{320}$& 37.1 &133.1 &$\sim$3.57&02h20m40.00s& -04$^{\circ}$49$^{\prime}$59$^{\prime\prime}{.}$0&$\rm{9.53\times5.66}$ 
\\
$\rm{370}$&32.4  &123.0 &$\sim$3.57&02h20m40.00s& -04$^{\circ}$49$^{\prime}$59$^{\prime\prime}{.}$0&$\rm{8.40\times4.89}$
\\
$\rm{420}$&23.5  &100.1 &$\sim$3.57&02h20m40.00s& -04$^{\circ}$49$^{\prime}$59$^{\prime\prime}{.}$0&$\rm{7.56\times4.57}$ 
\\
$\rm{460}$&26.3  &115.9 &$\sim$3.57&02h20m40.00s& -04$^{\circ}$49.59$^{\prime\prime}{.}$0&$\rm{7.16\times4.25}$ 
\\
$\rm{592}$&4.1  &50.0 &$\sim$2.71&02h20m40.00s& -04$^{\circ}$49$^{\prime}$59$^{\prime\prime}{.}$0&$\rm{4.73\times3.19}$ 
\\
$\rm{656}$& 8.9 &48.2 &$\sim$2.71&02h20m40.00s& -04$^{\circ}$49$^{\prime}$59$^{\prime\prime}{.}$0&$\rm{4.31\times3.19}$ 
\\
$\rm{720}$& 9.3 & 57.3&$\sim$2.71&02h20m40.00s& -04$^{\circ}$49$^{\prime}$59$^{\prime\prime}{.}$0&$\rm{4.02\times2.97}$ 
\\
$\rm{784}$&12.9  &100.2 &$\sim$2.71&02h20m40.00s& -04$^{\circ}$49$^{\prime}$59$^{\prime\prime}{.}$0&$\rm{3.79\times2.78}$ 
\\
$\rm{1284}$&1.5  &10.0 &$\sim$3.5&02h20m36.53s&-04$^{\circ}$50$^{\prime}$01$^{\prime\prime}{.}$32&$\rm{8.0\times8.0}$
\\
$\rm{1500}$&10.6  &16 &$\sim$5&02h21m59.97s&-04$^{\circ}$45$^{\prime}$10$^{\prime\prime}{.}$18&$\rm{4.5\times4.5}$ 
\\
 \hline
 \end{tabular}}
 \label{tab_data} 
 \end{table*}

\subsection{The \textit{super}MIGHTEE Pilot Project Data}
The \textit{super}MIGHTEE Pilot Project  (Lal et al., submitted) observations target the MIGHTEE XMM-LSS early science region. This programme is with the upgraded GMRT (\textit{u}GMRT) designed exploit  flux density levels about the depths of MIGHTEE  Survey \citep{2016mks..confE...6J} at lower frequencies to study  radio galaxies. The region is covered by a mosaic of 4 pointings at Band-3 and 19 pointings at Band-4. 
The calibration consisted of initial flagging, parallel-hand (Stokes I) calibration, a second round of flagging on the calibrated data in order to identify fainter radio frequency interference (RFI). This was followed by full-Stokes calibration, using a recipe suitable for circular feeds. The primary calibrator \texttt{3C147} was used to solve for polarisation leakages, and \texttt{3C286} was used as the absolute flux and polarisation angle calibrator.
The Band-3 RF covers the frequency range 300 to 500 MHz. Observations were taken in full Stokes mode with 8192 channels. 
The Band-4 RF covers the frequency range 550 to 950 MHz and was also observed in in 8194 channels in full Stokes mode. Above $\sim$800 MHz, we observed the gain is very low, and there are bands of strong persistent radio frequency interference (RFI) above 830 MHz. We processed 266 MHz of the RF from 550 to 816 MHz.
The mosaic images were created by combining images from each field using the \textsc{linearmosaic}
tool in \textsc{casa}. The mosaic weights and primary beam corrections were constructed using the circularly symmetric polynomial formulation of the \textit{u}GMRT  band-3 and band-4 wide-band primary beams provided by \cite{8738257}.
The wide-band images and sources lists provide our most sensitive flux density measurements and allow us to probe deeply into the spectral properties of the radio source population. However, each of the two band covers a broad RF and provides spectral information within the band. We can thus construct more detailed spectral energy distributions for the stronger sources by dividing the RF into sub-bands. To this end, each of the Band-3 and Band-4 RF have been divided into four sub-bands of approximately 50 MHz and mosaics created. 

\subsection{Archival Radio Survey Data}
\subsubsection{LOFAR 144 MHz data}

\citet{2019A&A...622A...4H}  presented observations of the XMM-LSS field observed with the LOFAR at 120-168 MHz. In order to remove areas with high primary beam corrections, where the telescope is less sensitive, their final image was generated over a reduced region of the primary beam corrected image. Regions with a primary beam correction of 2 or less (i.e. the primary beam response $\ge50\%$ or more of the response in the centre of the image) were used in the final image. This produced a $\rm{\sim27\,deg^{2}}$ elliptical image of the sky around the XMM-LSS field at \textbf{$\rm{7.5\times8.5\, arcsec^{2}}$ resolution}. The ellipse is aligned in the north-south direction. This image has a central frequency of $\rm{\nu\,=\,144\,MHz}$ and a central rms of 280 $\rm{\mu}$Jy beam$^{-1}$.

The image centred at a \textit{J2000} declination of -4.5$^{\circ}$ has a low elevation, reducing the effective collecting area of the telescope, thereby reducing sensitivity. 

\subsubsection{MIGHTEE 1.28 GHz data}

The description and validation of the MIGHTEE total intensity radio continuum imaging in the XMM-LSS Early Science field is outlined in \citet{2022MNRAS.509.2150H}. The MIGHTEE total intensity continuum image of the XMM-LSS consists of a three-pointing mosaic covering an area of $\sim3.5$ deg$^2$. The images are of both the low resolution / high sensitivity variants, imaged with a robust weighting value of 0.0, with angular resolutions of $\rm{8^{\prime\prime}.2}$ for XMM-LSS.
The processing includes the correction of direction-dependent effects, with the deepest combined part of the robust 0.0 XMM-LSS mosaic reaches 1.5 $\rm{\mu}$Jy beam$^{-1}$. 

The convolved image products for each pointing were corrected for primary beam attenuation by dividing them by a model of the Stokes I primary beam pattern, evaluated at 1284 MHz using the \textsc{eidos} \citep{2021MNRAS.502.2970A} package.  

\subsubsection{VLA 1.5 GHz data}
For our highest frequency data, we used the Karl G. Jansky Very Large Array (VLA) 1.5 GHz dataset from \citet{2020MNRAS.496.3469H}. This is a new survey undertaken with the VLA of the XMM-LSS/VISTA Deep Extragalactic Observations (VIDEO) survey field \citep{2013MNRAS.428.1281J} at 1–2 GHz using the B-configuration. This survey significantly extends the areal coverage over this field, which also incorporates the VVDS region \citep{2003A&A...403..857B} and the UKIDSS-UDS field \citep{2006MNRAS.372..741S}.

The image covers $\sim5$ deg$^2$ with a median depth of 16 $\rm{\mu}$Jy beam$^{-1}$, with almost 80 per cent of the mosaic area having a noise value of < 20 $\rm{\mu}$Jy 
beam$^{-1}$ and an angular resolution of $\rm{4.5^{\prime\prime}}$.  

In order to suppress error patterns from off-axis sources of even modest brightness, \citet{2020MNRAS.496.3469H} applied direction-dependent calibration and wide-field imaging.

\subsection{Optical/near-infrared data}
\subsubsection{The SPLASH-SXDF Multi-Wavelength Photometric Catalogue}\label{photo.sec}
We use the photometric catalogue for the \textit{Spitzer}
Large Area Survey with Hyper-Suprime-Cam (SPLASH) Subaru/\textit{XMM}-Newton Deep  Survey Field (SXDF), one of the deep fields with the largest contiguous area covered over a wide wavelength range produced by \citet{2018ApJS..235...36M}. The SPLASH-SXDF Catalogue combines Optical/Near-Infrared/IRAC data, and computes photometric redshifts as well as physical properties  using LePhare \citep{1999MNRAS.310..540A,2006A&A...457..841I} for about 1.17 million sources detected over 4.2 deg$^2$ centered at $\rm{\alpha, \delta = 02^{h}18^{m}00^{s}, -5^{\circ}00^{\prime}00^{\prime\prime}}$. This data is a publicly available resource, ideally suited for studying galaxies in the early universe and tracing their evolution through cosmic time. The catalogue is obtained by exploiting the extensive multi-wavelength coverage, hence measuring accurate photometric redshifts for all sources. The photometric redshifts are calibrated using $\sim$10,000 reliable spectroscopic redshifts available from various surveys. The magnitudes and fluxes in the final catalogue are corrected using these values and the Galactic reddening \textit{E(B - V )} values. Numerous spectroscopic surveys cover the SXDF and hence a large number of objects have spectroscopic redshifts available.  The catalogue assembles a sample of 12,342 galaxies with reliable spectroscopic redshifts covering the full range of $\rm{0 \,<\,\textit{z}\,<\,6}$. From the catalogue, \citet{2018ApJS..235...36M} selected 8647 galaxies that
are within the Hyper-Suprime-Cam - Ultra Deep (HSC-UD) area and have proper wavelength coverage (particularly in the \textit{NIR}) to use as the calibration sample for the photometric redshifts. Using the normalised median absolute deviation \citep[$\sigma_{\text{NMAD}}$,][]{1983ured.book.....H} and outlier fraction (O$_{f}$) to quantify the performance of the
photometric redshifts as a function of the i-band magnitude, they found excellent agreement over the full redshift range, with a
computed $\sigma_{\text{NMAD}}$ of 0.023 and O$_{f}$ of
$3.2\%$ for sources within the HSC-UD area.

\section{Methodology}\label{method.sec}

In this section, we discuss the methodology we follow for this work, mainly the prescription we apply in order to construct our sample. We furthur discuss the different selection criteria  we use for selecting our final sample of SFGs used in this investigation. We describe the stacking procedure we use to investigate the radio SED properties of galaxies at far greater sensitivity by combining many observations of individual galaxies at the expense of any specific knowledge of the individual galaxies. Stacking offers high sensitivity to the median signal of a class of radio sources. 

\subsection{Sample construction}\label{pres.sec}

Our methodology for compiling a sample of \textit{K$_{s}$}-selected galaxies, and classifying their activity status, is presented below. {We group galaxies according to redshift:
\begin{enumerate}
  \item[-] Using seven redshift bins: $\rm{0.1\,<\,\textit{z}_{1}\,<\,0.5}$, $\rm{0.5\,<\,\textit{z}_{2}\,<\,0.9}$, $\rm{0.9\,<\,\textit{z}_{3}\,<\,1.3}$, $\rm{1.3\,<\,\textit{z}_{4}\,<\,1.7}$,
  $\rm{1.7\,<\,\textit{z}_{5}\,<\,2.1}$,
  $\rm{2.1\,<\,\textit{z}_{6}\,<\,2.5}$
and $\rm{2.5\,<\,\textit{z}_{7}\,<\,3.0}$.
\end{enumerate}
Table~\ref{tabppts} presents the summary of the results of subsequent analysis of the median redshift, the number of galaxies, the mean stellar mass, and SFR for all the sources in each redshift range.
\citet{2018ApJS..235...36M} issue a caution that their \textit{SFR} values are computed without \textit{FIR} information and hence, values can be highly uncertain.
We then remove sources that could be spectroscopically and photometrically flagged as stars using the designated flag option from the \citet{2018ApJS..235...36M}. Objects were flagged as stars if the $\rm{\chi^{2}_{star}\,<\,\chi^{2}_{gal}}$ (i.e. $\rm{\chi^{2}_{star}}$ and $\rm{\chi^{2}_{gal}}$ represent the $\rm{\chi^{2}}$ minimisation of stars and galaxies, respectively), with a further restriction that the object is in the \textit{BzK} stellar sequence (i.e., $\rm{\textit{z}\,-\,\textit{Ks}\,<\,(\textit{B}\,-\,\textit{z})\times\,0.3\,-\,0.2}$, see \citealt{2018ApJS..235...36M}).
The data is separated into seven redshift bins from $\rm{0.1\leq\,\textit{z}\,\leq\,3.0}$ by using a 0.4 dex bins (with an exception to the last redshift bin  which is 0.5 dex). 

The redshifts in the catalogue are spectroscopic (spec-\textit{zs}) when available, otherwise, photometric (photo-\textit{zs}) are adopted.
\begin{figure}
\centerline{\includegraphics[width=0.5\textwidth]{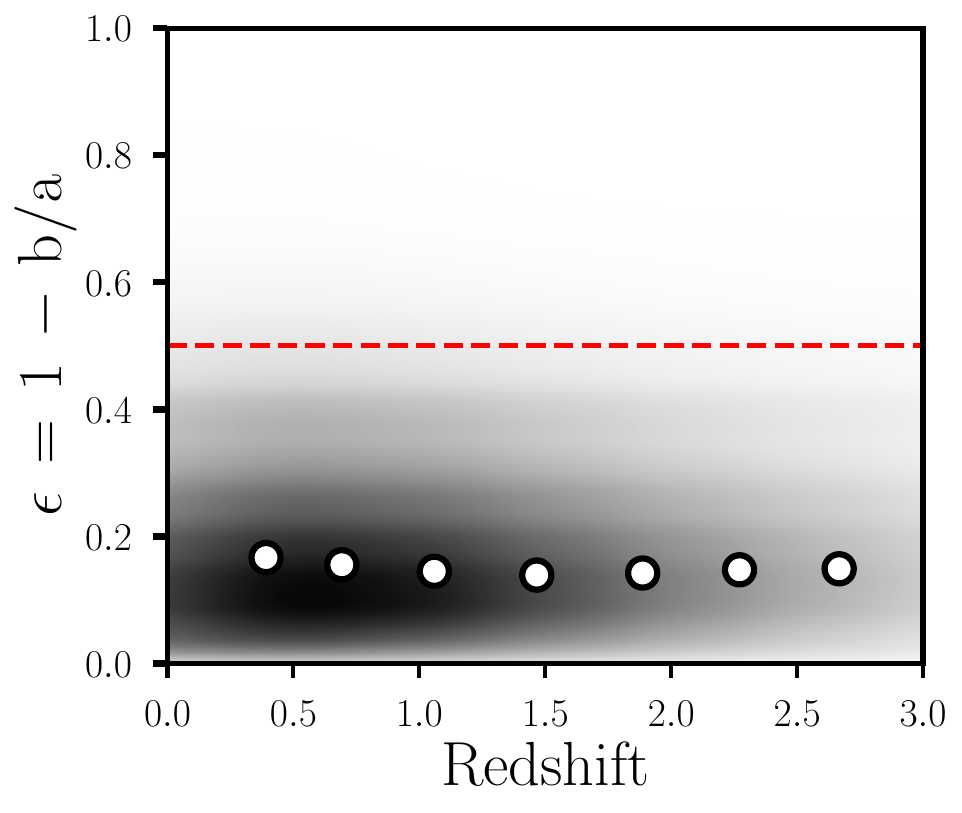}}
    \caption{The scatter plot of the ellipticity, $\rm{\epsilon\,=\,1\,-\,\textit{b}/\textit{a}}$, against redshift for galaxies $\rm{0.1\leq\,\textit{z}\,\leq\,3.0}$. The dashed horizontal red line shows $\rm{\epsilon\,=\,1\,-\,\textit{b}/\textit{a}\,>\,0.5}$, which is the threshold to select elongated galaxies. 
    The median value (white filled circles) of ellipticity in each redshift bin is also plotted for comparison.
   }
    \label{ellipticity.fig}
\end{figure}

We then applied the prescription below to select our sample:

\begin{multline}\label{eqn1}
\rm{\left(\frac{\textit{z}_{med,u68}\,-\,\textit{z}_{med,l68}}{1+\textit{z}_{med}}\right)\times\,0.5<0.1\,\&\,\left(\frac{\textit{S}_\textit{Ks}}{\textit{S}_\textit{Ks,err}}\right)>5\,\&\,}
\\
(0.1\,\leq\,\textit{z}_{phot}\,\leq 3.0)
\end{multline}

\footnote{$\rm{\textit{z}_{med}\,-\,Photo-z\,from\ median\, of\,\textit{P(z)}\,distribution}$
$\rm{\textit{z}_{med,u68}\,-\,Upper\,limit\,of\,68\%\,CI\,for\,the\, \textit{z}_{med}}$
$\rm{\textit{z}_{med,l68}\,-\,Lower\,limit\,of\,68\%\,CI\,for\,the\,\textit{z}_{med}}$
$\rm{\textit{z}_{phot}\,-\,Best\,estimate\,of\, redshift\,including\,spectroscopic\,
redshifts\,when\,available}$}.
The $\rm{\textit{S}_\textit{Ks}}$ and $\rm{\textit{S}_\textit{Ks,err}}$ represent  \textit{K}$_{S}$-band flux and flux error respectively.
By applying the prescription outlined above, we  select $\sim144,736$ galaxies in total. For our sample, we select galaxies with \textit{K}$_{S}$ mag  with  a 5$\sigma$ depth in 2$^{\prime\prime}$  aperture size $\leq$ 23.68 AB \citep[from VIDEO,][]{2013MNRAS.428.1281J}. 

We then adopt a  uniform ellipticity distribution from 0 to 0.5 for the sources, where this adoption is directly used to remove galaxies with observed elongated shapes \citep{2005A&A...442..405P,2009MNRAS.398.1298P,2009ApJ...707..472G}. \citet{2009ApJ...707..472G} confirmed that shape distribution of real galaxies has a dramatic effect on the lensing probability, where lensing probability is boosted by a factor of 2 relative to that based on simple assumptions of the shape distribution (e.g. random ellipticity between 0 and 0.5).
Figure~\ref{ellipticity.fig}
presents a scatter plot of the ellipticity, $\rm{\epsilon\,=\,1\,-\,\textit{b}/\textit{a}}$, against redshift for galaxies $\rm{0.1\leq\,\textit{z}\,\leq\,3.0}$. Where \textit{a} and \textit{b} represents the semi-major axis and semi-minor axis [px], respectively from the \citet{2018ApJS..235...36M} catalogue. The dashed horizontal red line shows $\rm{\epsilon\,=\,1\,-\,\textit{b}/\textit{a}\,>\,0.5}$, which is the threshold to select elongated galaxies. The median (white filled circles) of source below $\rm{\epsilon\,=\,1\,-\,\textit{b}/\textit{a}\,<\,0.5}$ are indicated in seven bins of redshift. From Figure~\ref{ellipticity.fig}, while it is true that the ellipticity \textit{1 - b/a} of most sources is less than 0.5, there is still a fraction of sources, with ellipticity larger than 0.5. The sources with large ellipticities, such as edge-on spiral galaxies, could be stretched much more easily, and could eventually affect the stack images by forming artefacts around the central bright source.

By applying the criteria for ellipticity and the prescription in Equation~\ref{eqn1}, we  select $\sim141, 994$ galaxies.
In the next subsection, we describe how we distinguish SFGs from quiescent galaxies (QGs) and also the removal of objects hosting an AGN.

\begin{table}
 \centering
 \caption{Table showing the summary of the results of subsequent analysis of the median redshift, the number of galaxies, the mean stellar mass and SFR  for the all sources in each redshift range.}
 \scalebox{0.98}{
 \begin{tabular}{ccccc}
 \hline
 \hline
\textit{z} range& $\rm{\textit{z}_{med}}$ &$\rm{N_{gal}}$& $\rm{\left(\log\frac{M_{\star}}{M_{\odot}}\right)}$ & $\rm{\left(\log \frac{SFR_{UV}}{M_{\odot}yr^{-1}}\right)}$
\\
\hline
$\rm{0.1<z<0.5}$ & $0.37^{+0.09}_{-0.15}$& 19267& $\rm{8.67^{+1.15}_{-0.71}}$& $-0.73^{+1.02}_{-1.31}$
\\ 
$\rm{0.5<z<0.9}$&$0.69^{+0.14}_{-0.13}$&35041&$\rm{9.18^{+0.98}_{-0.61}}$&$0.33^{+0.88}_{-0.93}$ 
\\
$\rm{0.9<z<1.3}$&$1.06^{+0.14}_{-0.10}$&35111&$\rm{9.46^{+0.87}_{-0.55}}$&$0.71^{+0.74}_{-0.59}$
\\
$\rm{1.3<z<1.7}$&$1.47^{+0.13}_{-0.09}$&22899&$\rm{9.81^{+0.72}_{-0.57}}$&$0.92^{+0.66}_{-0.56}$ 
\\
$\rm{1.7<z<2.1}$&$1.89^{+0.14}_{-0.14}$&11692&$\rm{9.96^{+0.65}_{-0.57}}$&$1.28^{+0.61}_{-0.57}$ 
\\
$\rm{2.1<z<2.5}$&$2.27^{+0.14}_{-0.13}$&10544&$\rm{9.80^{+0.72}_{-0.49}}$&$1.29^{+0.57}_{-0.56}$ 
\\
$\rm{2.5<z<3.0}$&$2.68^{+0.16}_{-0.14}$&7440&$\rm{9.90^{+0.62}_{-0.45}}$&$1.48^{+0.56}_{-0.55}$ 
\\
\\
 \hline
 \end{tabular}}
 \begin{tablenotes}
\item N.B. The stellar physical properties are derived from the \citet{2018ApJS..235...36M} SPLASH-SXDF  catalogue.

\end{tablenotes}
 \label{tabppts} 
\end{table}

\subsection{Source Classification}\label{src.sep}
\subsubsection{Star-forming Galaxy Selection}\label{sfg.sec}
Classification of extragalactic sources is fundamental for statistical analyses of large populations, as well as probing the properties of individual objects.
At high redshifts most
galaxies are forming stars, since the fraction of SFGs
increases rapidly with redshift \citep[e.g.,][and references therein]{2013A&A...556A..55I} and,
 even radio galaxies at $\rm{\textit{z}\,\gtrsim\,3}$ undergo vigorous star formation \citep[e.g.][]{2008A&ARv..15...67M}. However, disentangling the radio flux contribution from star formation (SF) and active-galactic-nuclei (AGN) activity is a long-standing
problem in extragalactic astronomy.
 The use of \textit{UV}-optical colors has emerged as a commonly employed proxy 
 \citep[see,][]{2011ApJ...735...53P,2012ApJ...748L..27P} in distinguishing quiescent galaxies from SFGs. The rest-frame \textit{u - V} vs \textit{V - J} diagram allows for a cleaner separation
of star-forming and quiescent galaxies because colors are
defined in the rest-frame rather than observed frame \citep[see,][]{2009ApJ...691.1879W,2009ApJ...706L.173B,2011ApJ...739...24B,2012ApJ...754L..29W,2013ApJ...767...39M,2014ApJ...783L..14S,2015A&A...575A..74S,2019ApJ...880L...9L}.

The \textit{UVJ} colors are efficient in distinguishing SFGs from quiescent ones, since \textit{U - V} colors generally exhibit unobscured star formation. Whilst, quiescent galaxies generally have relatively lower dust attenuation and tend to have bluer \textit{V - J} colors \citep[see][]{2019ApJ...880L...9L}.

We used the \citet{2015A&A...575A..74S} color-color definition, at all redshifts up to $\rm{\textit{z}\simeq3}$ to separate SFGs from QGs:

\begin{equation}
Quiescent = \begin{cases*}
  \textit{u} - V > 1.3 ,\\
  V - J < 1.6  ,\\
  \textit{u} - V > 0.88 \times (V - J) + 0.49 .
\end{cases*}
\end{equation}

\begin{figure*}
\includegraphics[width=0.75\textwidth]{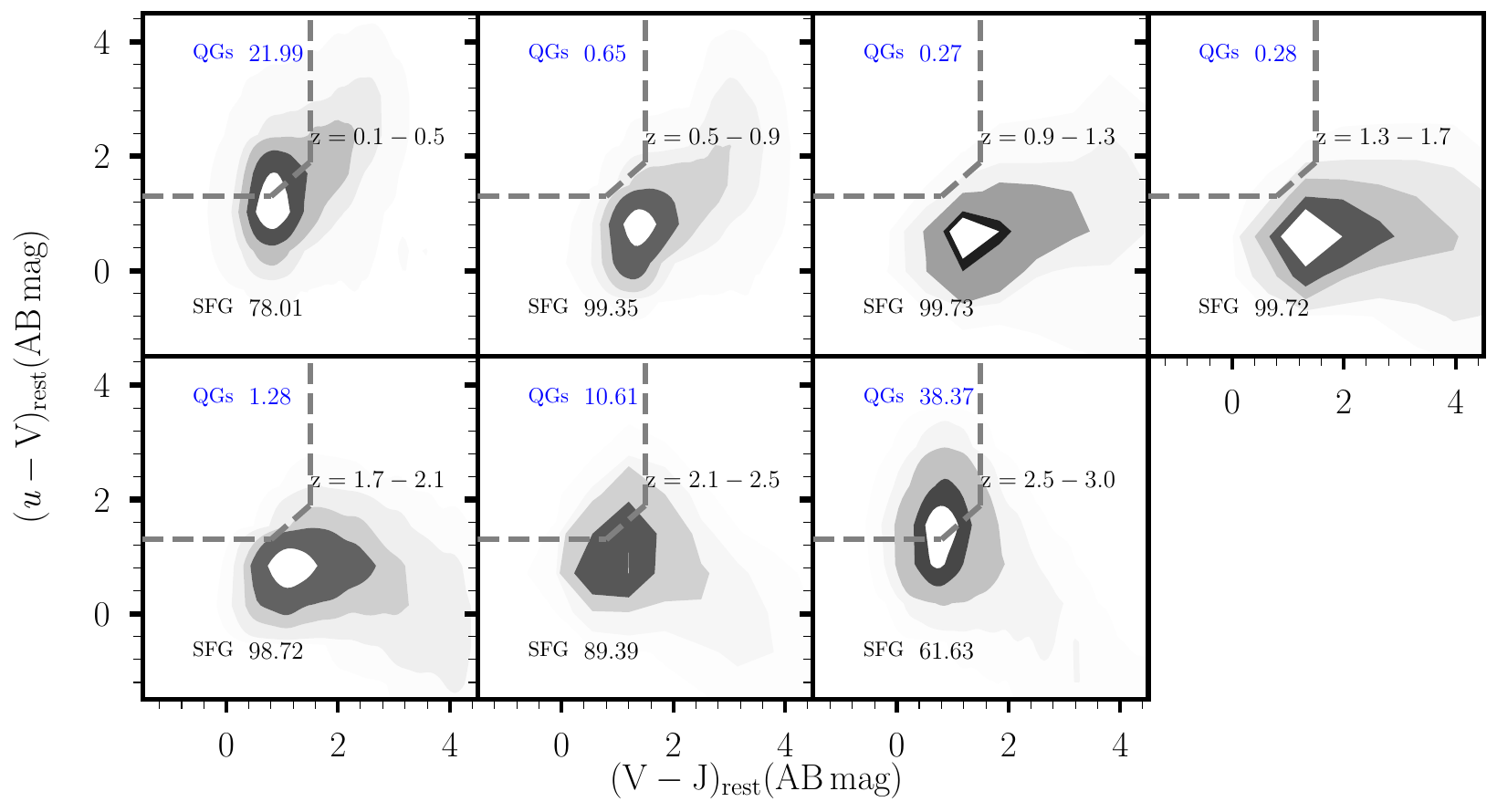}
    \caption{Rest-frame \textit{uVJ} diagram for galaxies in seven redshift bins. The dividing line between quiescent and star-forming galaxies is shown as a dashed black line on each plot, with quiescent galaxies located on the top-left corner. In each panel we show the percentage of quiescent (blue numbers), and that of SFGs (black numbers), respectively.}
    \label{uvj_z_bins.fig}
\end{figure*}

We use the rest-frame magnitude the \textit{u} \citep[CFHTLS,][]{2018PASJ...70S...8A}, \textit{V} \citep[Suprime-Cam,][]{2008ApJS..176....1F}, and \textit{J} \citep[VIDEO,][]{2013MNRAS.428.1281J} filters. 

Figure~\ref{uvj_z_bins.fig} presents the rest-frame \textit{UVJ} diagram for galaxies in seven redshift bins. Binning in redshift space allows us to have a good understanding of how these galaxies evolve.
The dividing line between quiescent and star-forming galaxies is shown as a dashed black line on each plot, with quiescent galaxies located on the top-left corner. The bimodality is not clearly evident in this diagram at all redshift bins. %
For each redshift bin (i.e. each panel in the plot), we write in text the percentage of QGs (blue text) and that of SFGs (black text). The number percentage of QGs decreases acutely from being $\sim22$ percent in the first redshift bin to $\sim0.27$ percent in the third bin. After this point, there is a  constant increase in the QGs population within the different redshift bins from $\sim0.28$ percent up to $\sim38.37$ percent at $\rm{2.5\,<\,\textit{z}\,<\,3.0}$. By applying the \textit{uVJ} color selection criterion, we  select $\rm{133,247}$ are SFGs out of the $141,994$ total number of galaxies.

\begin{figure}
\includegraphics[width=0.48\textwidth]{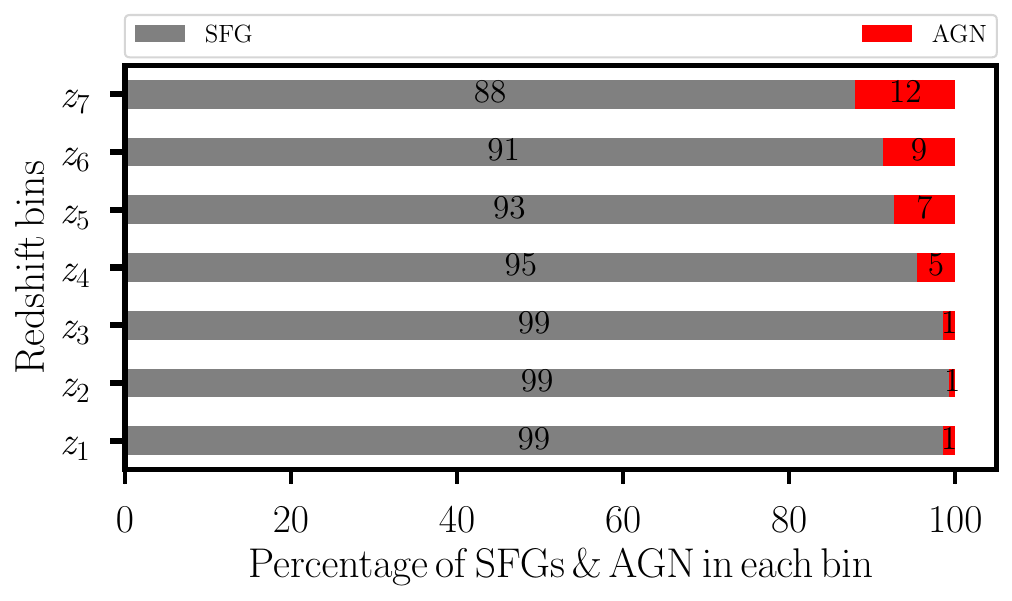}

    \caption{Stacked bar charts showing the discrete distributions (numbers) of the percentage of the different class of sources in each bin. SFGs and  AGN  are represented as grey horizontal bars and red horizontal bars, respectively. }
    \label{redshift.fig}
\end{figure}

We also removed potential AGN contaminated sources from the sample using the \citet{2012ApJ...748..142D} colour selection criterion. By applying this selection criterion to the $\rm{133,247}$ sources selected after the removal of QGS, we obtained 128,967 sources. Therefore, a total of 4280 sources were removed as potential AGN contaminated sources.

Figure~\ref{redshift.fig} presents stacked bar charts showing the discrete distributions (numbers) of the percentage of the different class of sources in each redshift bin.  The grey horizontal bars represent the percentage of SFGs in each redshift bin, whereas the red horizontal bars represents that of AGN,  respectively. It is important to note is that even after using the \textit{uVJ} colors to separate QGs from SFGs, applying the \citet{2012ApJ...748..142D} selection cuts, our subsample may still be contaminated by AGN.

Following the prescription in Equation~\ref{eqn1} and the adoptation of uniform ellipticity in subsection~\ref{pres.sec}, and  the   separation of quiescent/passively evolving galaxies from candidate SFGs as well as the removal of candidate AGN, we define our final sample selected for the subsequent analyses is a follows:

\begin{enumerate}
\item[1.] All Galaxies: the original 141,994 sources that satisfying Equation~\ref{eqn1} and the criteria for ellipticity separated into seven redshift bins.
\item[2.] SFGs: sources from the original 141,994 sample that are classified as SFGs based on the \textit{uVJ} galaxy color-color diagnostics and the \citet{2012ApJ...748..142D} selection cuts. Removing these flags, we obtain a subsample of 128,967 as SFGs.
\end{enumerate}

Table~\ref{tab_ppts} presents a summary of the results of subsequent analysis of the average galaxy properties in each redshift range when suspected QGS and AGN have been separated using \citet{2015A&A...575A..74S} and \citet{2012ApJ...748..142D}, respectively.

\subsubsection{Trends of the photometric redshifts for SFGs, AGN and QGs}
   
We verify the overall redshift accuracy of the photo-\textit{z} estimates as a function of $\textit{K}_{S}$ magnitude  for the selected sample  $\rm{0.1\leq\,\textit{z}\,\leq\,3.0}$, following the statistical metrics used to evaluate photometric redshift accuracy and quality 
\citet{2018MNRAS.473.2655D,2018MNRAS.477.5177D} for our selected QGs, SFGs and AGN. Appendix~\ref{AppendA} presents the subsets of the corresponding photo-\textit{z} quality metrics we use for each subsample. \citet{2018MNRAS.473.2655D} introduced the continuous ranked probability score ($\rm{CRPS}$) and the corresponding mean values, i.e. $\overline{\rm{CRPS}}$, for a given sample \citep[see,][for details and behaviour of the metric]{2000MWRv..128.3501H,2017IAUS..325..156P}. The main advantage of this metric is that it takes into account the full PDF rather than just a simple point value when evaluating a model prediction as noted by \citet{2018MNRAS.473.2655D,2018MNRAS.477.5177D}. 

Figure~\ref{z_quality.fig} illustrates the photometric redshift scatter ($\sigma_{\text{NMAD}}$), outlier fraction (O$_{f}$) and  $\overline{\rm{CRPS}}$ (see Appendix~\ref{AppendA}) performance redshift estimates in the range of this study. Across the entire $\textit{K}_{S}$ magnitude the scatter, $\sigma_{\text{NMAD}} \lesssim 0.1$ for both QG and SFG population but not the AGN population. We find that the measured O$_{f}$ for known AGN at a given $\textit{K}_{S}$ magnitude is higher than that of the QGs and SFG population. The highest $\sigma_{\text{NMAD}}$ and O$_{f}$ values measured for the photometric redshifts of our SFGs is 0.024 and 2.8$\%$ (0.028), respectively.

Comparing this to the original catalogue from \citet{2018ApJS..235...36M}, they quantified $\sigma_{\text{NMAD}}$ = 0.023 and O$_{f}$ of 0.032 for all galaxies within the HSC-UD area as described in subsection~\ref{photo.sec}.
The performance of the selected SFGs redshifts in the $\overline{\rm{CRPS}}$ is significantly better than that of the selected QGs and AGN redshift estimates. Overall, AGN have the worst photometric redshift performance as can be seen by numbers measured for each metric shown in red, blue and black
colors for AGN, QGS and SFGs, respectively, in each panel of Figure~\ref{z_quality.fig}.

\begin{figure}
\centering

\includegraphics[width=0.48\textwidth]{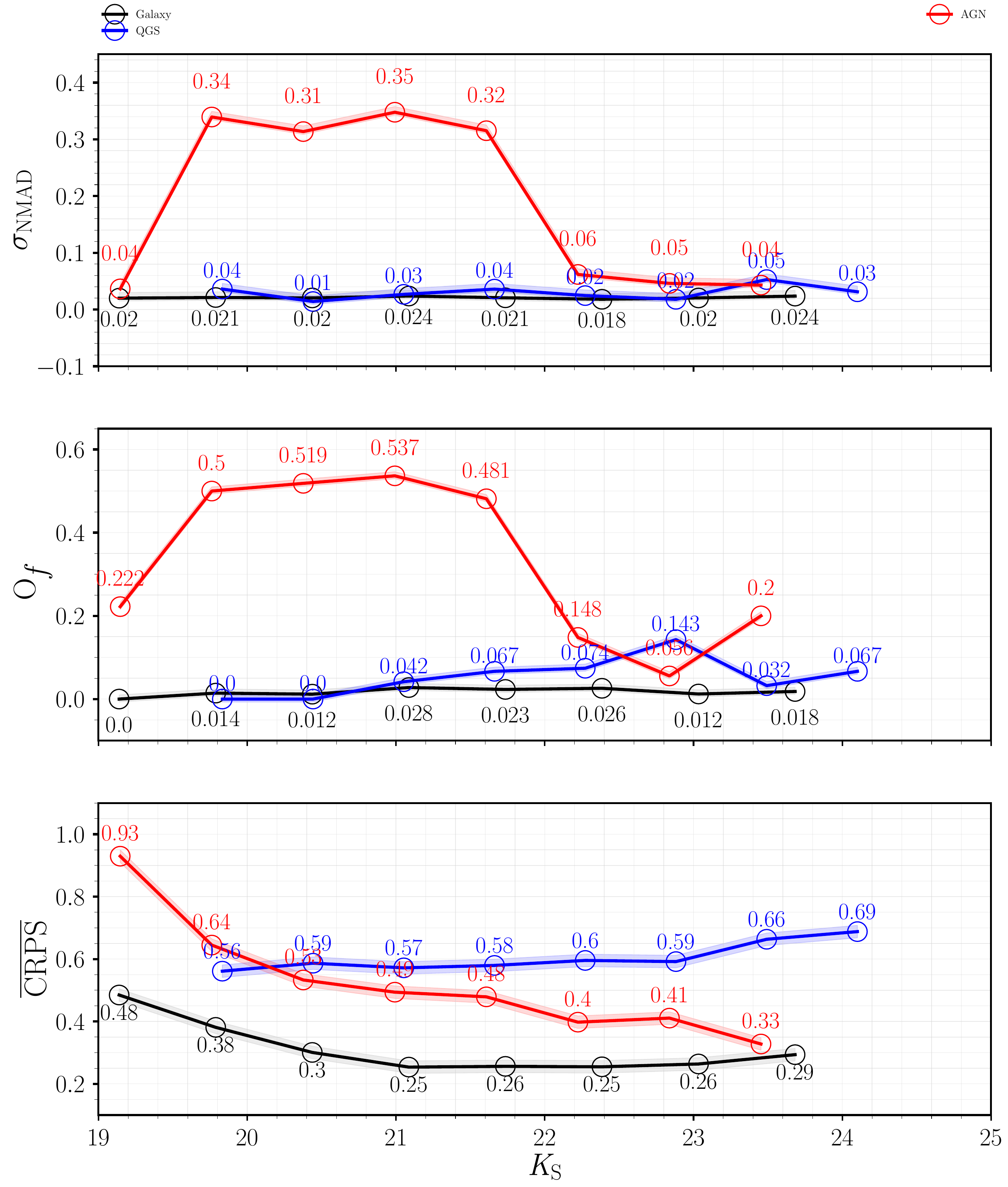}
    \caption{Photometric redshift scatter ($\rm{\sigma_{NMAD}}$), outlier fraction ($\rm{OLF}$) and mean continuous ranked probability score ($\rm{\overline{\rm{CRPS}}}$) as a function of \textit{K} magnitude. In all panels, solid black lines represent SFGs, solid red lines represent AGN, whereas solid blue lines represents QGs. At almost all redshift ranges, the photo-\textit{z} performance of SFGs is better to than that of the  photo-\textit{zs} of QGS. The numbers measured for each metric are shown in red, blue and black colors for AGN, QGS and SFGs, respectively, in each panel.
}
\label{z_quality.fig}
\end{figure}

\begin{table*}
 \centering
 \caption{Table showing the summary of the results of subsequent analysis of the \text{median} galaxy properties  of classified sources in each redshift range. These avaerage galaxy physical properties are obtained from  the \citet{2018ApJS..235...36M} SPLASH-SXDF  catalogue.}
 \scalebox{0.81}{
 \begin{tabular}{c ccccc| ccccc| ccccc}
 \hline
 \hline
  &  \multicolumn{5}{c} {SFGs}  &  \multicolumn{5}{c} {QGs} &  \multicolumn{5}{c} {AGN}\\ 
\textit{z} bin &$\rm{z_{med}}$&$\rm{N}$&$\rm{R_{LUM}}$&$\rm{K_{LUM}}$&$\rm{NUV_{LUM}}$ & $\rm{z_{med}}$&$\rm{N}$&$\rm{R_{LUM}}$&$\rm{K_{LUM}}$&$\rm{NUV_{LUM}}$ &
$\rm{z_{med}}$&$\rm{N}$&$\rm{R_{LUM}}$&$\rm{K_{LUM}}$&$\rm{NUV_{LUM}}$\\
\\
 \hline
$\rm{\textit{z}_{1}}$ &$0.39^{+0.07}_{-0.14}$&  14806&$8.57^{+1.00}_{-0.67}$ &$7.71^{+1.12}_{-0.66}$&$8.44^{+0.94}_{-0.80}$ & 

$0.22^{+0.12}_{-0.07}$ & 4236 &$7.85^{+0.60}_{-0.58}$ & $7.00^{+0.59}_{-0.58}$&$7.73^{+0.67}_{-0.98}$ &

$0.35^{+0.10}_{-0.10}$ & 225 & $8.34^{+1.25}_{-0.49}$ & $7.49^{+1.24}_{-0.54}$ & $8.55^{+1.41}_{-0.65}$

\\

$\rm{\textit{z}_{2}}$&$0.69^{+0.14}_{-0.13}$& 34555 &$9.16^{+0.71}_{-0.54}$&$8.22^{+0.84}_{-0.53}$&$9.27^{+0.76}_{-0.74}$ &

$0.65^{+0.12}_{-0.13}$ & 228&$8.61^{+0.67}_{-0.39}$&$7.68^{+0.56}_{-0.35}$ & $8.77^{+0.90}_{-0.61}$&

$0.69^{+0.12}_{-0.11}$&258 & $9.28^{+1.31}_{-0.49}$& $8.18^{+1.45}_{-0.61}$ & $9.88^{+1.28}_{-1.32}$

\\

$\rm{\textit{z}_{3}}$&$1.06^{+0.14}_{-0.09}$& 34515&$9.44^{+0.62}_{-0.43}$ &$8.54^{+0.70}_{-0.46}$&$9.61^{+0.66}_{-0.54}$ &

$1.04^{+0.12}_{-0.09}$ & 95&$9.22^{+0.75}_{-0.42}$&$8.19^{+0.87}_{-0.36}$&$9.68^{+0.63}_{-0.67}$ &

$1.11^{+0.11}_{-0.10}$&501 & $9.53^{+1.15}_{-0.47}$& $8.59^{+1.17}_{-0.51}$ & $10.06^{+1.20}_{-0.73}$

\\

$\rm{\textit{z}_{4}}$&$1.47^{+0.12}_{-0.09}$ &21754&$9.73^{+0.57}_{-0.42}$ &$8.86^{+0.60}_{-0.47}$&$9.84^{+0.63}_{-0.50}$ &

$1.49^{+0.14}_{-0.08}$ & 64&$9.47^{+0.70}_{-0.22}$&$8.60^{+0.67}_{-0.34}$&$9.82^{+0.66}_{-0.40}$ &

$1.48^{+0.14}_{-0.08}$&1081 & $9.69^{+0.64}_{-0.41}$& $8.81^{+0.63}_{-0.44}$ & $9.91^{+0.75}_{-0.52}$

\\

$\rm{\textit{z}_{5}}$&$1.86^{+0.16}_{-0.13}$ &10690&$9.97^{+0.46}_{-0.37}$ &$9.06^{+0.50}_{-0.42}$&$10.19^{+0.53}_{-0.52}$ &

$2.01^{+0.08}_{-0.17}$ & 150& $9.75^{+0.33}_{-0.25}$&$8.81^{+0.32}_{-0.25}$&$10.10^{+0.42}_{-0.40}$ &

$1.92^{+0.14}_{-0.08}$&852 & $10.04^{+0.59}_{-0.39}$& $9.17^{+0.55}_{-0.43}$ & $10.31^{+0.74}_{-0.53}$

\\

$\rm{\textit{z}_{6}}$&$2.26^{+0.14}_{-0.12}$& 8620&$9.91^{+0.44}_{-0.31}$&$8.95^{+0.52}_{-0.36}$&$10.19^{+0.45}_{-0.47}$ &

$2.37^{+0.10}_{-0.15}$ & 1119&$9.86^{+0.31}_{-0.26}$&$8.88^{+0.30}_{-0.27}$ &$10.15^{+0.43}_{-0.41}$&

$2.27^{+0.13}_{-0.12}$&805 & $10.05^{+0.50}_{-0.37}$& 
$9.20^{+0.52}_{-0.45}$ & $10.29^{+0.49}_{-0.48}$

\\

$\rm{\textit{z}_{7}}$&$2.66^{+0.15}_{-0.12}$&4027&$10.06^{+0.41}_{-0.30}$&$9.09^{+0.47}_{-0.35}$&$10.34^{+0.48}_{-0.50}$ &

$2.75^{+0.15}_{-0.19}$ & 2855 &$10.02^{+0.30}_{-0.26}$&$9.01^{+0.31}_{-0.26}$&$10.40^{+0.39}_{-0.41}$ &

$2.70^{+0.15}_{-0.15}$&558 & $10.27^{+0.46}_{-0.35}$& 
$9.45^{+0.44}_{-0.47}$ & $10.42^{+0.47}_{-0.45}$

\\

\\
 \hline
 \end{tabular}}
  \begin{tablenotes}
\item $\rm{R_{LUM}}$ - $\rm{LUM{\_}R{\_}BEST}$ in the \citet{2018ApJS..235...36M} catalogue, decimal logarithm of R-band luminosity from best-fit \textsc{LEPHARE} model [solar luminosities]
\item $\rm{K_{LUM}}$ - $\rm{LUM{\_}K{\_}BEST}$ in the \citet{2018ApJS..235...36M} catalogue, decimal logarithm of K-band luminosity from best-fit \textsc{LEPHARE} model [solar luminosities]
\item $\rm{NUV_{LUM}}$ - $\rm{LUM{\_}NUV{\_}BEST}$ in the \citet{2018ApJS..235...36M} catalogue, decimal logarithm of NUV luminosity from best-fit \textsc{LEPHARE} model [solar luminosities]
\end{tablenotes}
 \label{tab_ppts} 
 \end{table*}

\subsection{Stacking into the LOFAR, \textit{u}GMRT, MeerKAT and VLA radio images}\label{stack.sec}

To explore statistically the potential presence of
signals from the fainter population below the respective survey thresholds, we undertook a stacking analysis.
Stacking of radio images is a powerful statistical approach to measure the mean or median flux density of a class of sources that cannot be detected individually in a survey to reveal the broad picture of the faint radio sky  (see \citealt{2007ApJ...654...99W,2014ApJ...787...99S}).
\citet{2015aska.confE.172Z} defined stacking as the co-addition of maps at the positions of sources detected in another map or catalogue. If the position of a sample of sources is known from a catalogue, one can do the co-addition of maps at the positions of sources detected in another map. Hence, the intensities at the recorded positions can be combined by taking their mean or their median \citep{2014ApJ...787...99S}.

\begin{table*}
 \centering
 \caption{Table showing the summary of the properties of the multi-frequency samples used in this work. We list
the median \textit{S} flux densities for each stacked image for the different frequencies,
the number of sources included in each stack $\rm{(N_{stack})}$.}
 \scalebox{0.76}{
 \begin{tabular}{cccccccccccccccccccccccc}
 \hline
 \hline
\textit{z} bin&  $\rm{\textit{S}_{144}}$&$\rm{N_{144}}$&$\rm{\textit{S}_{320}}$&$\rm{N_{320}}$&$\rm{\textit{S}_{370}}$&$\rm{N_{370}}$&$\rm{\textit{S}_{420}}$&$\rm{N_{420}}$&$\rm{\textit{S}_{460}}$&$\rm{N_{460}}$&$\rm{\textit{S}_{592}}$&$\rm{N_{592}}$&$\rm{\textit{S}_{656}}$&$\rm{N_{656}}$&$\rm{\textit{S}_{720}}$&$\rm{N_{720}}$&$\rm{\textit{S}_{784}}$&$\rm{N_{784}}$&$\rm{\textit{S}_{1280}}$&$\rm{N_{1280}}$&$\rm{\textit{S}_{1500}}$&$\rm{N_{1500}}$\\

 \hline
$\rm{\textit{z}_{1}}$ &$\rm{25.20}$&14801&$\rm{9.25}$&14806&$\rm{6.68}$&14806&$\rm{6.11}$&14806&${6.13}$&14806&${4.71}$&10683&${4.10}$&10683&${4.45}$&10683&${3.30}$&10683&${2.66}$&13631&${2.41}$&14806\\ 

$\rm{\textit{z}_{2}}$&nan&0&$\rm{6.53}$&34555&$\rm{5.42}$&34555&$\rm{5.13}$&34555&${5.52}$&34555&${4.43}$&25464&${3.82}$&25464&${3.99}$&25464&${4.15}$&25464&${2.62}$&32527&${2.68}$&34555\\

$\rm{\textit{z}_{3}}$&$\rm{16.33}$&34507&$\rm{5.82 }$&34515&$\rm{5.72}$&34515&$\rm{5.11}$&34515&${6.77}$&34515&${5.06}$&25427&${4.26}$&25427&${4.19}$&25427&${4.42}$&25427&${2.93}$&32822&${2.57}$&34515\\

$\rm{\textit{z}_{4}}$&$\rm{16.37}$&21751&$\rm{8.72}$&21754&$\rm{6.55}$&21754&$\rm{5.87}$&21754&${6.27}$&21754&${5.03}$&16530&${4.98}$&16530&${4.89}$&16530&${4.74}$&16529&${3.11}$&20735&${3.08}$&21754\\

$\rm{\textit{z}_{5}}$&$\rm{24.25}$&10690&$\rm{9.75}$&10690&$\rm{8.71}$&10690&$\rm{7.42}$&10690&${7.19}$&10690&${6.63}$&7905&${6.59}$&7905&${5.44}$&7905&${6.12}$&7904&${3.79}$&10150&${3.68}$&10690\\

$\rm{\textit{z}_{6}}$&$\rm{13.71}$&8618&$\rm{6.67}$&8620&$\rm{7.67}$&8620&$\rm{4.06}$&8620&${6.34}$&8620&${4.28}$&6298&${3.46}$&6298&${3.65}$&6298&${5.05}$&6298&${2.49}$&8197&${3.03}$&8620\\

$\rm{\textit{z}_{7}}$&$\rm{27.53}$&4025&$\rm{5.98}$&4027&$\rm{6.72}$&4027&$\rm{6.13}$&4027&${5.86}$&4027&${4.96}$&3048&${5.43}$&3048&${2.54}$&3048&${4.31}$&3048&${3.11}$&3813&${3.10}$&4027\\
\\
 \hline
 \end{tabular}}
\begin{tablenotes}
\item N.B. the flux densities are quoted in $\mu$Jy/beam.

\end{tablenotes}
 \label{tab_stacked_ppts} 
 \end{table*}

\newsavebox\mybox
\savebox\mybox
{
\centering
  \begin{minipage}[t]{0.4\linewidth}
    \includegraphics[width=0.2\linewidth]{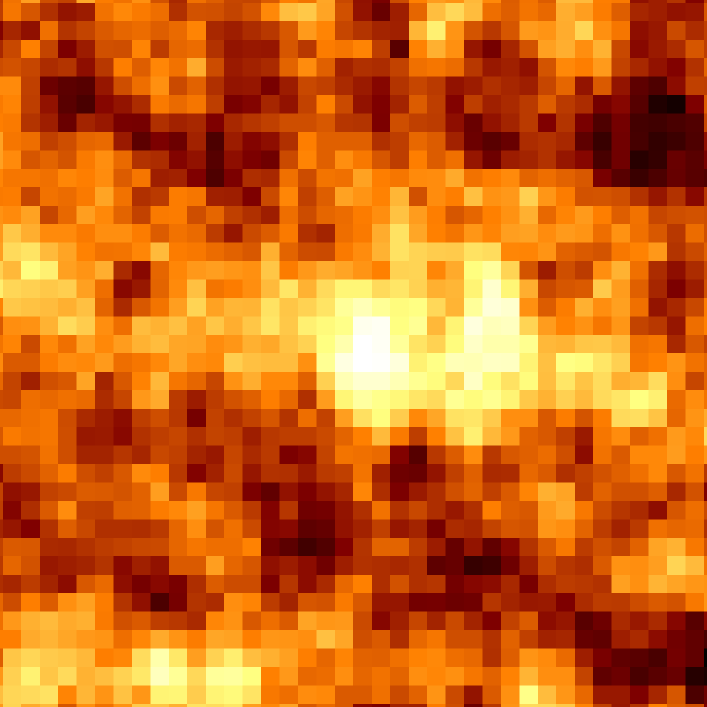}
  \end{minipage}
}
\newlength{\ImageHt}
\setlength\ImageHt{\ht\mybox}

\begin{figure*}
\begin{tabular}{@{}c@{ }c@{ }c@{ }c@{ }c@{ }c@{ }c@{ }c@{}}
&\textbf{\textit{z}$_{1}$=0.1-0.5} & \textbf{\textit{z}$_{2}$=0.5-0.9} & \textbf{\textit{z}$_{3}$=0.9-1.3} & \textbf{\textit{z}$_{4}$=1.3-1.7} & \textbf{\textit{z}$_{5}$=1.7-2.1} & \textbf{\textit{z}$_{6}$=2.1-2.5} & \textbf{\textit{z}$_{7}$=2.5-3.0}\\
\centering
\rotatebox[origin=b]{90}{\makebox[\ImageHt]{\normalsize $\rm{144\,MHz}$}}&

\subfloat[\large \textit{SNR} = 7.06]{\includegraphics[width=.12\linewidth]{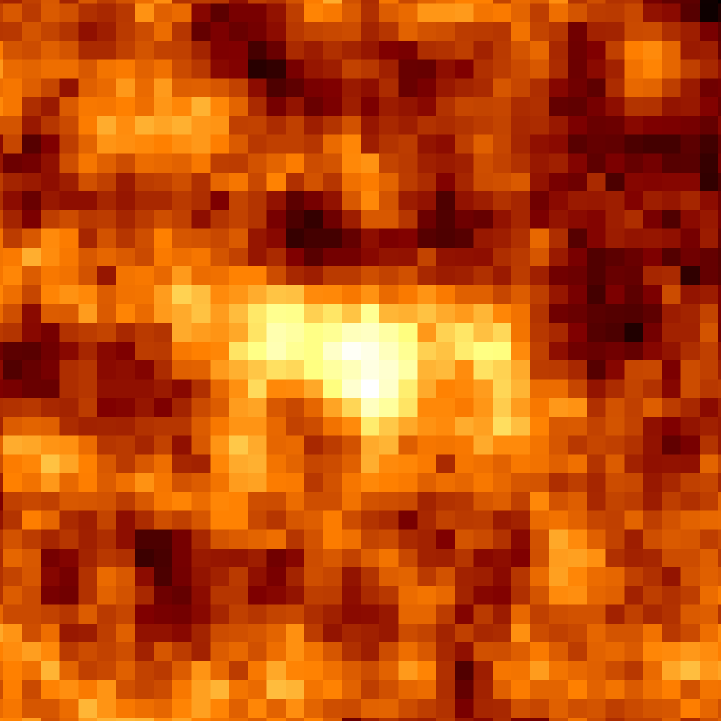}}&
\hspace*{0.05\textwidth}&%
\subfloat[\large \textit{SNR} = 6.78]{\includegraphics[width=.12\linewidth]{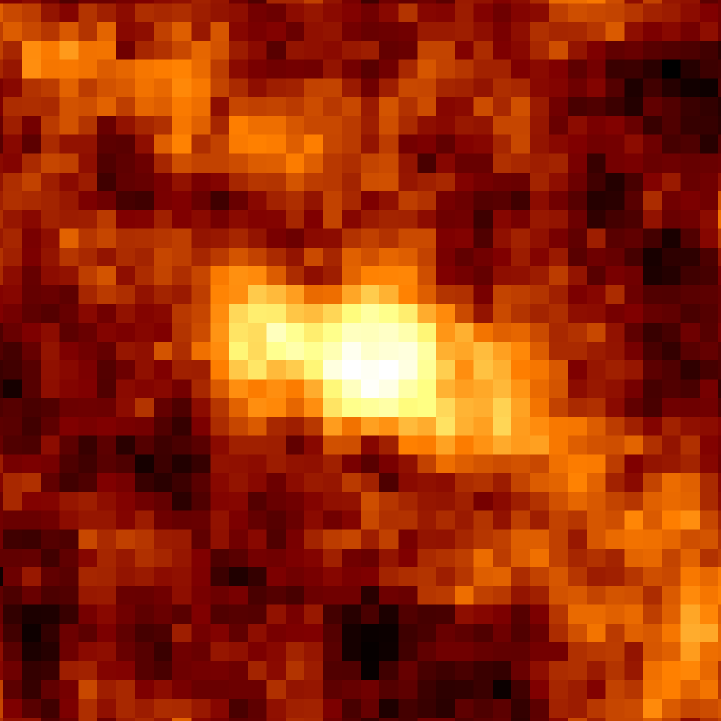}}\hfill&
\subfloat[\large \textit{SNR} = 6.04]{\includegraphics[width=.12\linewidth]{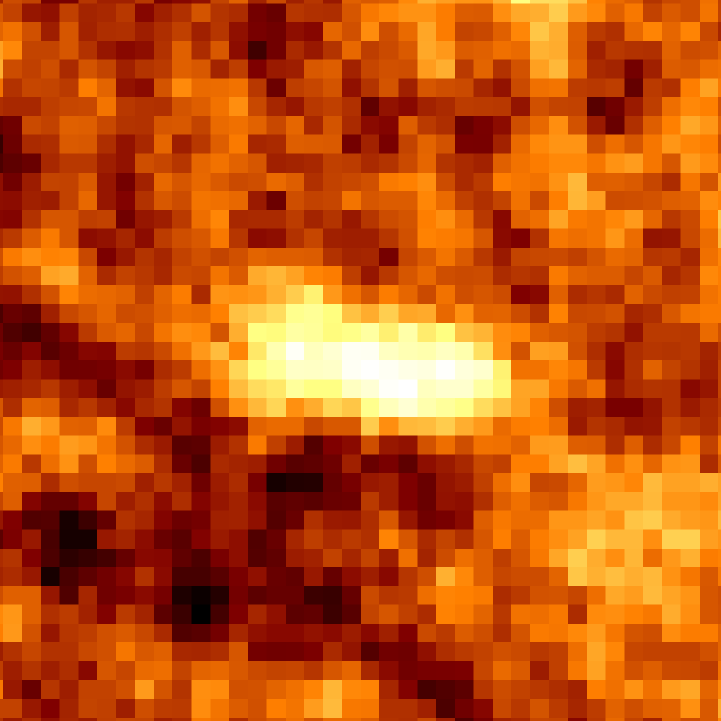}}\hfill&
\subfloat[\large \textit{SNR} = 5.77]{\includegraphics[width=.12\linewidth]{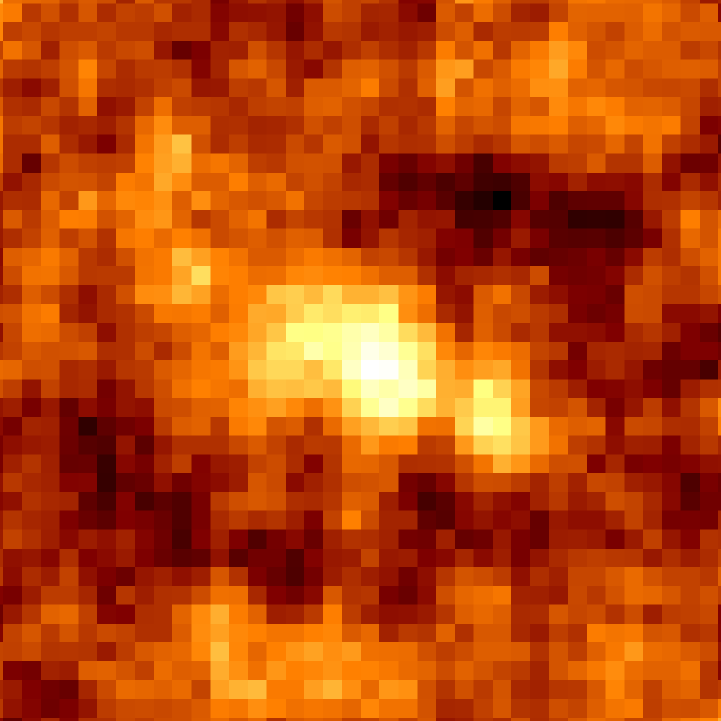}}\hfill&
\subfloat[\large \textit{SNR} = 3.61]{\includegraphics[width=.12\linewidth]{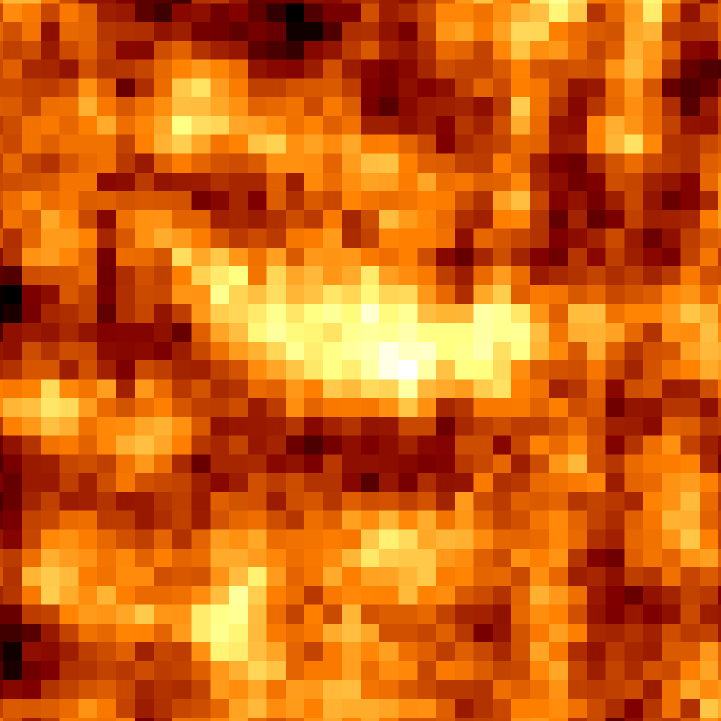}}\hfill&
\subfloat[\large \textit{SNR} = 4.14]{\includegraphics[width=.12\linewidth]{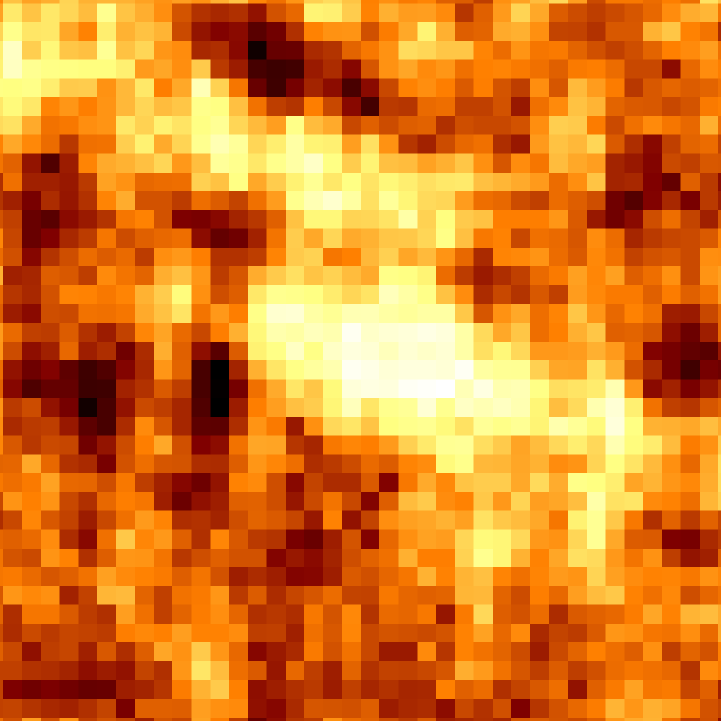}}
\\

\rotatebox[origin=b]{90}{\makebox[\ImageHt]{\normalsize $\rm{320\,MHz}$}}&
\subfloat[\large \textit{SNR} = 7.98]{\includegraphics[width=.12\linewidth]{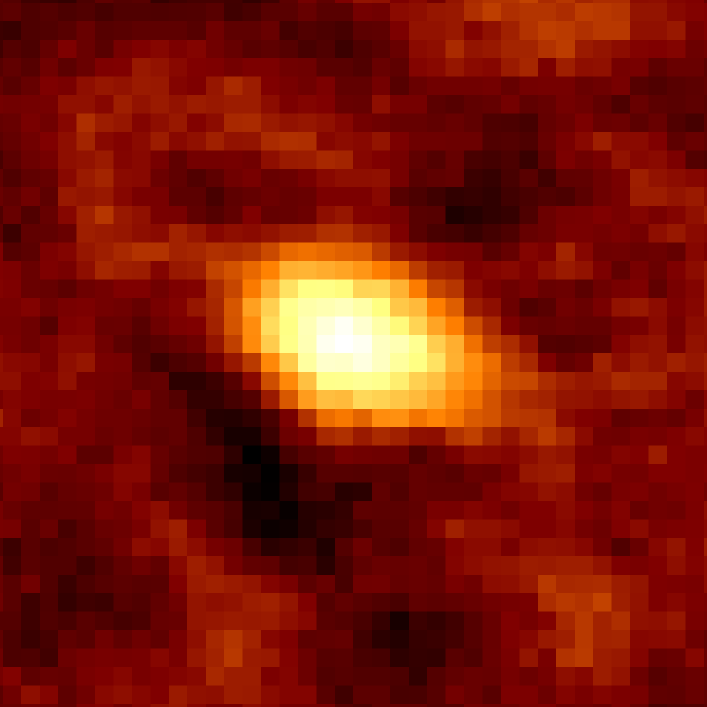}}\hfill&
\subfloat[\large \textit{SNR} = 6.56 ]{\includegraphics[width=.12\linewidth]{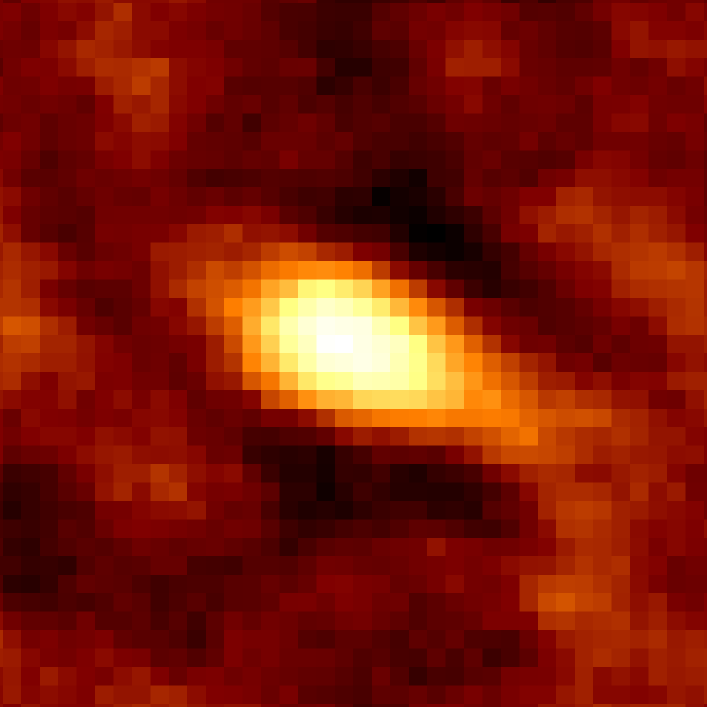}}\hfill&
\subfloat[\large \textit{SNR} = 6.25]{\includegraphics[width=.12\linewidth]{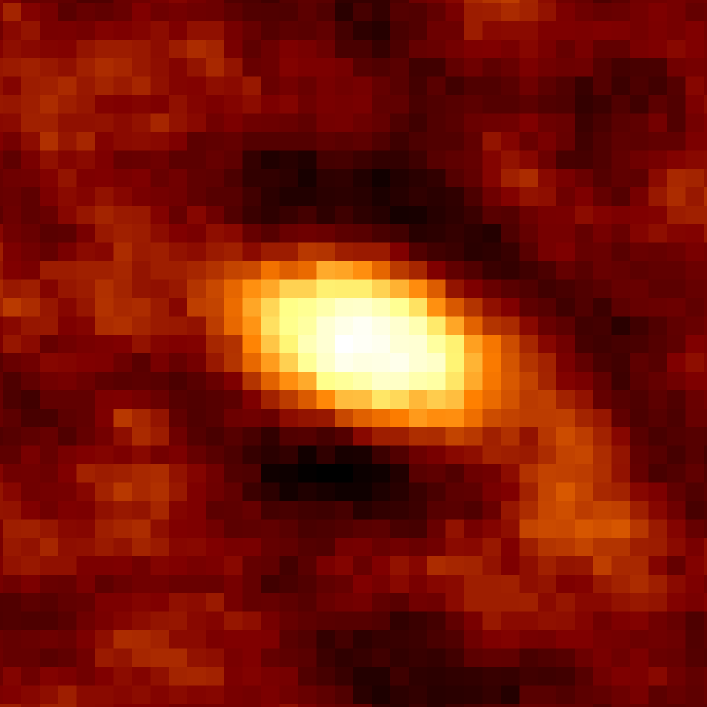}}\hfill&
\subfloat[\large \textit{SNR} = 7.44]{\includegraphics[width=.12\linewidth]{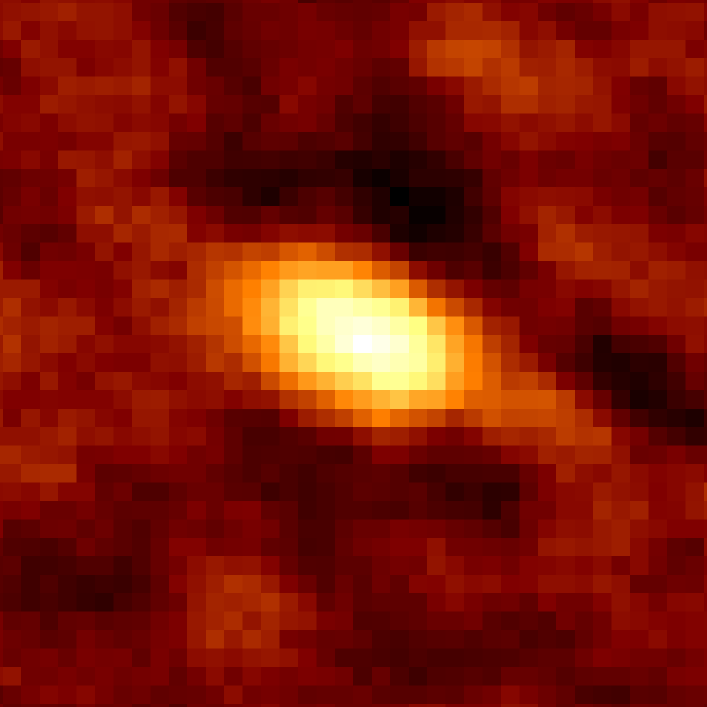}}\hfill&
\subfloat[\large \textit{SNR} = 5.98]{\includegraphics[width=.12\linewidth]{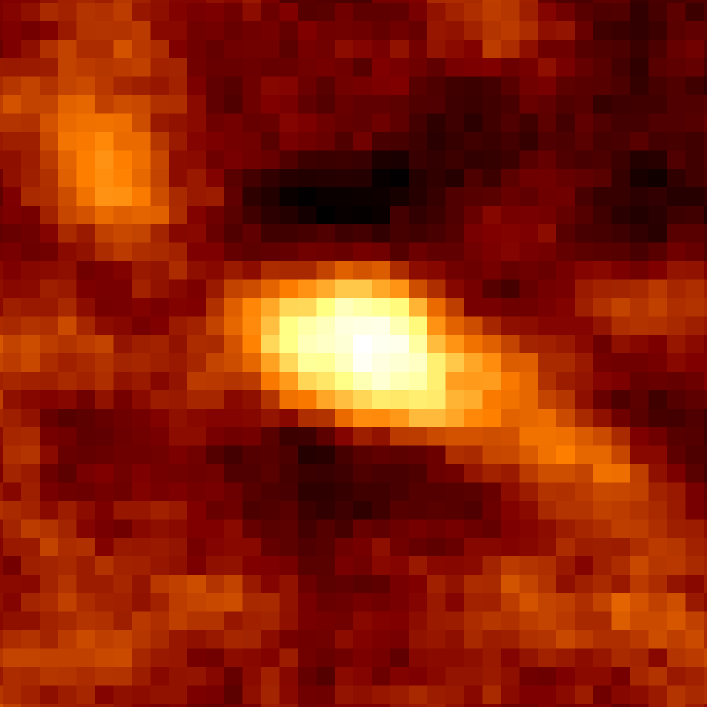}}\hfill&
\subfloat[\large \textit{SNR} = 4.10]{\includegraphics[width=.12\linewidth]{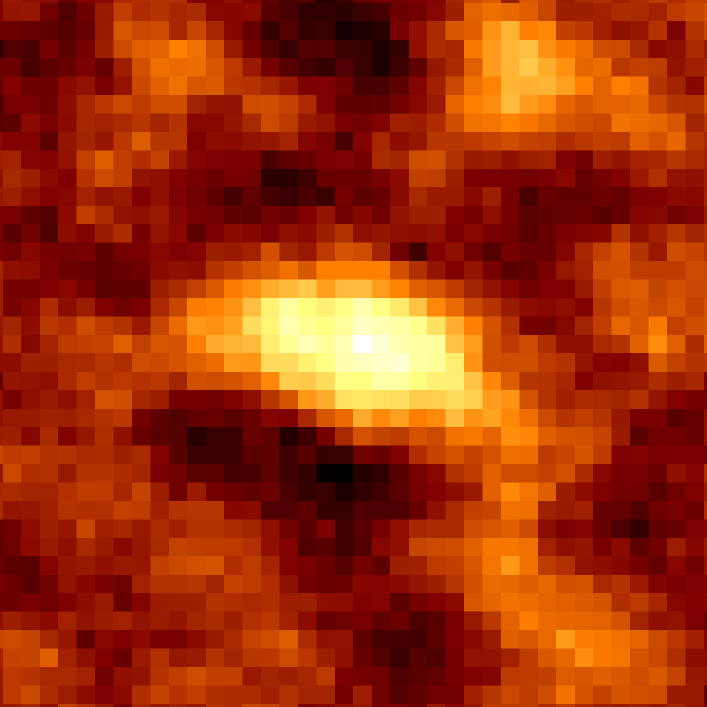}}\hfill&
\subfloat[\large \textit{SNR} = 3.40]{\includegraphics[width=.12\linewidth]{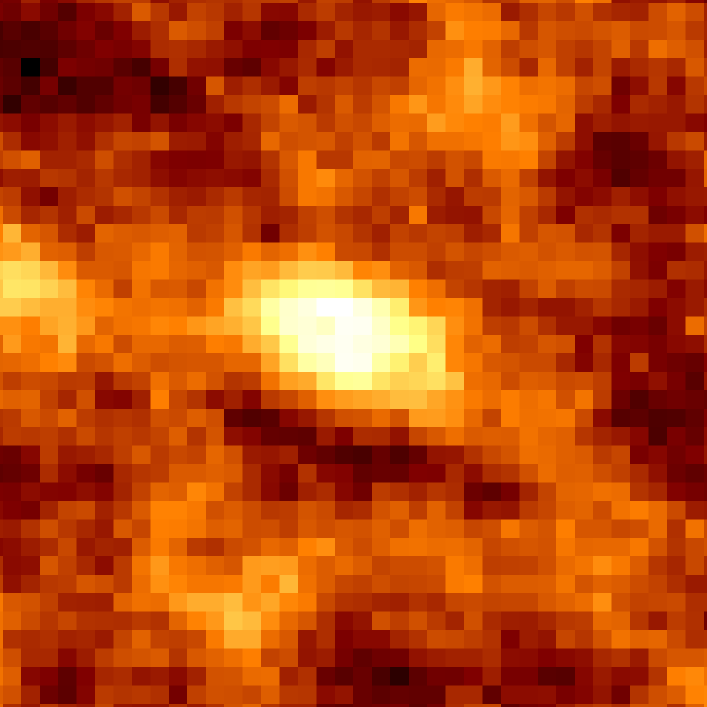}}\\

\rotatebox[origin=b]{90}{\makebox[\ImageHt]{\normalsize $\rm{370\,MHz}$}}&
\subfloat[\large \textit{SNR} = 6.85]{\includegraphics[width=.12\linewidth]{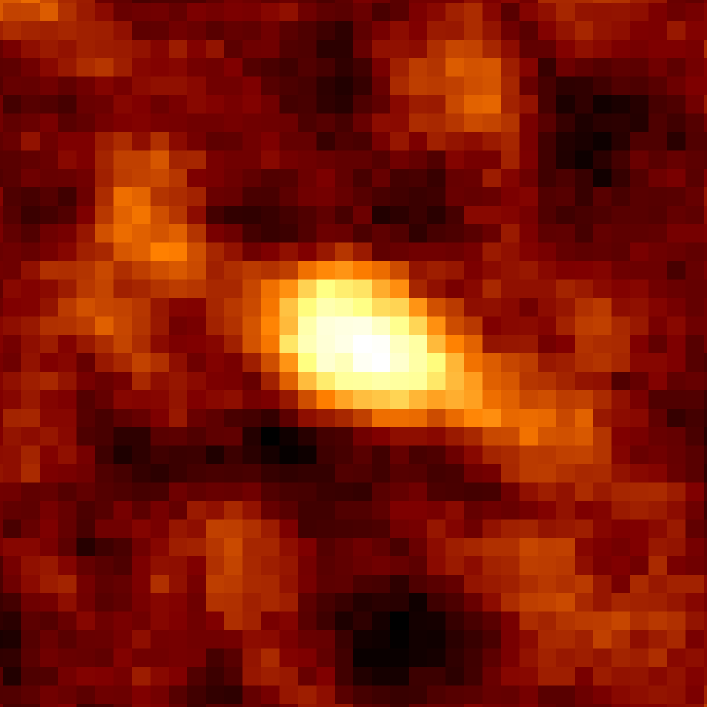}}&
\subfloat[\large \textit{SNR} = 7.29]{\includegraphics[width=.12\linewidth]{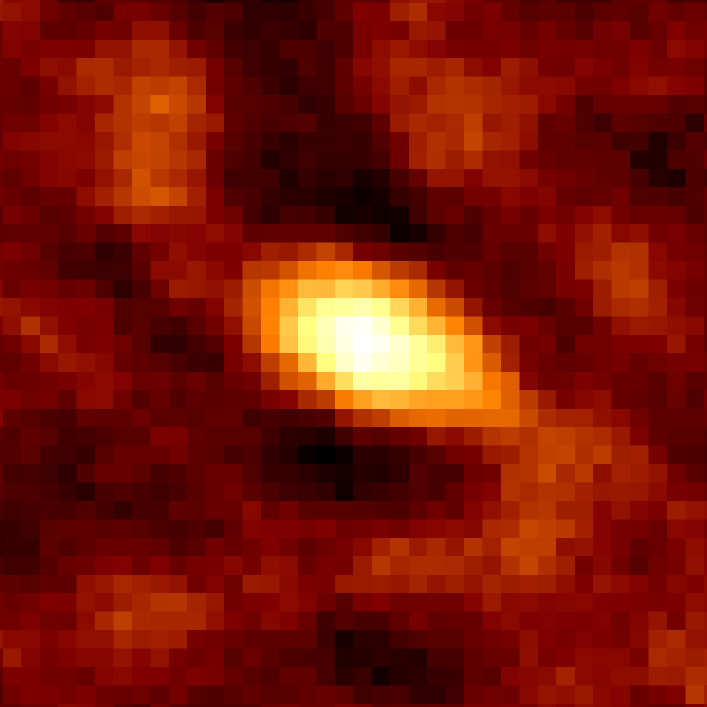}}&
\subfloat[\large \textit{SNR} = 8.15]{\includegraphics[width=.12\linewidth]{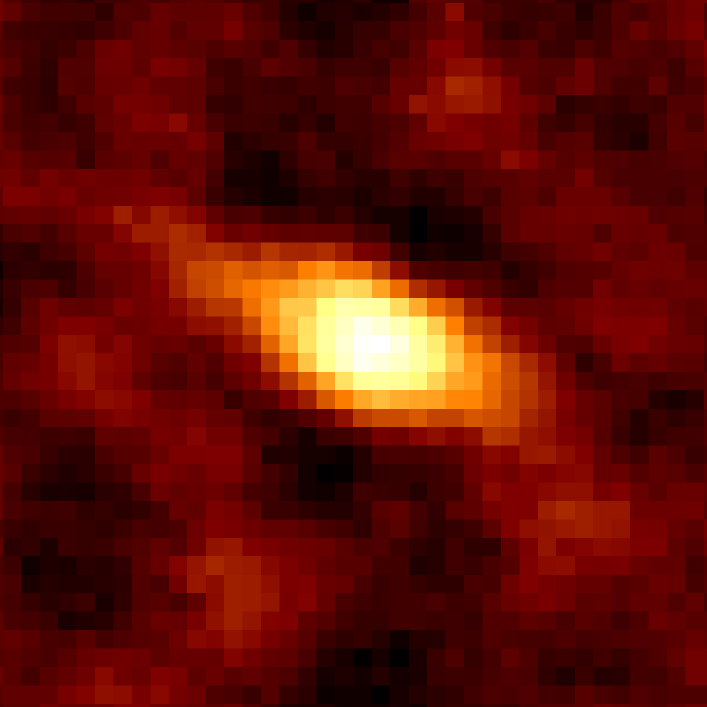}}&
\subfloat[\large \textit{SNR} = 7.61]{\includegraphics[width=.12\linewidth]{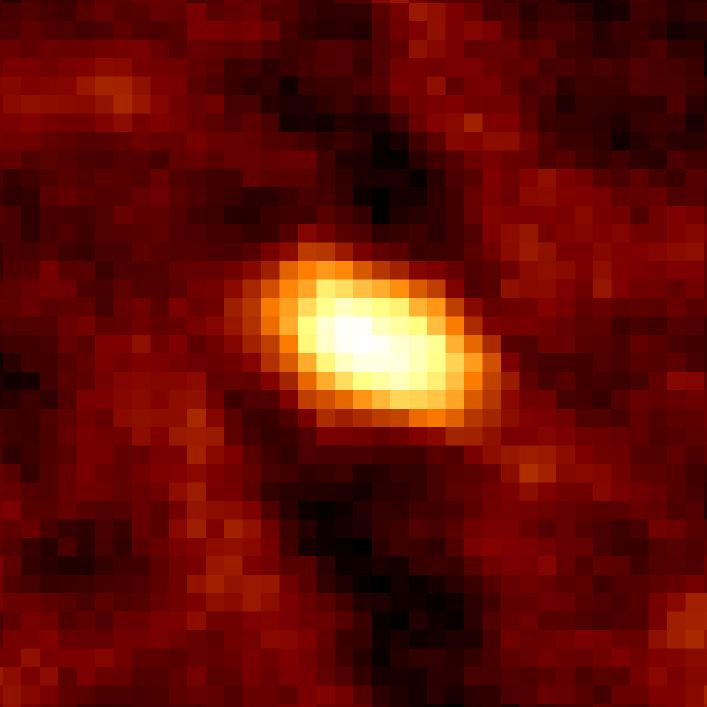}}&
\subfloat[\large \textit{SNR} = 8.16]{\includegraphics[width=.12\linewidth]{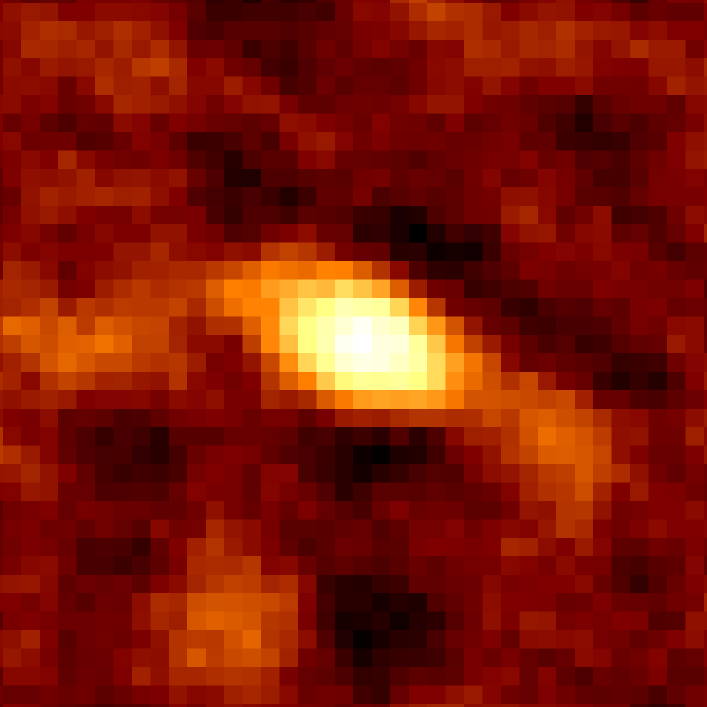}}&
\subfloat[\large \textit{SNR} = 6.88 ]{\includegraphics[width=.12\linewidth]{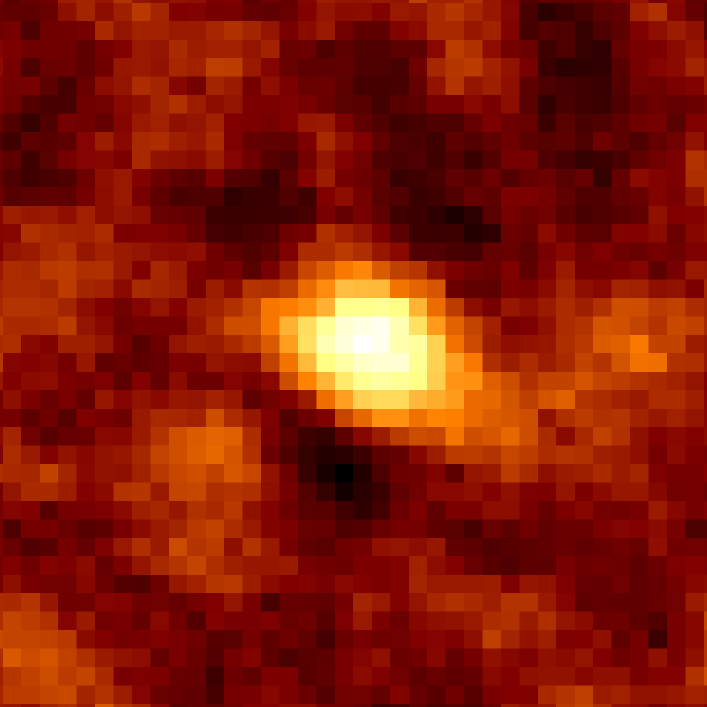}}&
\subfloat[\large \textit{SNR} = 3.77]{\includegraphics[width=.12\linewidth]{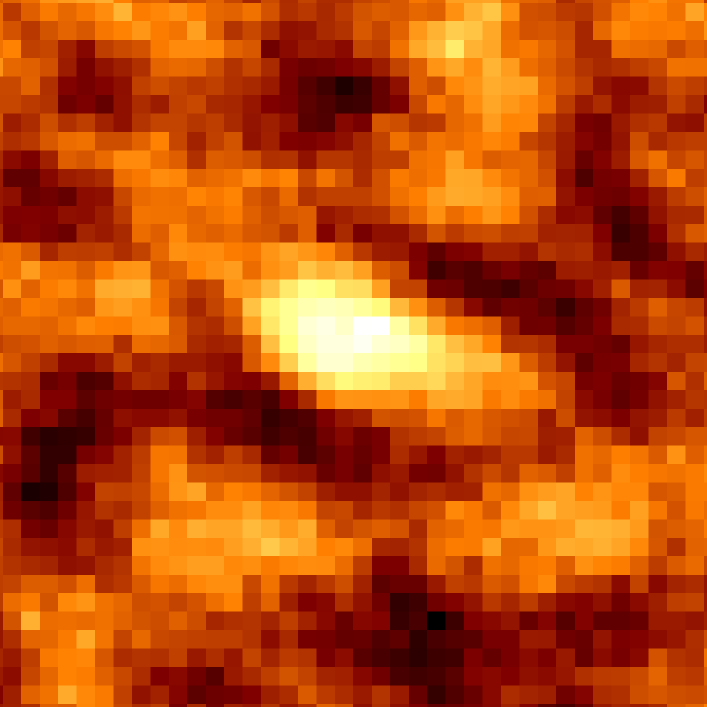}}\\

\rotatebox[origin=b]{90}{\makebox[\ImageHt]{\normalsize $\rm{420\,MHz}$}}&
\subfloat[\large \textit{SNR} = 8.36]{\includegraphics[width=.12\linewidth]{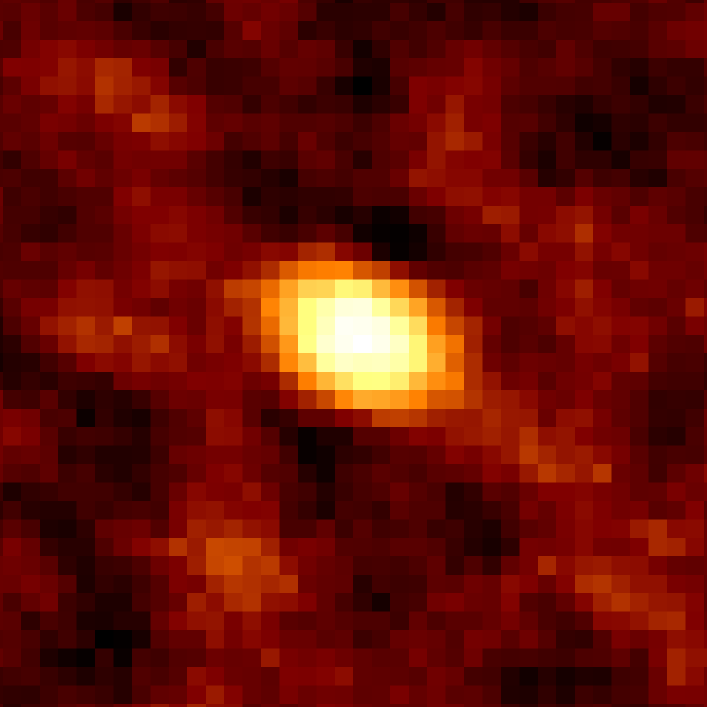}}&
\subfloat[\large \textit{SNR} = 8.71]{\includegraphics[width=.12\linewidth]{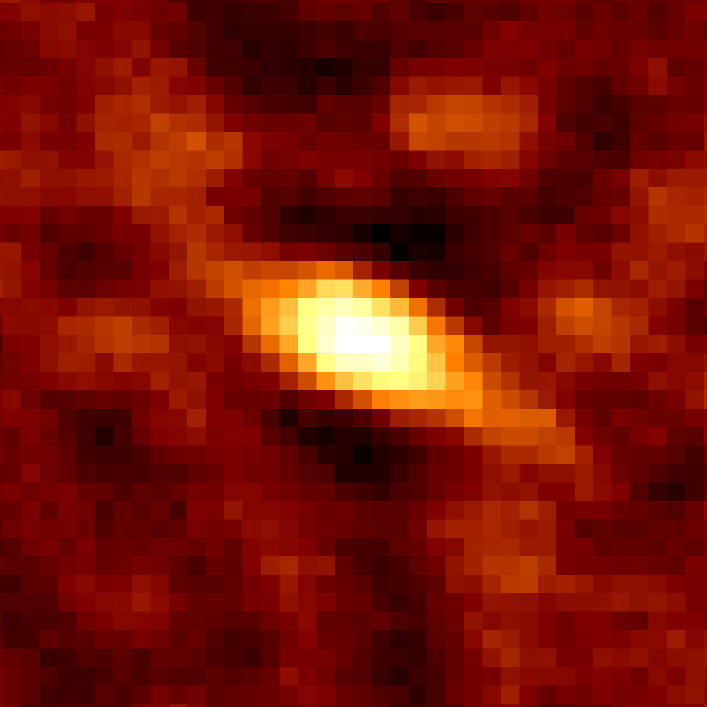}}&
\subfloat[\large \textit{SNR} = 8.99]{\includegraphics[width=.12\linewidth]{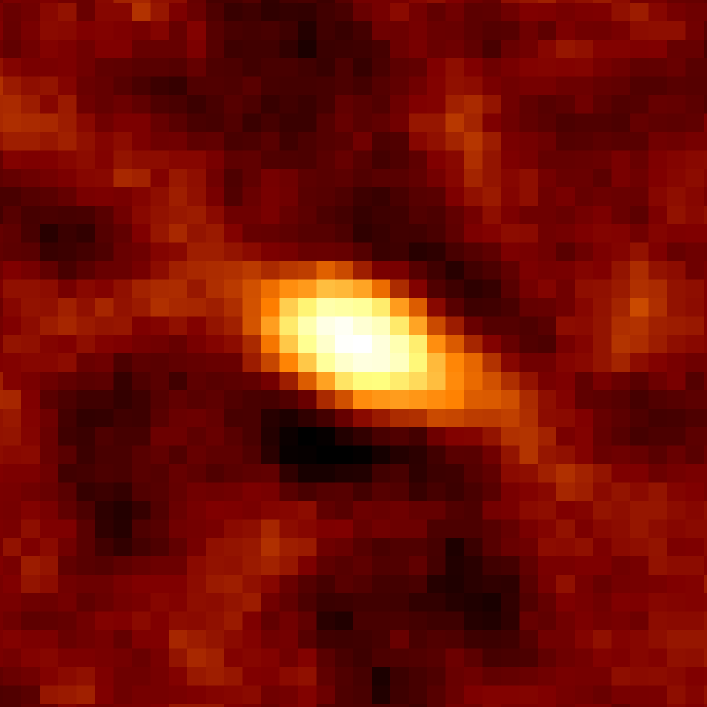}}&
\subfloat[\large \textit{SNR} = 9.33]{\includegraphics[width=.12\linewidth]{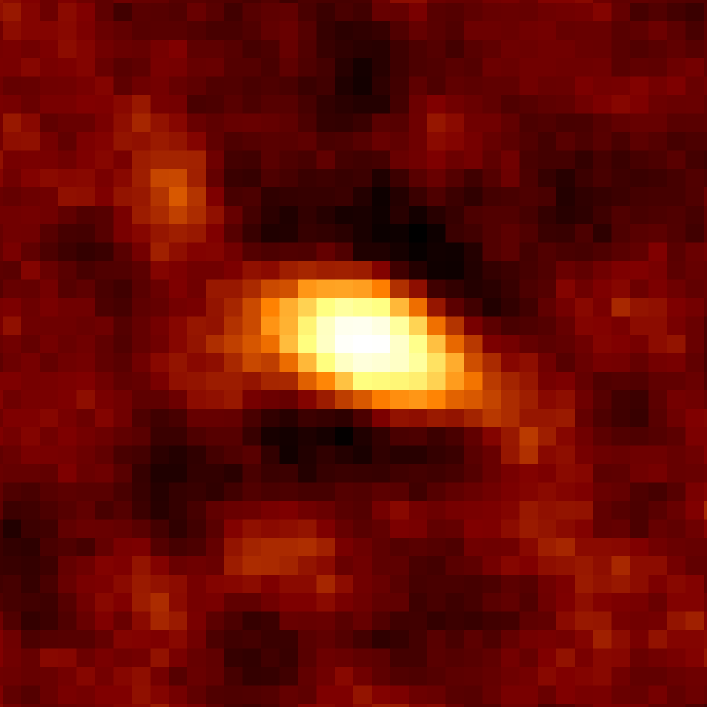}}&
\subfloat[\large \textit{SNR} = 8.20]{\includegraphics[width=.12\linewidth]{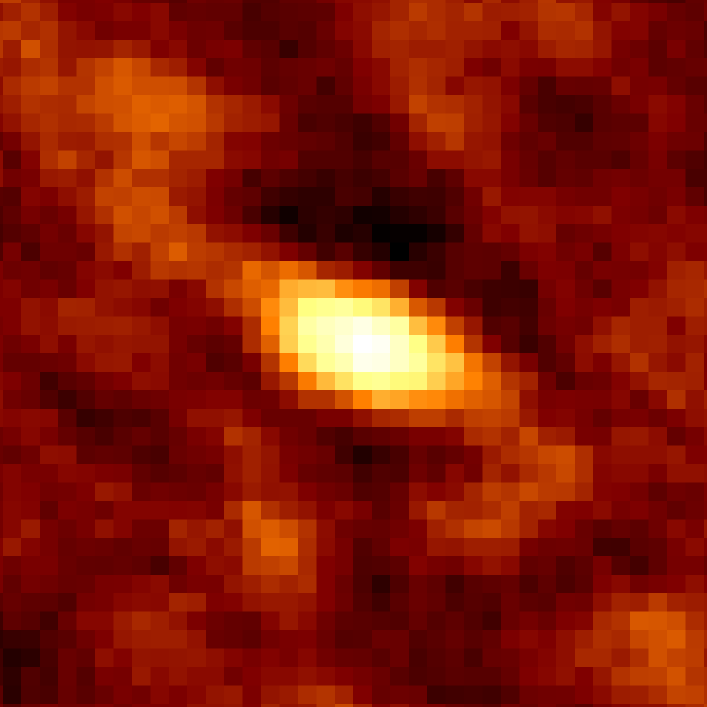}}&
\subfloat[\large \textit{SNR} = 4.85]{\includegraphics[width=.12\linewidth]{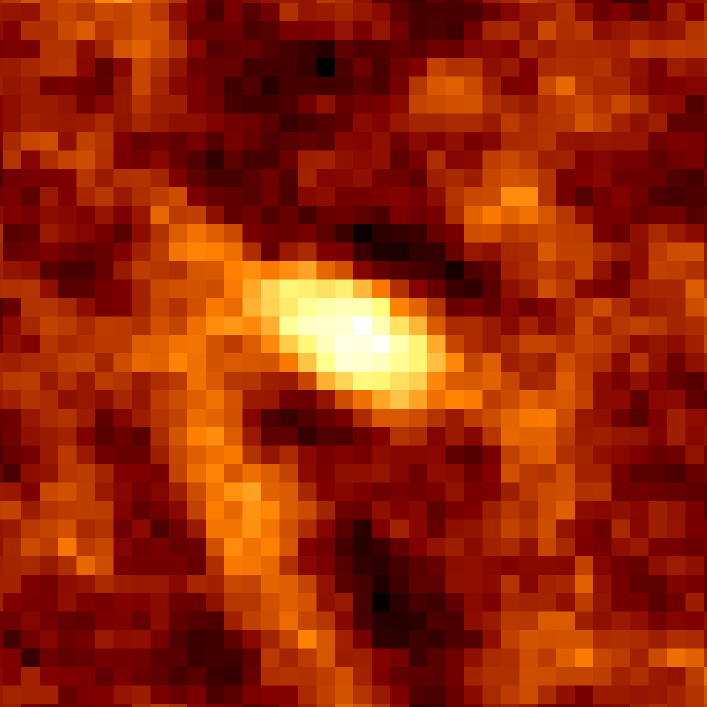}}&
\subfloat[\large \textit{SNR} = 5.10]{\includegraphics[width=.12\linewidth]{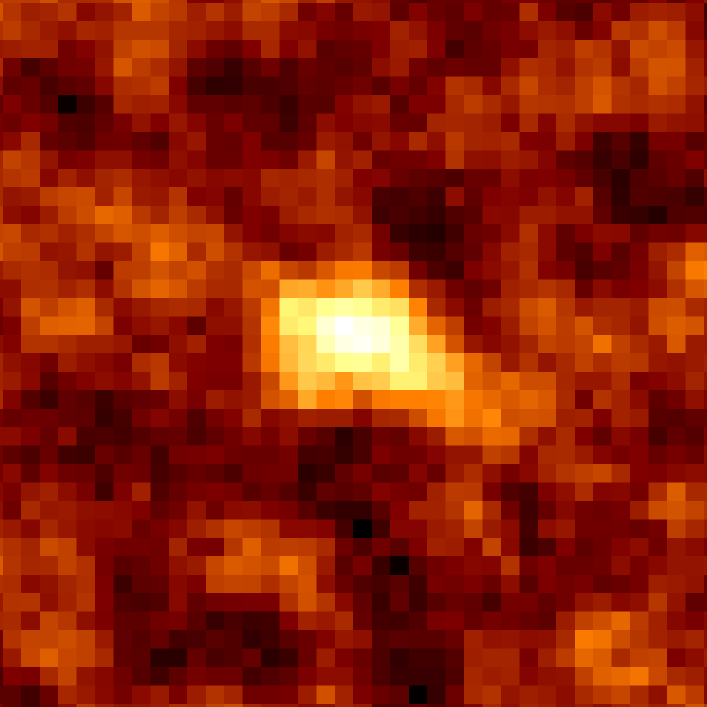}}\\
\rotatebox[origin=b]{90}{\makebox[\ImageHt]{\normalsize $\rm{460\,MHz}$}}&
\subfloat[\large \textit{SNR} = 7.57]{\includegraphics[width=.12\linewidth]{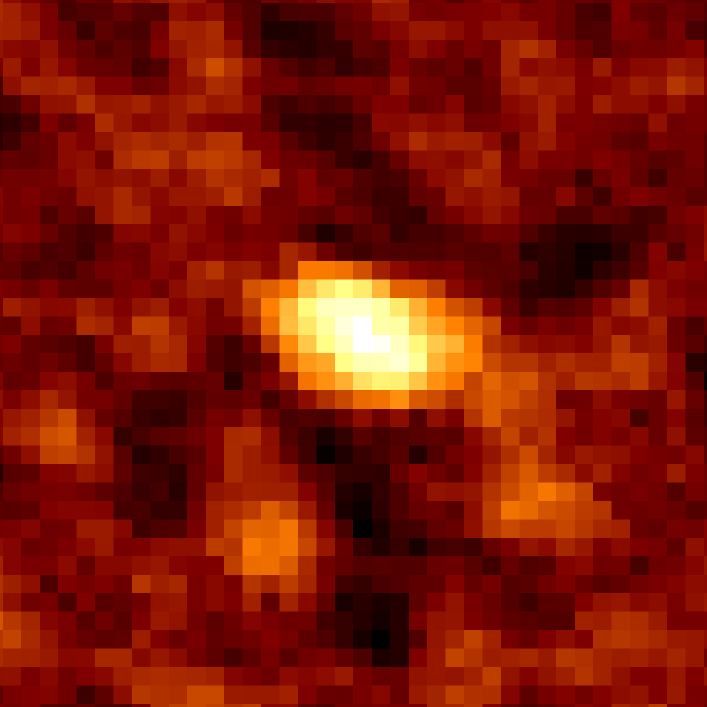}}&
\subfloat[\large \textit{SNR} = 9.76]{\includegraphics[width=.12\linewidth]{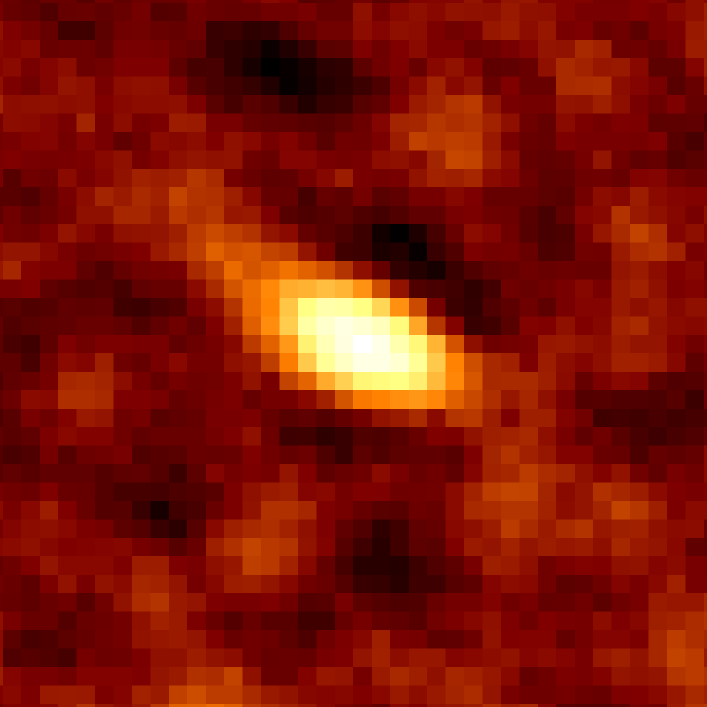}}&
\subfloat[\large \textit{SNR} = 12.34]{\includegraphics[width=.12\linewidth]{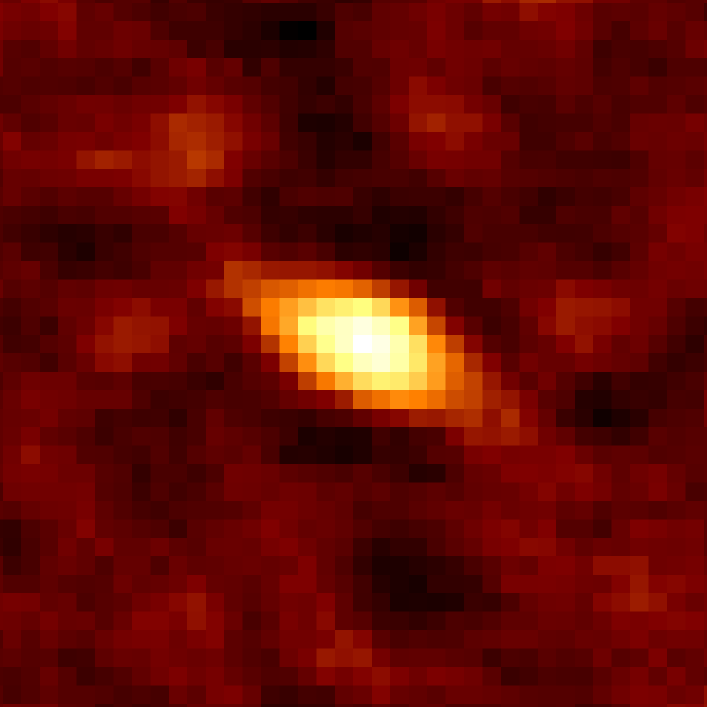}}&
\subfloat[\large \textit{SNR} = 9.43]{\includegraphics[width=.12\linewidth]{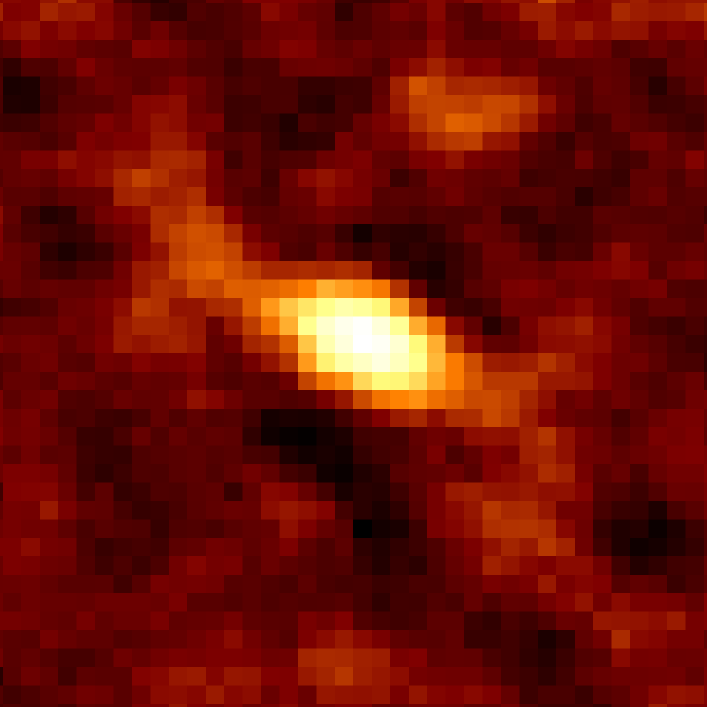}}&
\subfloat[\large \textit{SNR} = 6.97]{\includegraphics[width=.12\linewidth]{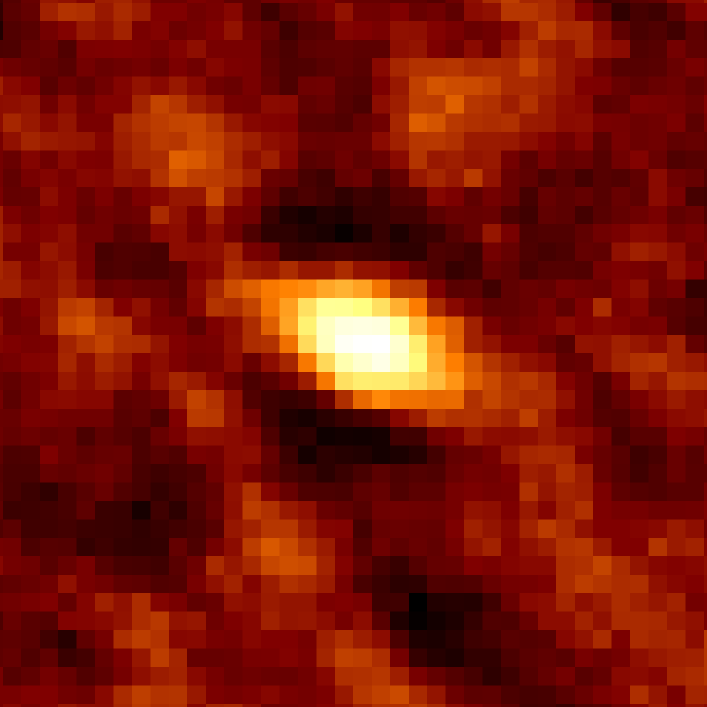}}&
\subfloat[\large \textit{SNR} = 6.70]{\includegraphics[width=.12\linewidth]{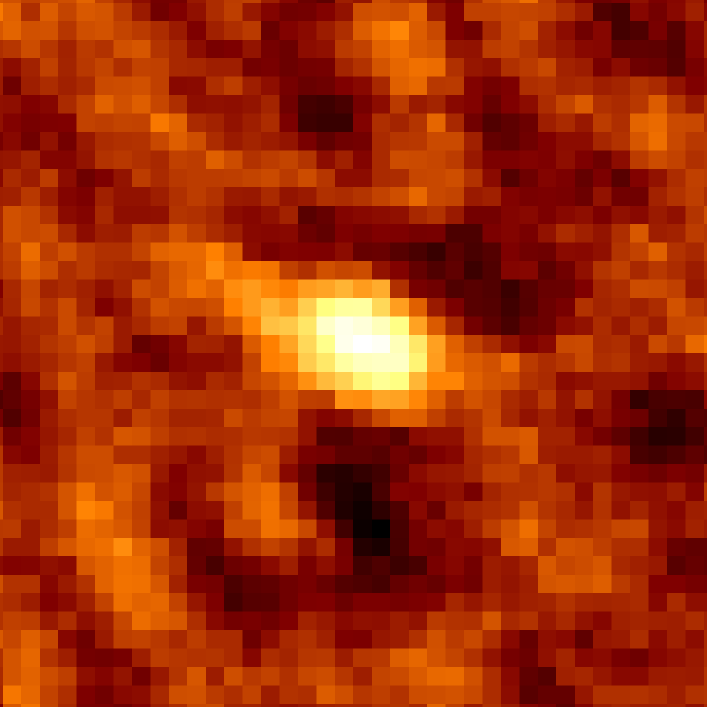}}&
\subfloat[\large \textit{SNR} = 4.56]{\includegraphics[width=.12\linewidth]{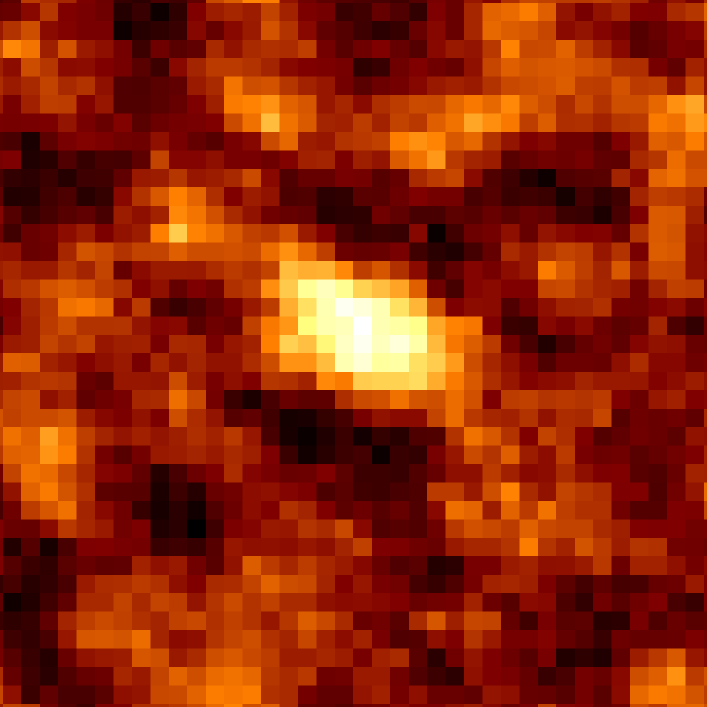}}\\

\rotatebox[origin=b]{90}{\makebox[\ImageHt]{\normalsize $\rm{592\,MHz}$}}&
\subfloat[\large \textit{SNR} = 7.18] {\includegraphics[width=.12\linewidth]{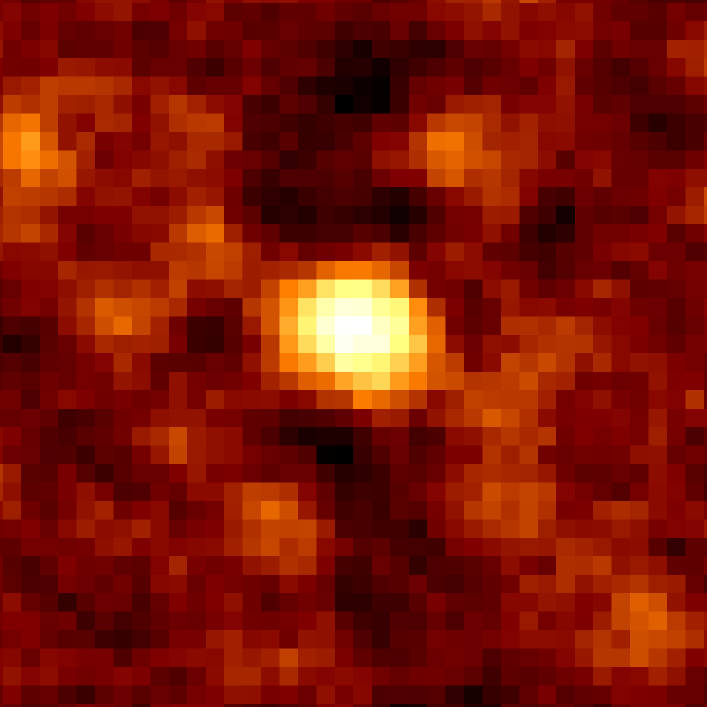}}&
\subfloat[\large \textit{SNR} = 10.64]{\includegraphics[width=.12\linewidth]{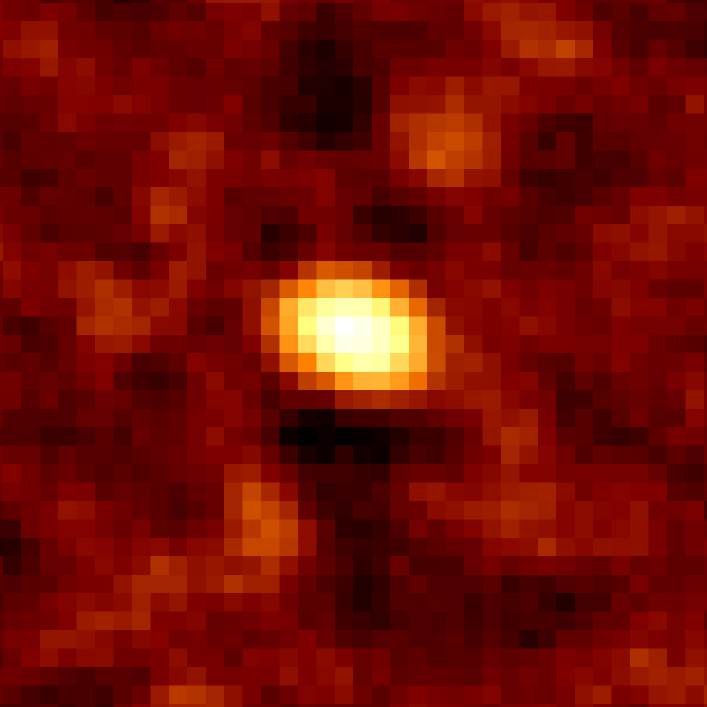}}&
\subfloat[\large \textit{SNR} = 12.26]{\includegraphics[width=.12\linewidth]{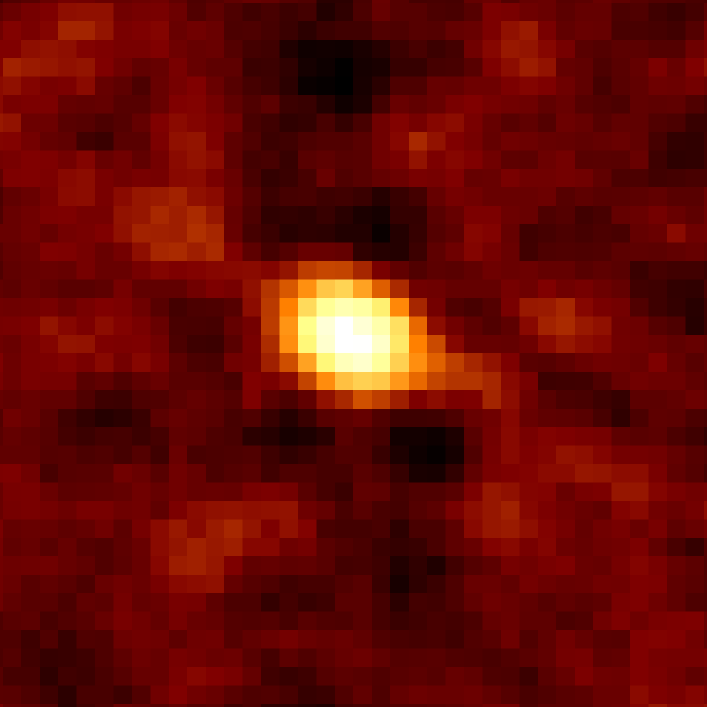}}&
\subfloat[\large \textit{SNR} = 10.04]{\includegraphics[width=.12\linewidth]{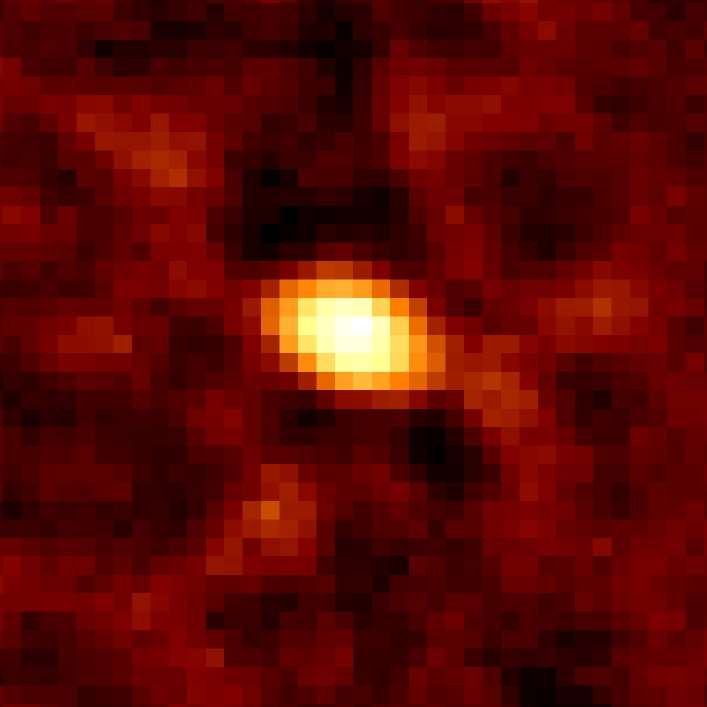}}&
\subfloat[\large \textit{SNR} = 9.99]{\includegraphics[width=.12\linewidth]{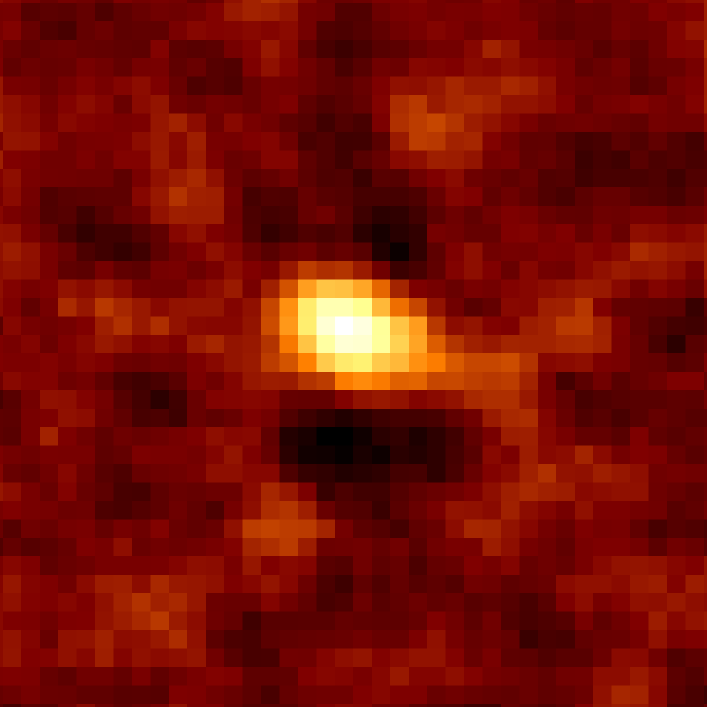}}&
\subfloat[\large \textit{SNR} = 6.05]{\includegraphics[width=.12\linewidth]{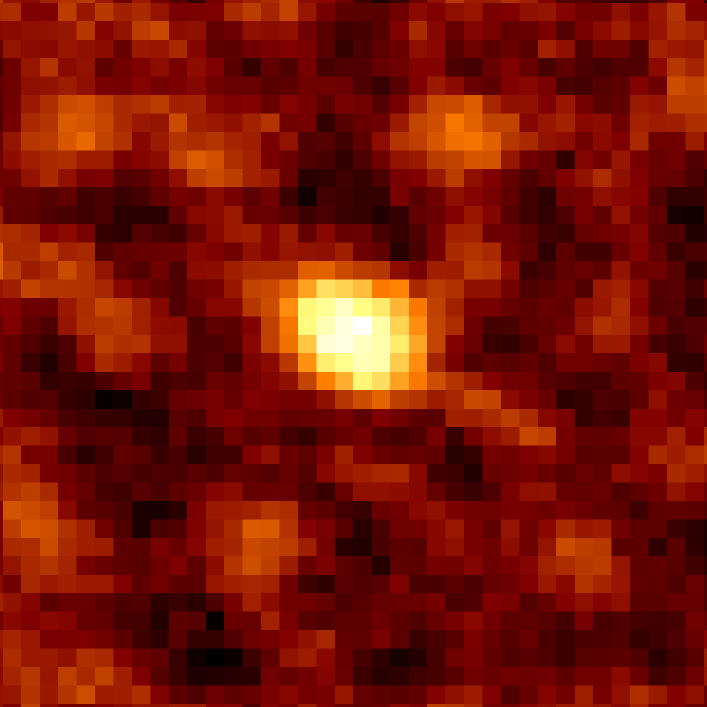}}&
\subfloat[\large \textit{SNR} = 5.14]{\includegraphics[width=.12\linewidth]{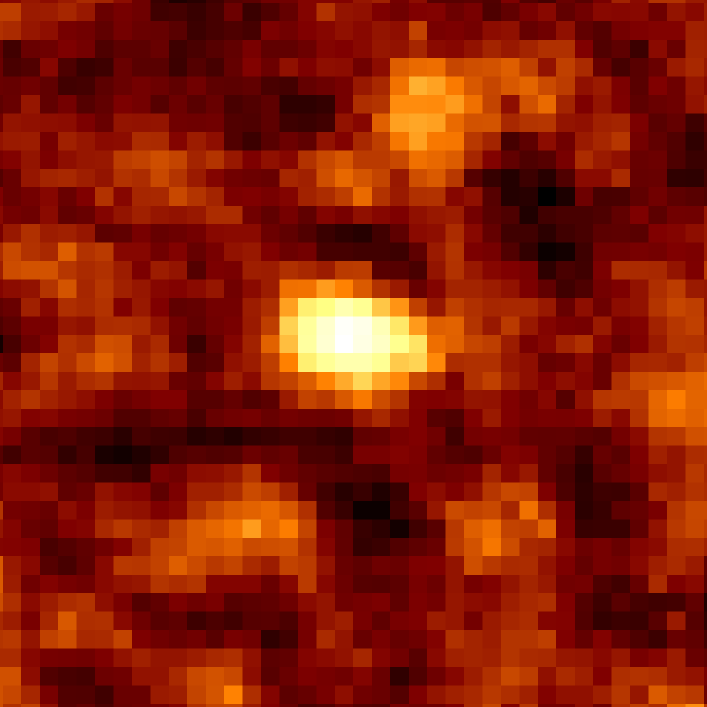}}\\

\rotatebox[origin=b]{90}{\makebox[\ImageHt]{\normalsize $\rm{656\,MHz}$}}&
\subfloat[\large \textit{SNR} = 7.89]{\includegraphics[width=.12\linewidth]{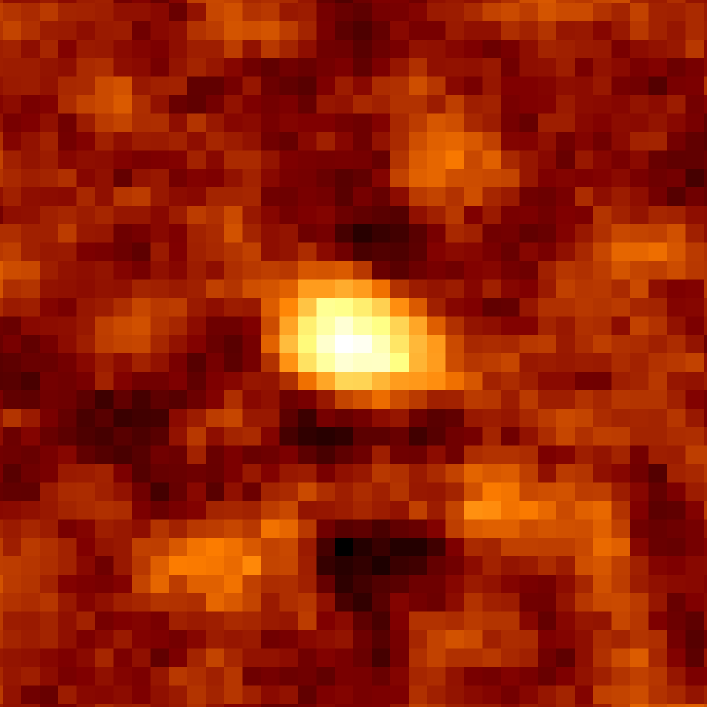}}&
\subfloat[\large \textit{SNR} = 10.33]{\includegraphics[width=.12\linewidth]{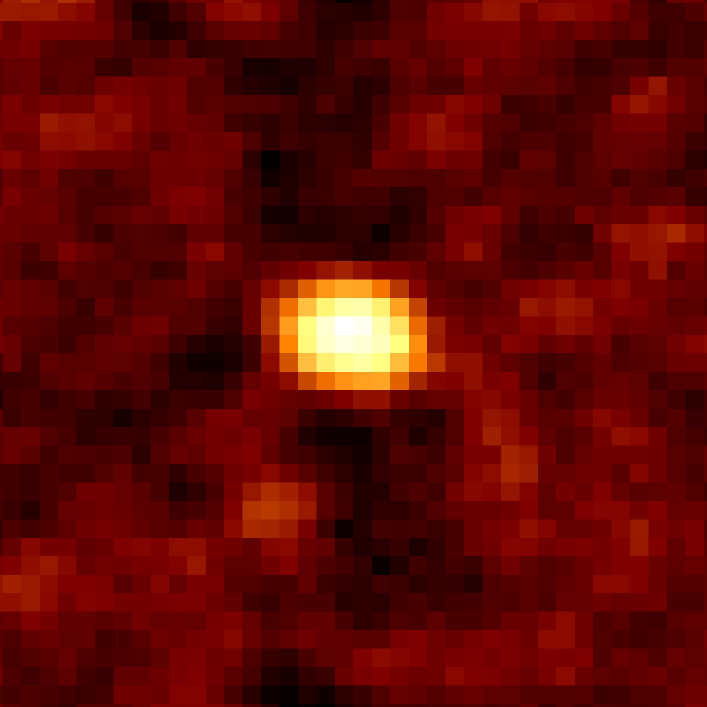}}&
\subfloat[\large \textit{SNR} = 12.16]{\includegraphics[width=.12\linewidth]{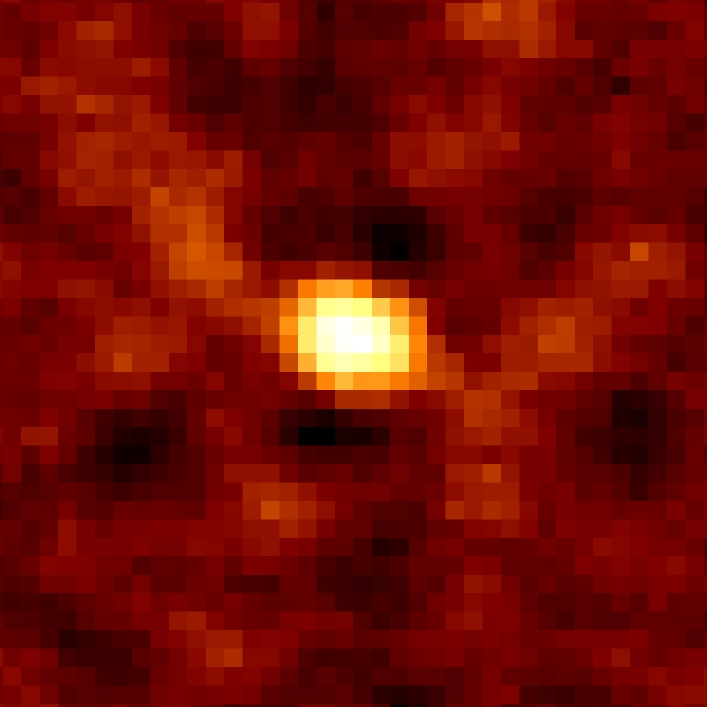}}&
\subfloat[\large \textit{SNR} = 10.76]{\includegraphics[width=.12\linewidth]{figures/stack_656_kz_sfg00.pdf}}&
\subfloat[\large \textit{SNR} = 10.60]{\includegraphics[width=.12\linewidth]{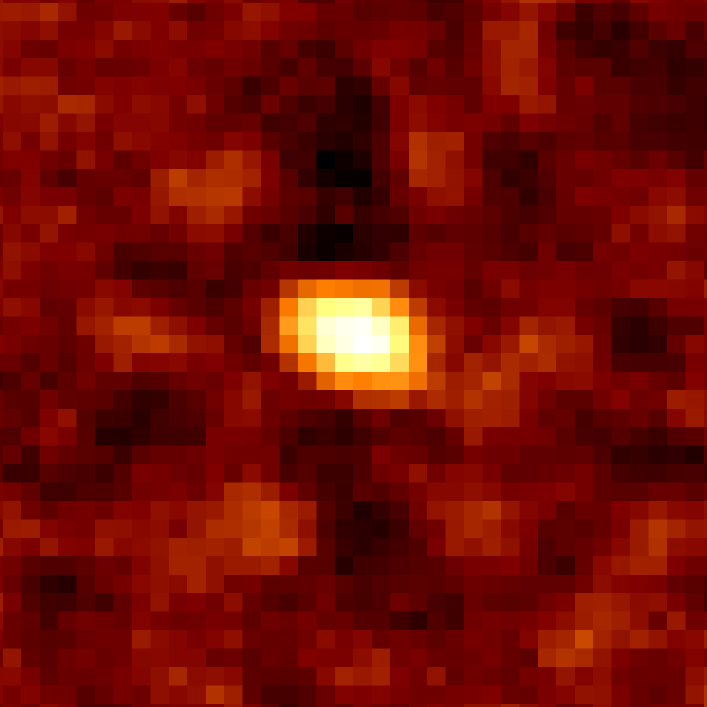}}&
\subfloat[\large \textit{SNR} = 6.16]{\includegraphics[width=.12\linewidth]{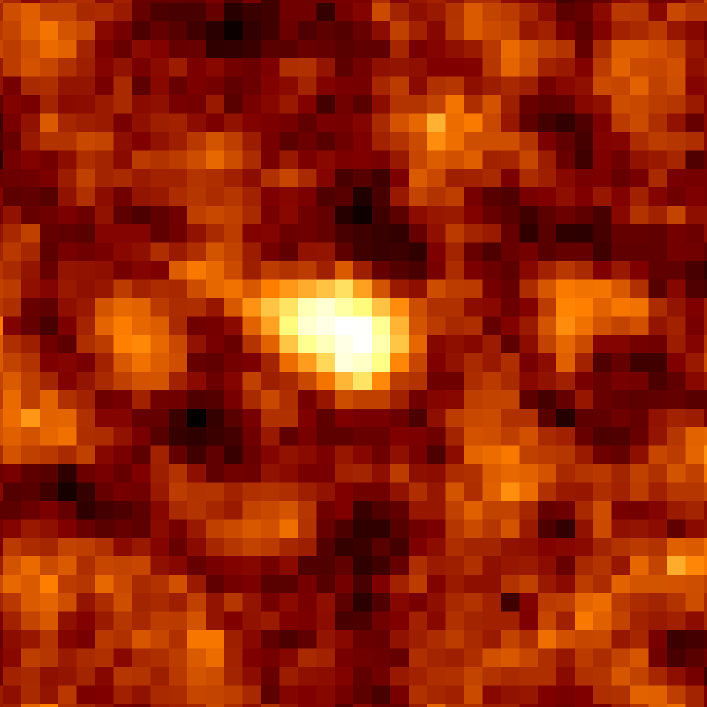}}&
\subfloat[\large \textit{SNR} = 6.08]{\includegraphics[width=.12\linewidth]{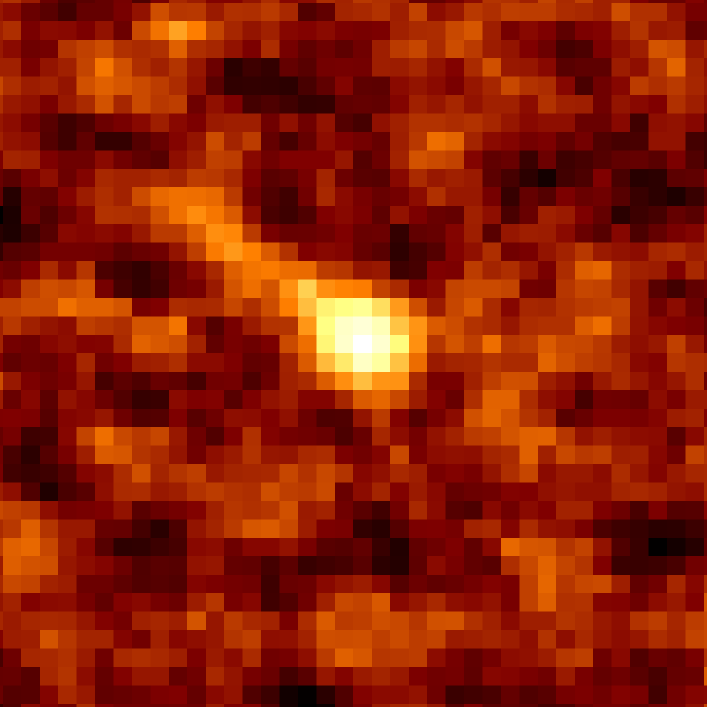}}\\

\rotatebox[origin=b]{90}{\makebox[\ImageHt]{\normalsize $\rm{720\,MHz}$}}&
\subfloat[\large \textit{SNR} = 8.34]{\includegraphics[width=.12\linewidth]{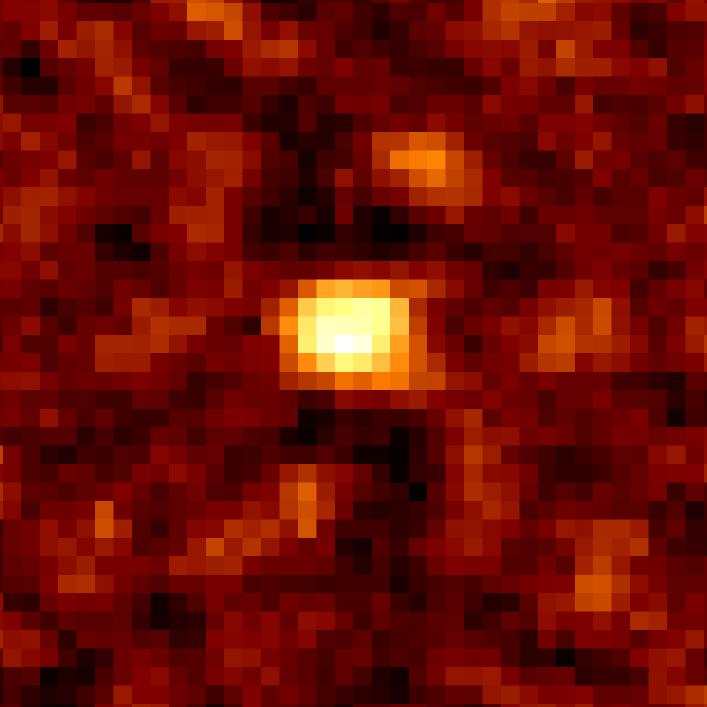}}&
\subfloat[\large \textit{SNR} = 10.45]{\includegraphics[width=.12\linewidth]{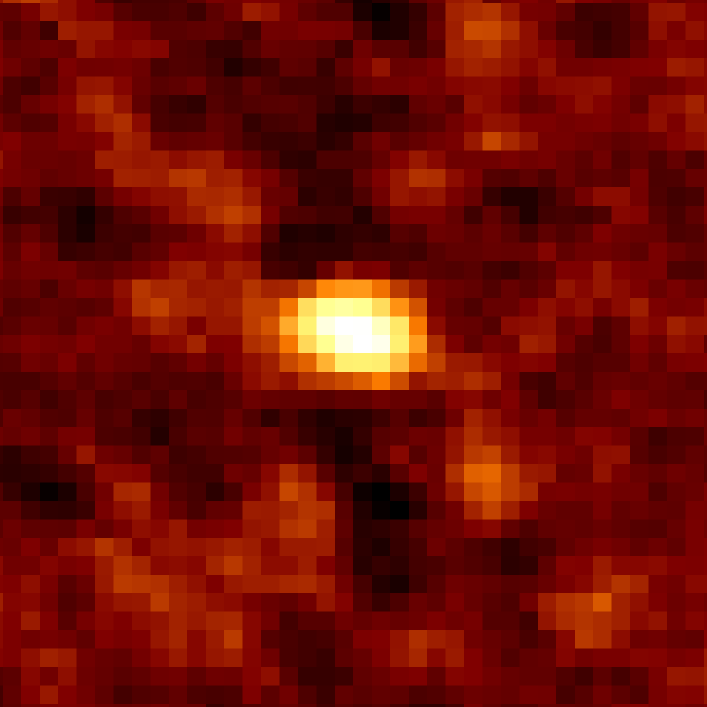}}&
\subfloat[\large \textit{SNR} = 11.91]{\includegraphics[width=.12\linewidth]{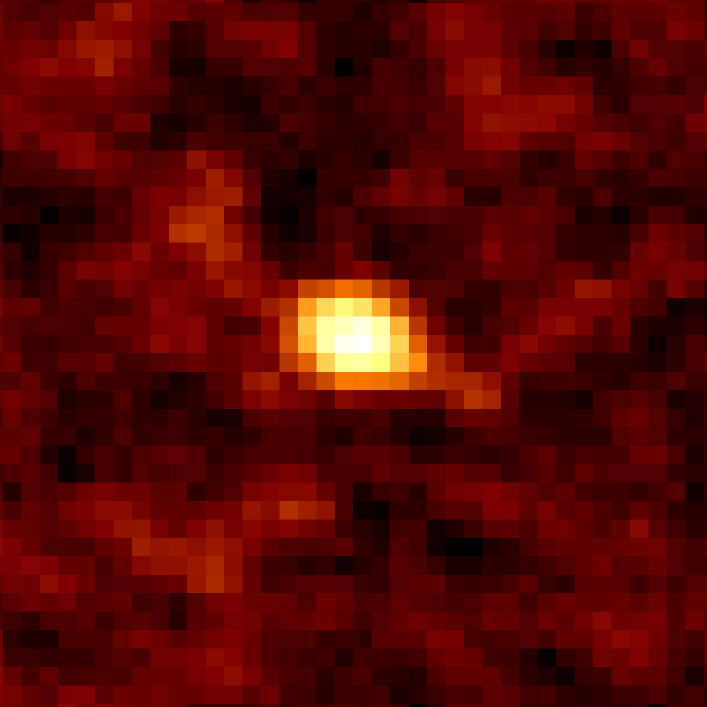}}&
\subfloat[\large \textit{SNR} = 12.12]{\includegraphics[width=.12\linewidth]{figures/stack_720_kz_sfg00.pdf}}&
\subfloat[\large \textit{SNR} = 9.39]{\includegraphics[width=.12\linewidth]{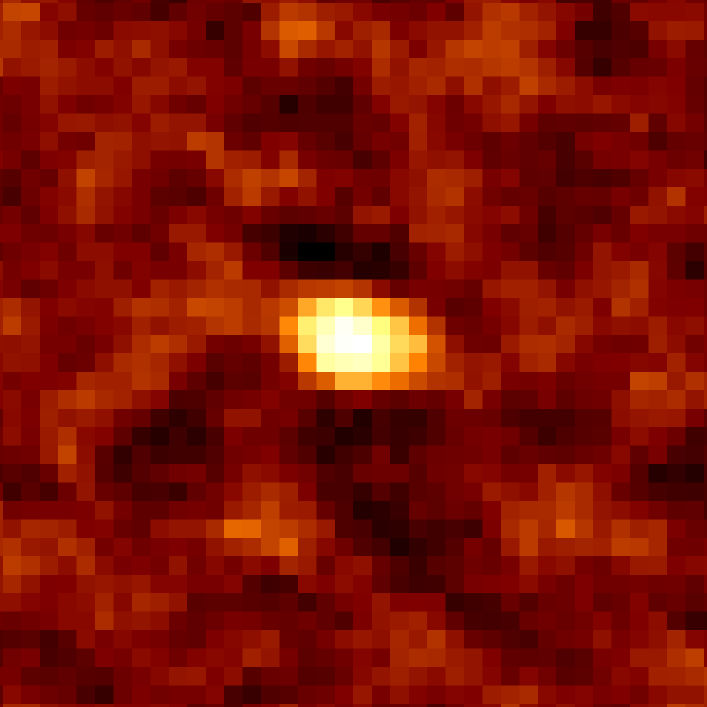}}&
\subfloat[\large \textit{SNR} = 5.95]{\includegraphics[width=.12\linewidth]{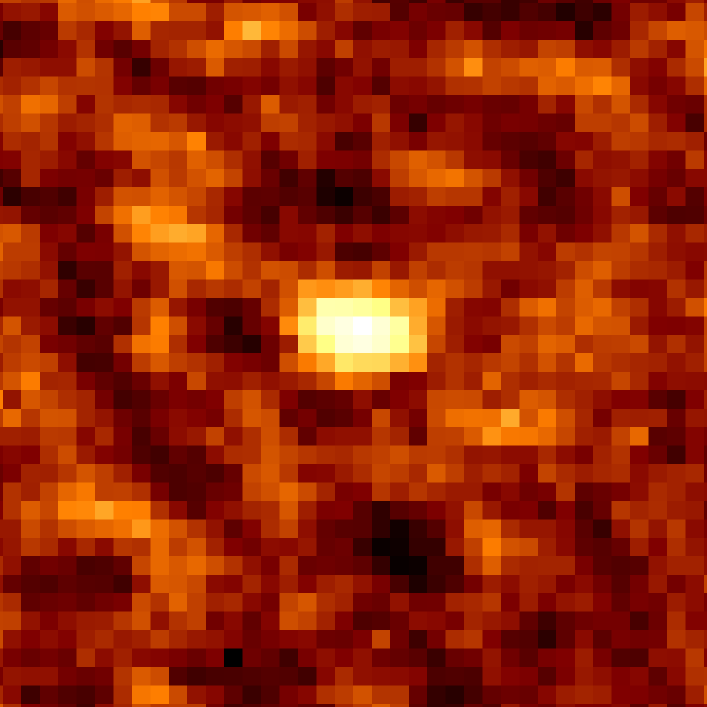}}&
\subfloat[\large \textit{SNR} = 2.76]{\includegraphics[width=.12\linewidth]{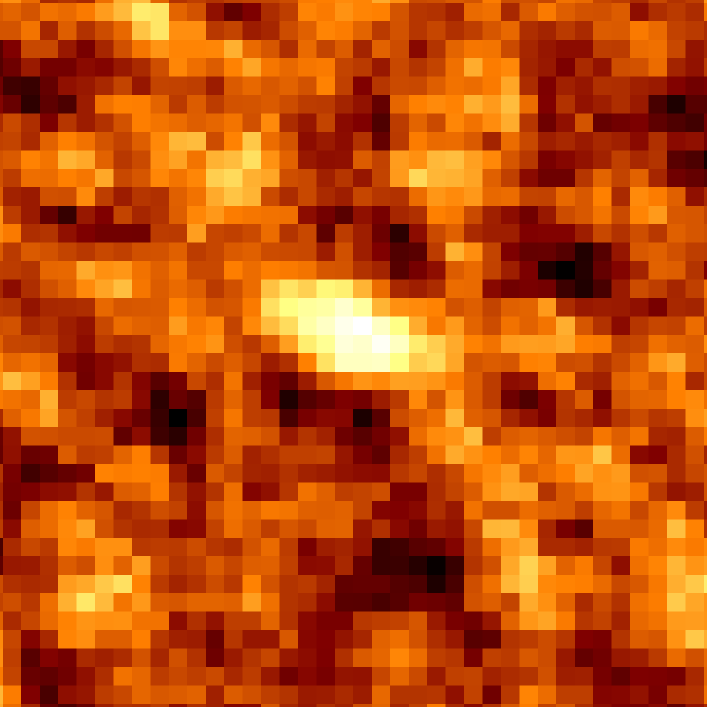}}\\

\end{tabular}
\caption{Median stacked thumbnail radio images from 144 - 720 MHz as a function of frequency and redshift for the SFGs. All image-scale ranges between 1 and 100 $\rm{\mu}$Jy beam$^{-1}$. All images have a size of $\rm{\sim 36\times36}$ arcsec$^2$. The \textit{SNR} values corresponding to each stack are written below each thumbnail image. Missing image represents a bin where no galaxies were stacked in the 144 MHz image. }%
\label{med_stack_fig_append}
\end{figure*}

\begin{figure*}
\begin{tabular}{@{}c@{ }c@{ }c@{ }c@{ }c@{ }c@{ }c@{ }c@{}}
&\textbf{\textit{z}$_{1}$=0.1-0.5} & \textbf{\textit{z}$_{2}$=0.5-0.9} & \textbf{\textit{z}$_{3}$=0.9-1.3} & \textbf{\textit{z}$_{4}$=1.3-1.7} & \textbf{\textit{z}$_{5}$=1.7-2.1} & \textbf{\textit{z}$_{6}$=2.1-2.5} & \textbf{\textit{z}$_{7}$=2.5-3.0}\\
\centering

\rotatebox[origin=b]{90}{\makebox[\ImageHt]{\normalsize $\rm{784\,MHz}$}}&
\subfloat[\large \textit{SNR} = 4.30]{\includegraphics[width=.12\linewidth]{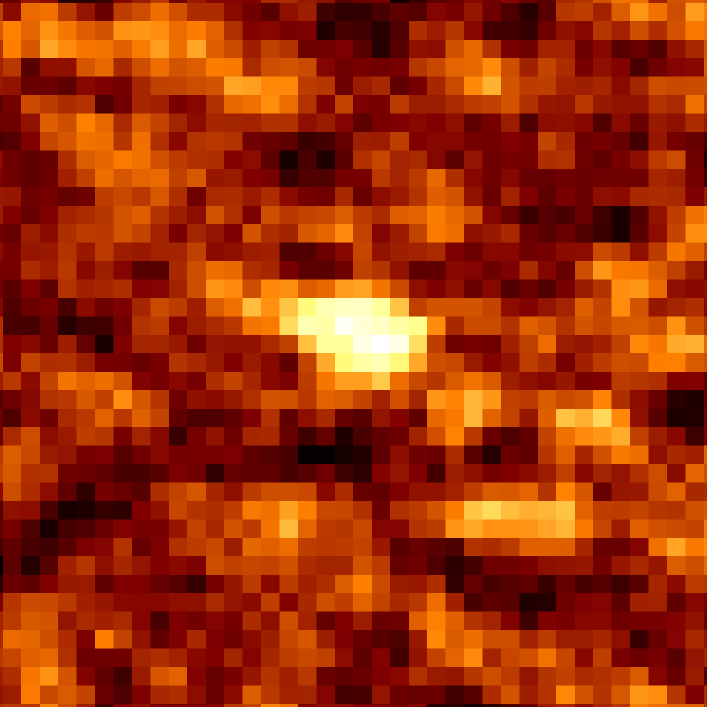}}&
\subfloat[\large \textit{SNR} = 8.60]{\includegraphics[width=.12\linewidth]{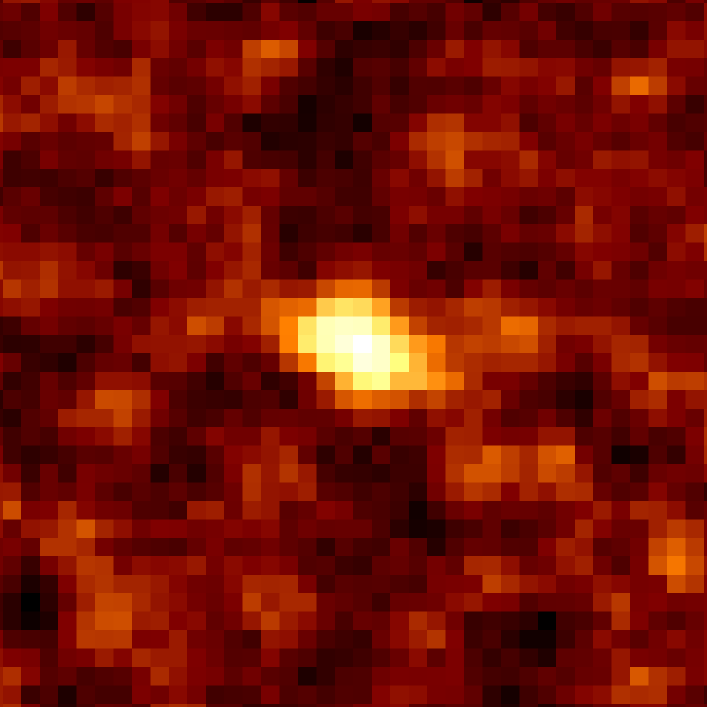}}&
\subfloat[\large \textit{SNR} = 10.29]{\includegraphics[width=.12\linewidth]{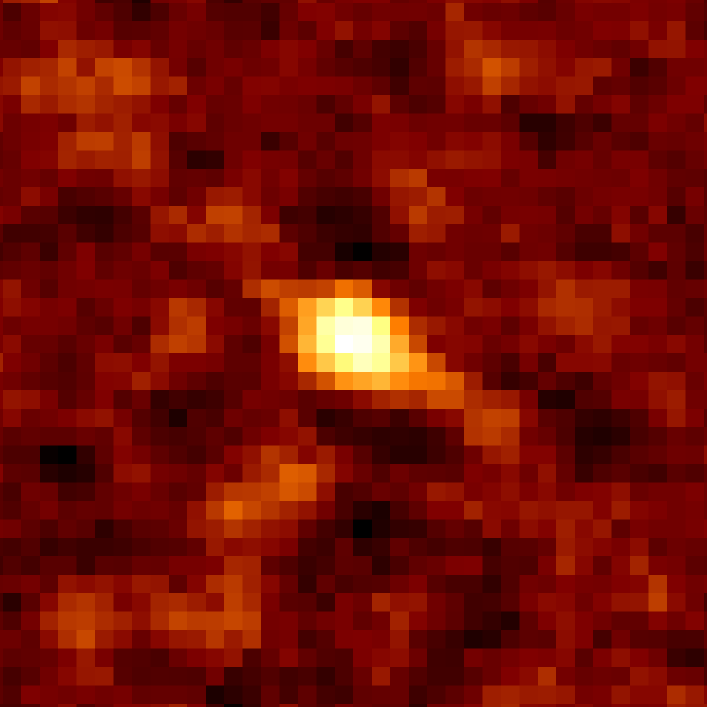}}&
\subfloat[\large \textit{SNR} = 8.44]{\includegraphics[width=.12\linewidth]{figures/stack_784_kz_sfg00.pdf}}&
\subfloat[\large \textit{SNR} = 8.98]{\includegraphics[width=.12\linewidth]{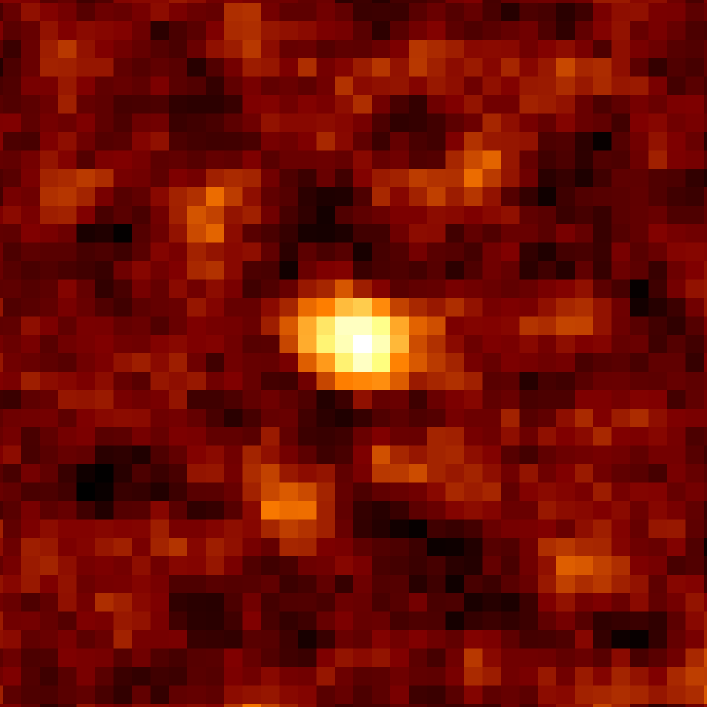}}&
\subfloat[\large \textit{SNR} = 5.81]{\includegraphics[width=.12\linewidth]{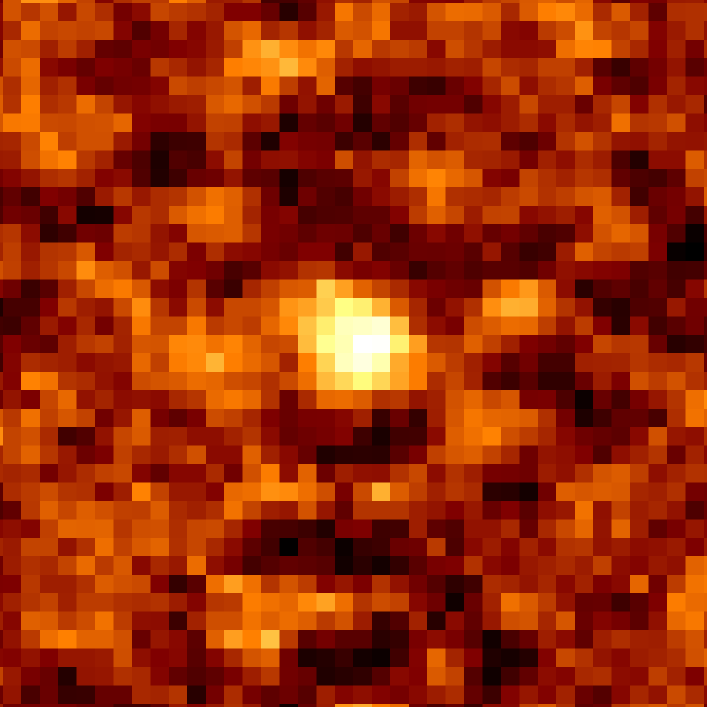}}&
\subfloat[\large \textit{SNR} = 3.33]{\includegraphics[width=.12\linewidth]{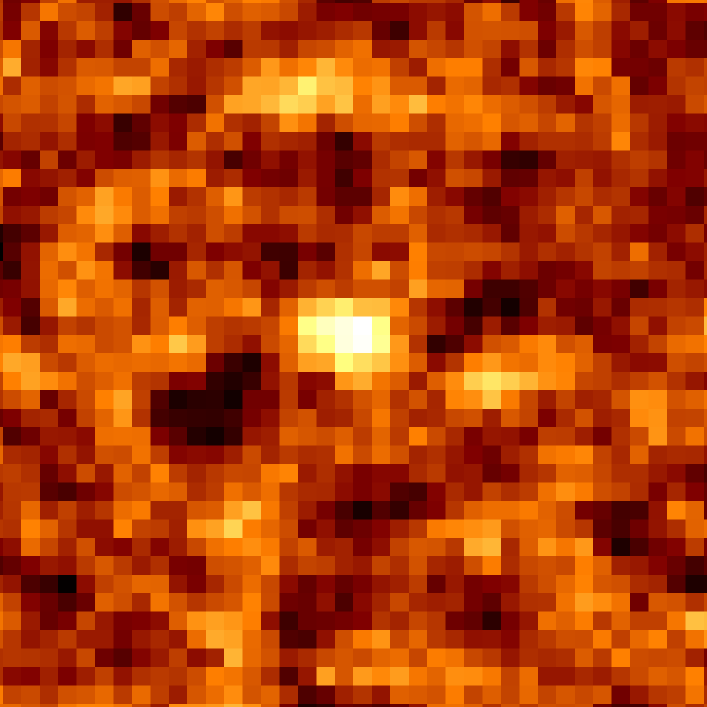}}\\

\rotatebox[origin=b]{90}{\makebox[\ImageHt]{\normalsize $\rm{1284\,MHz}$}}&
\subfloat[\large \textit{SNR} = 23.76]{\includegraphics[width=.12\linewidth]{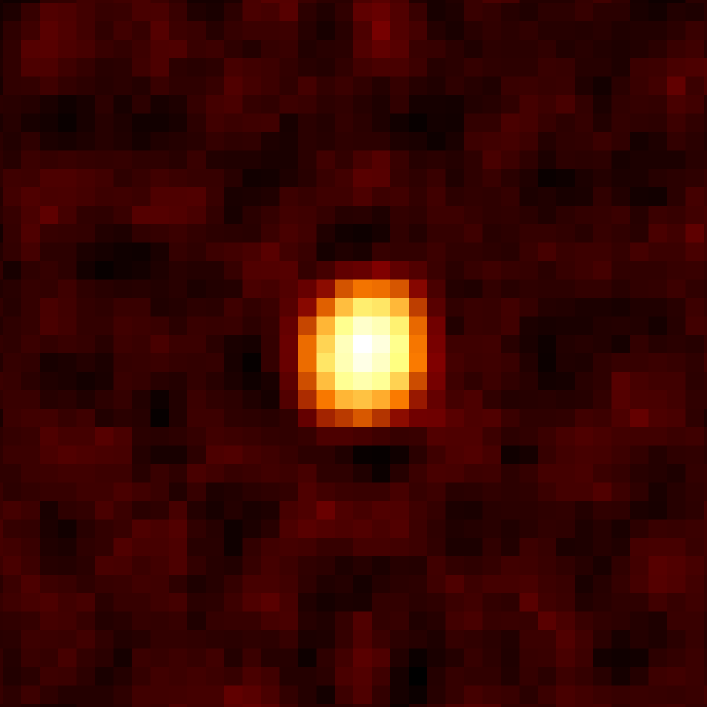}}&
\subfloat[\large \textit{SNR} = 34.99]{\includegraphics[width=.12\linewidth]{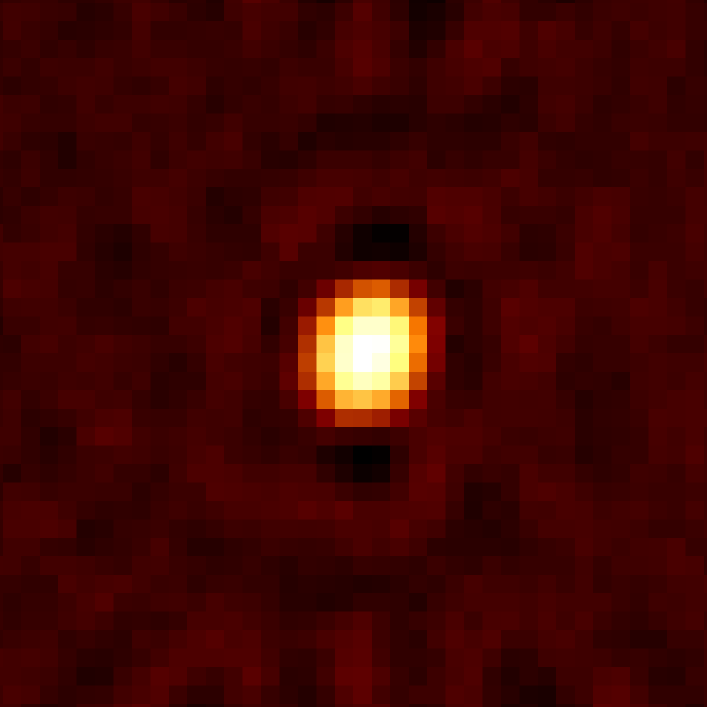}}&
\subfloat[\large \textit{SNR} = 38.53]{\includegraphics[width=.12\linewidth]{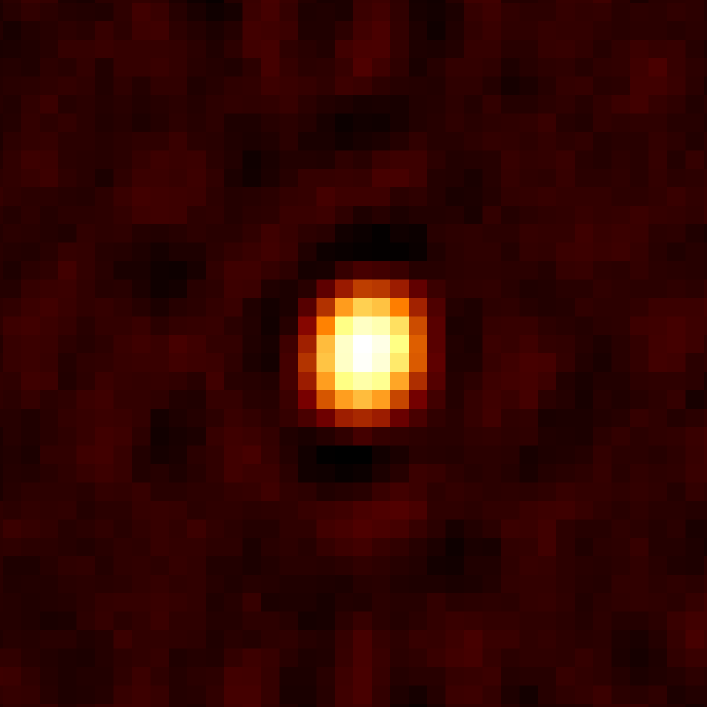}}&
\subfloat[\large \textit{SNR} = 34.16]{\includegraphics[width=.12\linewidth]{figures/stack_1280_kz_sfg00.pdf}}&
\subfloat[\large \textit{SNR} = 33.01]{\includegraphics[width=.12\linewidth]{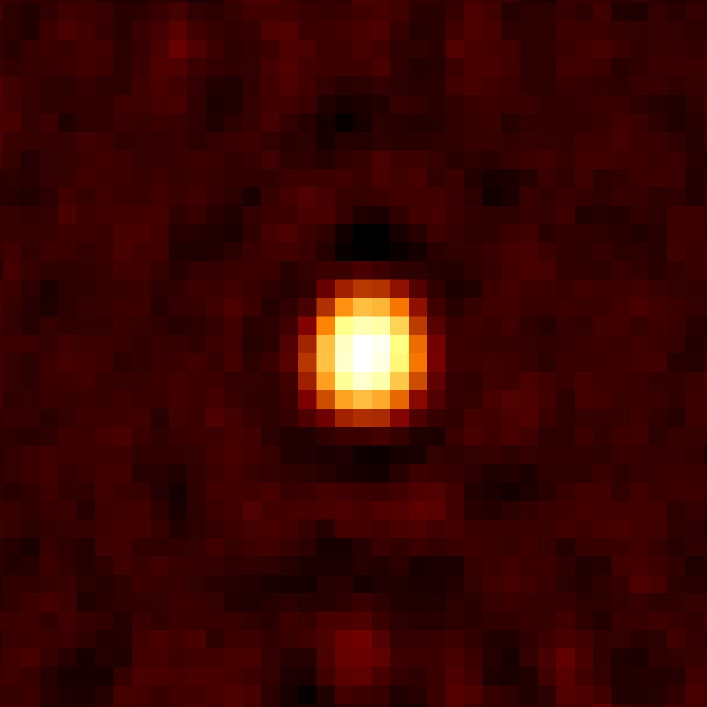}}&
\subfloat[\large \textit{SNR} = 19.47]{\includegraphics[width=.12\linewidth]{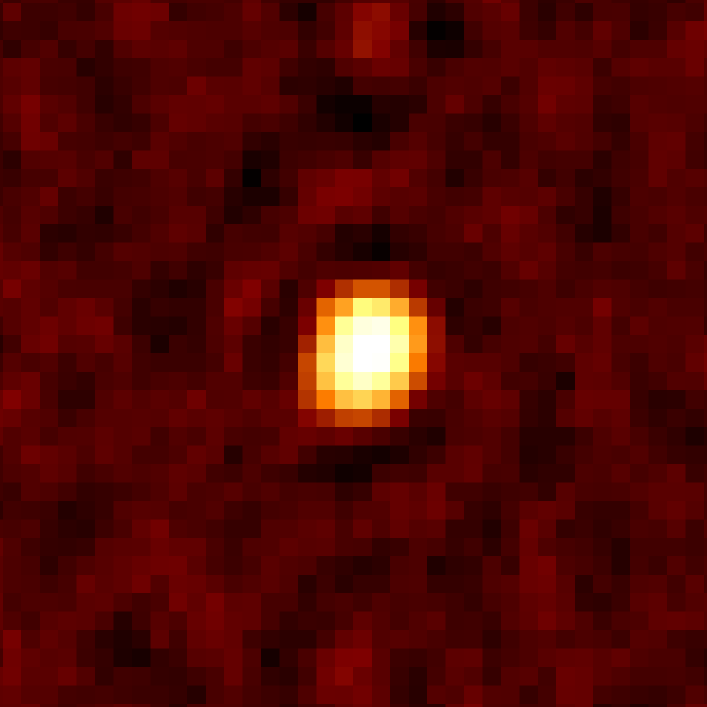}}&
\subfloat[\large \textit{SNR} = 15.75]{\includegraphics[width=.12\linewidth]{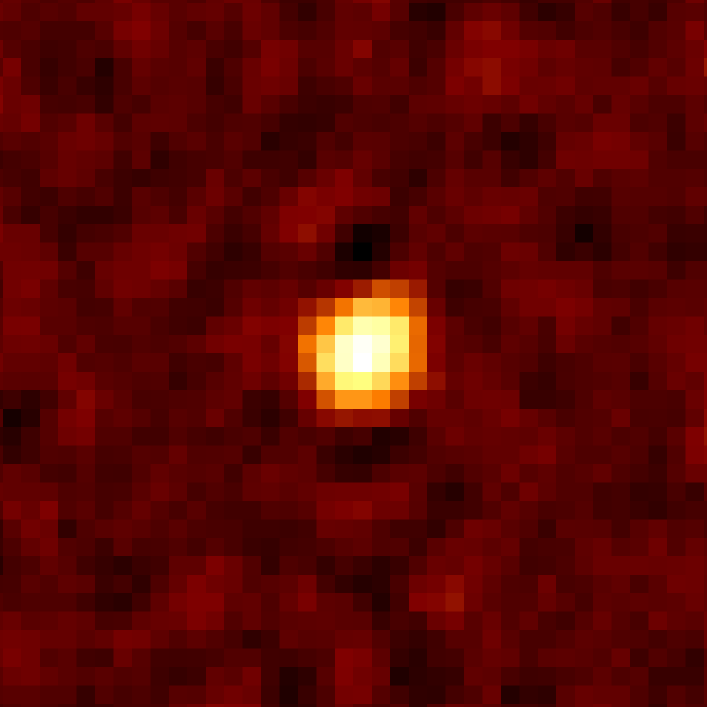}}\\

\rotatebox[origin=b]{90}{\makebox[\ImageHt]{\normalsize $\rm{1500\,MHz}$}}&
\subfloat[\large \textit{SNR} = 9.89]{\includegraphics[width=.12\linewidth]{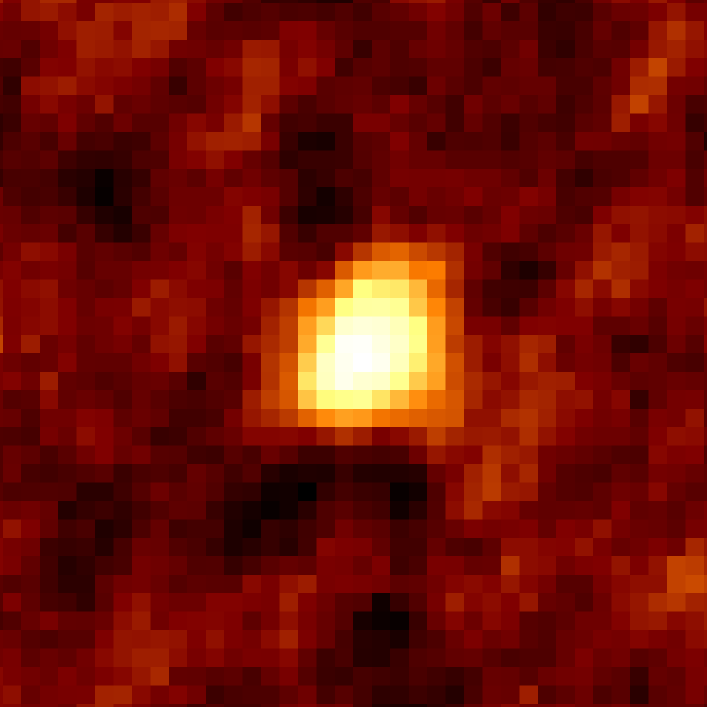}}&
\subfloat[\large \textit{SNR} = 17.29]{\includegraphics[width=.12\linewidth]{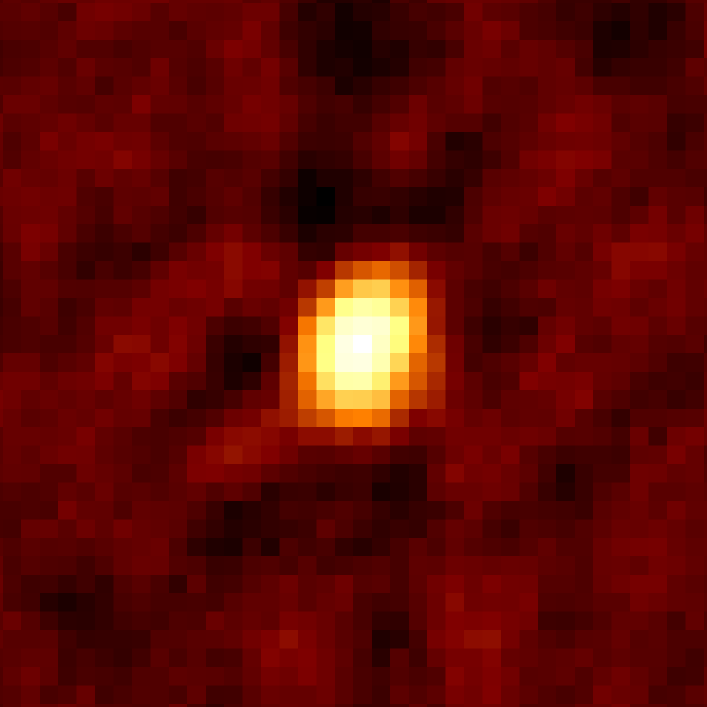}}&
\subfloat[\large \textit{SNR} = 15.84]{\includegraphics[width=.12\linewidth]{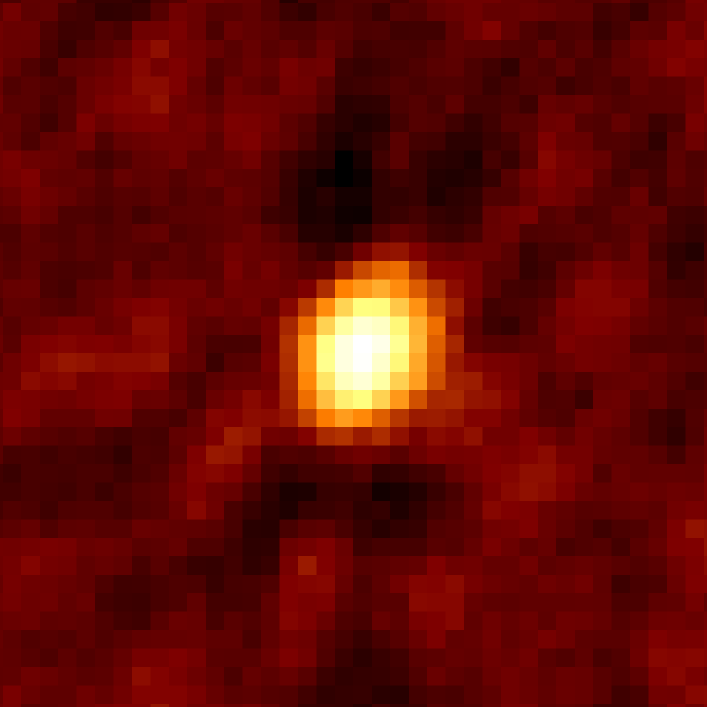}}&
\subfloat[\large \textit{SNR} = 16.79]{\includegraphics[width=.12\linewidth]{figures/stack_1500_kz_sfg00.pdf}}&
\subfloat[\large \textit{SNR} = 14.21]{\includegraphics[width=.12\linewidth]{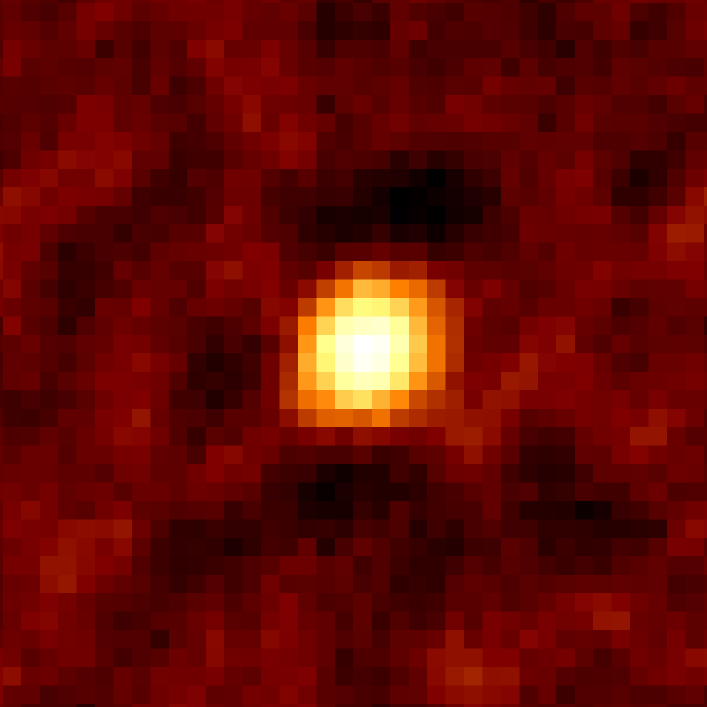}}&
\subfloat[\large \textit{SNR} = 11.26]{\includegraphics[width=.12\linewidth]{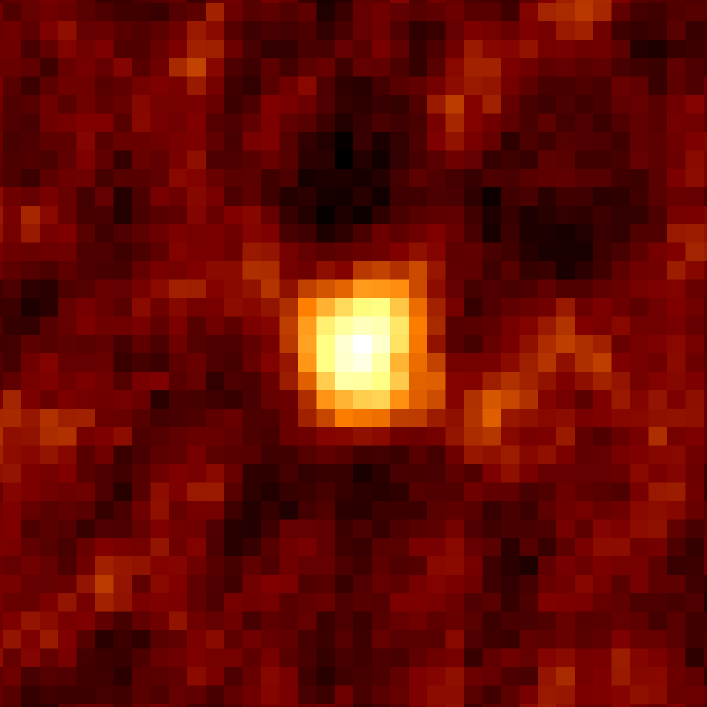}}&
\subfloat[\large \textit{SNR} = 7.67]{\includegraphics[width=.12\linewidth]{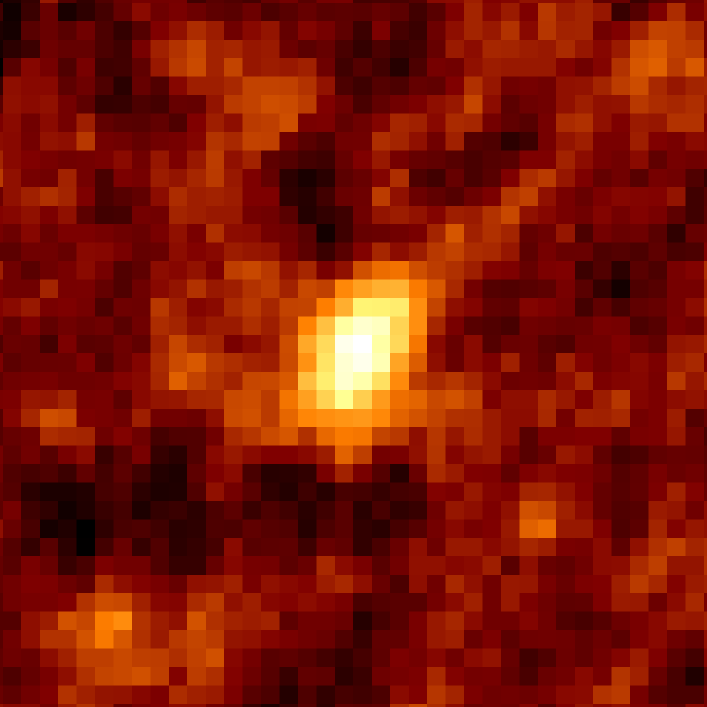}}\\

\end{tabular}
\caption{Same as Figure~\ref{med_stack_fig_append}. The median stacked thumbnail radio images shown here are from 784 - 1500 MHz as a function of frequency and redshift for the SFGs.}%
\label{med_stack_fig_append_}
\end{figure*}

In this work, we use the radio images described in Section~\ref{data.sec} (see Table~\ref{tab_data}). In the stacking of these radio images, we utilise the Python Astronomical Stacking Tool Array (\texttt{PASTA}) \citep{2018ascl.soft09003K} programme \footnote{\url{https://github.com/bwkeller/PASTA}}. The stacking algorithm entails cutting out square patches (thumbnails) of the maps around the (RA, DEC) coordinate of each galaxy in each redshift bin and then computing the median stacked image of all thumbnails in a bin. Stacking allows the investigation of bulk properties of objects above thresholds and in catalogues in one waveband, but generally (not exclusively) below survey thresholds in another band. 
Radio images have pixel units of Janskys (Jy) per beam and adhere
to the convention whereby each pixel value is equal to the flux
density of a point source located at that position \citep[e.g. see,][]{2011MNRAS.410.1155B}.

We stack at different radio frequencies using the LOFAR, \textit{u}GMRT, MeerKAT, and  VLA maps in the XMM-LSS field in seven redshift bins, to investigate the spectral properties of average radio sources below the survey threshold. The bins are divided such that there is a large number of SFGs in each bin. For each redshift bin, the stacked signal is constructed by fitting Gaussian source components to the stacked radio thumbnails using the \textsc{PYBDSF} source finder \citep{2015ascl.soft02007M}, and using the peak or integrated flux density in the median stacked image (see Appendix~\ref{stack_analysis} for more details) as the representative value for the population of the thumbnails centered on all SFGs in that redshift bin. 

Consequently, the stacked sources can be classified into resolved and unresolved by comparing $\rm{\textit{S}_{int}}$ to $\rm{\textit{S}_{peak}}$, where $\rm{\textit{S}_{int}}$ is the integrated flux density and $\rm{\textit{S}_{peak}}$ is the peak flux density, respectively, taking the errors on the flux measurements into account.
Following  \citet{2015MNRAS.453.4020F,2019PASA...36....4F,2020MNRAS.495.4071M} we compute our total flux uncertainty by combining the two measurements in quadrature:
\begin{equation}\label{rms}
\centering
\rm{\sigma_{R}\,=\,\sqrt{\left(\frac{\sigma_{int}}{\textit{S}_{int}}\right)^{2}+\left(\frac{\sigma_{peak}}{\textit{S}_{peak}}\right)^{2}}}
\end{equation}
where $\rm{\sigma_{int}}$  and $\rm{\sigma_{peak}}$ are the uncertainties in the $\rm{\textit{S}_{int}}$ and the $\rm{\textit{S}_{peak}}$, respectively. We classify the stacked source as resolved if $\rm{\textit{ln}(\textit{S}_{int}/\textit{S}_{peak})\, >\, 3\sigma_{R}}$ \citep[see,][]{2015MNRAS.453.4020F}. This criterion
is elaborated in Appendix~\ref{stack_analysis}, where we compare to the conventional empirical definition of the envelope we used in \cite{2020MNRAS.491.1127O}. Table~\ref{tab_stacked_ppts} lists
the median \textit{S}, flux densities for each stacked image for the different frequencies and the number of sources included in each stack $\rm{(N_{stack})}$ in each redshift range. The images we analyse have varying coverage areas and sensitivity levels. Sources located too close to the image edge where a full stamp cannot be extracted are excluded from the stamp extraction process. Similarly, sources that fall within or near blank regions of the FITS image are also omitted. Consequently, the number of sources included in each stack (or redshift bin) differs across frequency bands. However, these variations have minimal to no impact on the resulting stacked radio SED. For the stacks with resolved emission, we prefer to use their integrated flux as the most accurate estimate. Instead, for unresolved stacks we use the peak flux.
From the Gaussian fitting error calculated from \textsc{PYBDSF},
 we obtain the error on the measure peak or integrated flux, $\rm{\sigma_{peak\,|int}}$. Additionally, e estimate the stacked
background noise by taking the robust standard deviation
in a 2D stack of $\rm{15^{\prime\prime}\,\times\,15^{\prime\prime}}$ region around each source. For our total flux uncertainty we combine the two measurements in
quadrature, i.e. $\rm{\sigma^{2}_{Tot}\,=\,\sigma^{2}_{peak\,|int}\,+\,\sigma^{2}_{BG}}$. 

Median image stacking technique is best applied in the radio regime where the angular resolution is high and the fraction of direct detections is comparatively low such that blending of sources is negligible \citep[see,][]{2011ApJ...730...61K}.  

In practice, we hardly ever use the mean image because it is sensitive to interloping sources in crowded fields and to image artefacts around bright sources.
We test for potential bias in the stacked flux via a \enquote{null} stacking analysis: stacking randomly drawn positions distributed equally over the sky coverage of each of the survey fields. We draw null stacks with equal numbers of random sky positions as there are SFGs in all  redshift bins for each band. All of the stack procedures remain the same as for stacking on the SFG locations. We repeated the  process several times by randomizing the positions of the SFGs in each redshift bin.The medians of these random stacked fluxes for each band are scattered around zero in an unbiased manner. \citet{2017MNRAS.472.2221S} demonstrated that random stacks fluctuating around around zero indicates that the contribution to stacked fluxes from source confusion due to the blending of faint sources is negligible. All median stacked flux densities reported in Table~\ref{tab_stacked_ppts} incorporate corrections for median boosting effects. We investigate the effect of median boosting by testing our stacking routine on realistically generated mock sources in  subsection~\ref{boosting.sec} of Appendix~\ref{stack_analysis}. The stacked median is boosted relative to the true median when radio map noise matches or exceeds the typical flux density of the underlying galaxy population \citep[see,][]{2007ApJ...654...99W,2022ApJ...924...76A}.  

Figure~\ref{med_stack_fig_append} and ~\ref{med_stack_fig_append_} presents the median stacked thumbnail radio images centered on the \textit{NIR} position of each input galaxy as a function of frequency and redshift. for the SFGs. Also included in this plot is the SNR of each stack printed below each image.  In the case of 720 MHz stacked image at redshift $\rm{2.5\leq\textit{z}\leq3.0}$ we measure an \textit{SNR}  = $\sim$2.76. All image-scale ranges between 1 and 100 $\rm{\mu}$Jy beam$^{-1}$. The median image provides a compelling visual impression of the statistical significance of the sample median compared to nearby off positions.  All images have a size of $\rm{\sim 36\times36}$ arcsec$^2$.
The empty space in the first row of Figure~\ref{med_stack_fig_append} represent redshift range where no median stacked image was produced (i.e $\rm{\textit{z}\in[0.5-0.9]}$ for 144 MHz). The Stamp contains NaN pixels, and hence it is rejected. This can occur when a source is in or near a blank area of the FITS image.

\section{Spectral Modelling}\label{results.sec}
In this section we constrain the shape of the average radio SED. We describe the SEDs of the stacked SFG subsample in the different redshift  bins.

\subsection{Stacked radio spectral properties at observer-frame 144 – 1500 MHz}\label{SED_ppts}

Understanding the spectral properties of sources is crucial for the characterization of the radio source population. Since the main goal of this paper is to investigate the radio SED properties of these stacked SFGs, in this section we detail the models used for fitting.

A simpler and common, approach to characterising radio spectra is to use flux densities measured at three widely separated frequencies to define two radio colors \citep{2006MNRAS.371..898S,2010MNRAS.402.2403M,2022ApJ...934...26P}. Hence, we compute  the representative two-point median stacked spectral index using $\rm{\alpha = \frac{\log (\textit{S}_{1} / \textit{S}_{2})}{\log (\nu_{1} / \nu_{2})}\,\pm\,\delta{\alpha}}$ and an associated uncertainty $\rm{\delta{\alpha} = \frac{1}{ln (\nu_{1} / \nu_{2})}}\sqrt{\left(\frac{\Delta \textit{S}_{1}}{\textit{S}_{1}}\right)^{2}+\left(\frac{\Delta \textit{S}_{2}}{\textit{S}_{2}}\right)^{2}}$. The $\rm{\Delta \textit{S}_{1}}$ and $\rm{\Delta \textit{S}_{2}}$ represents the $\rm{\sigma_{Tot}}$ (see subsection~\ref{stack.sec}) at the corresponding redshift of the respective frequency channel.

Also, we divide the SEDs into three sections, low frequency (144 MHz - 320 MHz, referred to herein as $\rm{\alpha_{low}}$), mid-frequency (320 MHz  - 656 MHz, referred to herein as $\rm{\alpha_{mid}}$),  and high frequency (656 MHz - 1500 MHz, referred to herein as $\rm{\alpha_{high}}$). This choice of dividing frequency (i.e. 656 MHz) averages over the part of the spectrum where we might see the effects of spectral flattening most prominently \citep[see,][]{2021A&A...648A..14R}.  The terms "low frequency" and "high frequency" are purely relative our case. In addition, since our radio data span the range 144-1500 MHz, we also considered $\rm{\alpha_{range}}$ (i.e. 144-1500 MHz).

\subsection{Radio Continuum Models}
We investigate, the stacked radio spectral properties of the sources in deep extragalactic radio survey presented in Section~\ref{data.sec} by fitting models to our median stacked measurements.
Given the moderate coverage of our radio continuum data,
which covers 144 MHz to 1500 MHz in the observed frame, and
size of our sample we elected to fit a series of simple models to the stacked SFG spectra.  All modelling was performed in the observed frame with a reference frequency, unless stated
otherwise, of $\rm{\nu_{0}\,= \frac{1}{2}\times(\nu_{min}+\nu_{max})}$ MHz.

\subsubsection{Power Law (\textit{PL})}\label{pl}
The frequency behaviour of the dominant emission mechanism is typically synchrotron emission most commonly modelled as a power law in frequency,

\begin{equation}\label{spec_indx.eqn}
 \rm{\textit{S}_{\nu} = \textit{S}_{o}\left(\frac{\nu}{\nu_{o}}\right)^{\alpha}}
\end{equation}

or  in log-log space,

\begin{equation}\label{spec_indx_log.eqn}
 \rm{\log \textit{S}_{\nu} = log \textit{S}_{o} + \alpha \log \left(\frac{\nu}{\nu_{o}}\right)}
\end{equation}

where $\rm{\textit{S}_{\nu}}$ is the flux density as a function of frequency $\nu$, $\rm{\textit{S}_{o}}$ is the flux density at the reference frequency, $\nu_{o}$ and $\alpha$ is the spectral index. 
 
Defining $\rm{\textit{x}\,=\,\log \left(\frac{\nu}{\nu_{o}}\right)}$ and $\rm{\textit{y}\,=\,\log \textit{S}_{\nu}}$ allows us to compute the spectral index $\alpha$ from N multi-frequency measurements using standard linear regression. 

\subsubsection{Curved Power Law (\textit{CPL})}\label{cpl}

Given that we do not see evidence of a curvature 

in the stacked spectra,
especially over such a large frequency range, we  fit a generic curved power law model. 
The curved power law is defined by:
\begin{equation}\label{curvedPL.eqn}
S_{\nu} = S_{o}\left(\frac{\nu}{\nu o}\right)^{\alpha} exp \left(q\ln\left(\frac{\nu}{\nu o}\right)^{2}\right)
\end{equation}

This includes the additional free parameter \textit{q} to capture any higher order spectral curvature features, where increasing |\textit{q}| captures stronger deviations from a simple power law. 
The $\textit{q}$ term, offers a parameterisation of the curvature in the spectrum. Where $\textit{q}\,<\,0$ describes a concave spectrum, and suggests the curve is opening downward. If \textit{q} is positive, the curve is opening upward (convex). 
For optically thin synchrotron emitted from radio lobes, $\textit{q}$ typically ranges within $\rm{-0.2\,\leq\,\textit{q}\,\leq\,0}$ \citep[see,][]{2021PASA...38....8Q,2021Galax...9..121S}.  A significant
curvature is represented by values of |\textit{q}| > 0.2. The
spectral curvature flattens towards a standard power law as
\textit{q} approaches zero or near-zero.

This model is not physically motivated, and may not appropriately describe sources with different power law slopes in the optically thin and thick regimes, but provides a useful filter to identify interesting sources.

\subsubsection{Synchrotron and Free-Free  Emission (FFE) (Double Power Law, \textit{DPL})}\label{sync_free.sec}
Studies have shown that SFGs may display complex radio SEDs \citep[see,][]{1992ARA&A..30..575C,2008A&A...477...95C,2017MNRAS.468..946S,2018A&A...619A..36C}. \citet{2010MNRAS.405..887C} illustrated that a wide variety of SEDs for SFGs show turnovers at low frequency.
We can model the radio continuum as the sum of two distinct power laws \citep[see,][]{2015ApJ...809..168C,2017ApJ...836..174C,2018MNRAS.474..779G}. One representing the steep spectrum nonthermal synchrotron emission, and the second describing the
flat spectral thermal free-free emission, following the form:
\begin{equation}\label{syn_freefree.eqn}
\rm{S_{\nu}\,=\,\textit{A}\left(\frac{\nu}{\nu_{0}}\right)^{-0.1}}\,+\,\textit{B}\left(\frac{\nu}{\nu_{0}}\right)^{\alpha_{NT}}
\end{equation}

where \textit{A} and \textit{B} are treated as free parameters, and represent the free-free (thermal) and  synchrotron (non-thermal) normalisation components respectively.  The free parameter $\rm{\alpha_{NT}}$ represents the synchrotron spectral index, and we fix the free-free spectral index $\rm{\alpha_{FF}}$   to be -0.1. This model describes both synchrotron and free-free emission components as being completely optically thin (i.e. no curvature at low frequencies), and the spectra are the superposition of two power-laws \citep[see,][]{2018MNRAS.474..779G,2021ApJ...912...73A}.
The superposition of thermal and synchrotron radiation thus produces a flattening of the total radio spectrum toward high frequencies, with the asymptotic value of -0.1. This, however, can hardly be seen, owing to the onset of thermal radiation from dust, which becomes more significant above about 10 GHz \citep[see,][]{2018A&A...611A..55K}.

\subsection{Model Fitting}

We applied the Levenberg-Marquardt non-linear least-squares regression algorithm using the \textsc{CURVE\textunderscore FIT} as implemented in the \textsc{Scipy} python module; \citealt{2020NatMe..17..261V}).
We provide columns in Table~\ref{tab_slope_alpha} to show the measurements of these statistics for the stacked SFGs. The errors in the fits  are the estimated covariance of the optimal values for the parameters given by \textsc{CURVE\textunderscore FIT} taking into account the flux errors. The diagonals provide the variance of the parameter estimates. To compute one standard deviation errors  on the parameters we use the  square root of the diagonal covariance matrix 

\subsubsection{Model Comparison/Selection}

We computed the values of the chi-square ($\rm{\chi^{2}}$) , reduced chi-squared ($\rm{\chi^{2}_{red}}$) ,and the corrected Akaike information criterion  \citep[\textit{AICc};][]{1974ITAC...19..716A} . 

The standard (uncorrected) \textit{AIC} \citep{1974ITAC...19..716A} takes into account the goodness of the fit and the complexity of the model at the same time. The "uncorrected" Akaike information criterion is defined by \citet{2018MNRAS.476.2717H} as, $\rm{\textit{AIC}\,=\,\chi^{2}\,+\,2\textit{k}}$, where \textit{k} is the number of fitted parameters. Among the different models, the one that best represents the data with the minimum amount of parameters corresponds to the model with the smallest \textit{AICc} value. 

Model comparison requires a metric that accounts for a different number of fit parameters.
Thus, to compare the three models we fit to the stacked SFG radio SEDs, we use the corrected form of the \textit{AICc} which was introduced by \citet{ref1} and defined as:
\begin{equation}
\rm{\textit{AICc}\,=\,\chi^{2}\,+2\textit{k}\,+\,\frac{2\textit{k}(k+1)}{\textit{N}\,-\,\textit{k}\,-\,1} }
\label{eq:AICc}
\end{equation}
where \textit{N} is the number of data points. The more parameters, the higher  \textit{AICc}.
The \textit{AICc} accounts for the number of degrees of freedom in a more detailed manner than the  $\rm{\chi^{2}_{red}}$ \citep[see,][]{2018MNRAS.476.2717H}, and can compare non-nested models. The \textit{AICc} is also used when the number of data points is small compared with the number of fitted parameters. Generally, \textit{AICc} is preferred for comparing models when a true model may not be among the candidates. The model which results in the lowest \textit{AICc} is most likely to be the model that most accurately describes the SFGs spectra.

Additionally, we also compute a Bayesian Information Criterion (BIC) for each model, which in the case of Gaussian distributed model errors is given by:
\begin{equation}
\rm{\textit{BIC}\,=\,\chi^{2}\,+\,\textit{k}\ln N}
\label{eq:BIC}
\end{equation}

We calculate $\rm{\Delta \textit{BIC}}$, where $\rm{\Delta \textit{BIC} = \textit{BIC}_{1}\,-\,\textit{BIC}_{2}}$. When $\rm{\Delta \textit{BIC}\,>\,0}$, this suggests a preference towards the second model. Similarly, if $\rm{\Delta \textit{BIC}\,<\,0}$, then there is a preference towards the first model. Weak model preference is implied by the following range $\rm{0\,<\,|\Delta \textit{BIC}|\,<\,2}$, whereas a model is strongly preferred if $\rm{|\Delta \textit{BIC}|\,>\,6}$ \citep[see][]{kass1995bayes}.

Hence, we use these information criteria as a scoring system for model comparison as commonly done in classical statistics when dealing with models with different numbers of free parameters.

\section{Results \& Discussions}\label{res_disc.sec}

The multi-frequency nature of this study allows the estimation of the stacked radio spectral properties at observer-frame 144 – 1500 MHz. 
In this section, we discuss our findings, and the global astrophysical implications of our sample of stacked SFGs. We discuss the possible physical mechanisms that determine the radio spectra over the frequency range of this study.

\subsubsection{Radio Colour-Colour Plot (RCC Diagram)}

\begin{figure*}
\centerline{\includegraphics[width=0.8\textwidth]{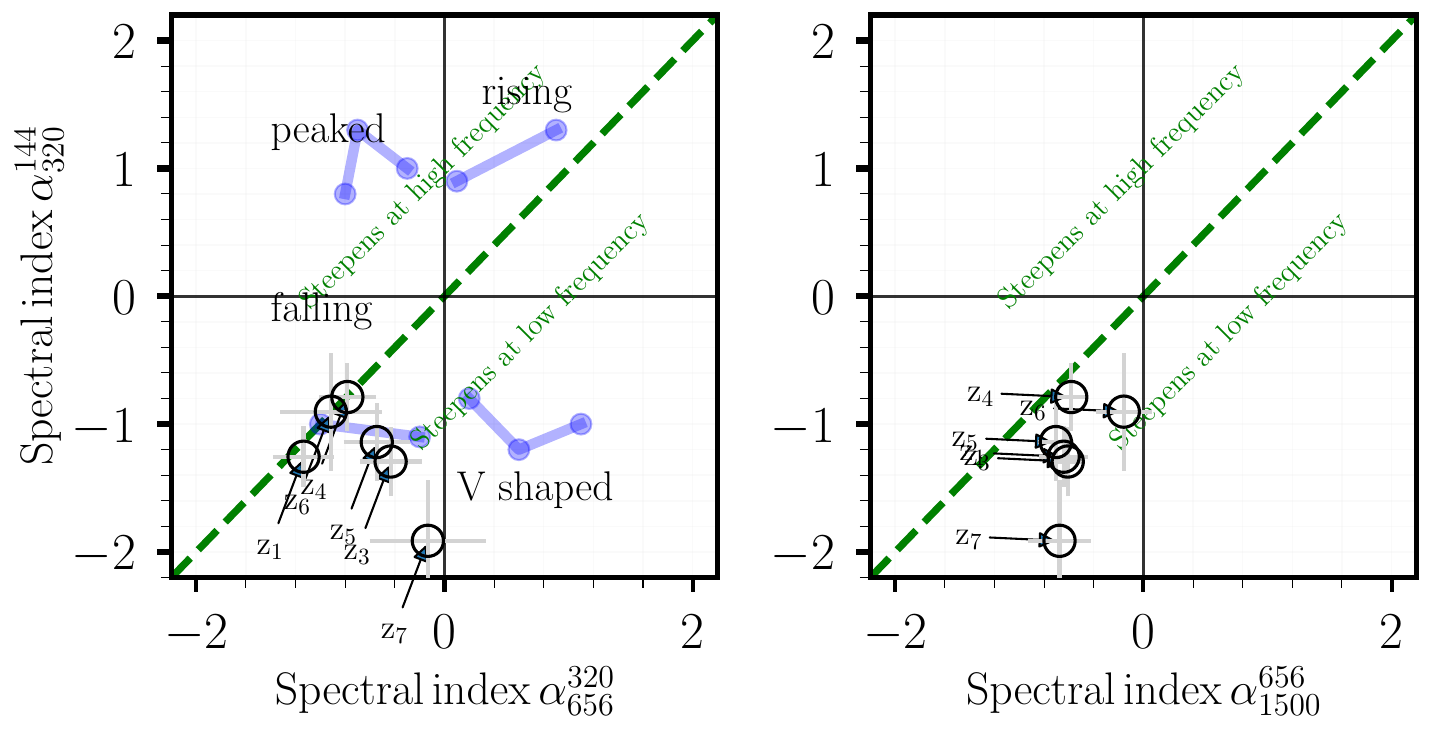}}
    \caption{Left: Stacked representative radio colour-colour plot, $\rm{\alpha_{[144\,-\,320\,MHz]}}$ vs $\rm{\alpha_{[320\,-\,656\,MHz]}}$. 
    The blue symbols in each outer corner of the plot represent the type of source in each quadrant of the plot, e.g. steep, inverted, peaked or upturn. Right: colour-colour diagram comparing the $\rm{\alpha_{[144\,-\,320\,MHz]}}$ vs $\rm{\alpha_{[656\,-\,1500\,MHz]}}$ spectral indices color in each redshift bin. The different classes of sources are represented by open black circles for the SFG population. The errors in the spectral indices were derived from the noise in the images using standard propagation..
    The dashed green diagonal line corresponds to $\rm{\alpha^{144}_{320}}$ = $\rm{\alpha^{320}_{656}}$ and $\rm{\alpha^{144}_{320}}$ = $\rm{\alpha^{656}_{1500}}$.
   }
    \label{alpha_color_spec.fig}
\end{figure*}

The radio colour-colour diagram \citep{1977AJ.....82..541K} usually require large samples of galaxies with multiple flux density measurements. We consider four classes of sources in our radio colour-colour diagram:

\begin{itemize}
\item[(i)] Steep-spectrum: objects from this class have a steep spectrum, $\rm{\alpha < -0.5}$
(the lower - left quadrant of Figure~\ref{alpha_color_spec.fig})

\item[(ii)] Peaked sources, where a turnover is detected within the observed spectral range (the upper - left quadrant of Figure~\ref{alpha_color_spec.fig}),

\item[(iii)] Inverted (or rising) sources, where the radio flux density increases with increasing frequency (the upper - right quadrant of Figure~\ref{alpha_color_spec.fig}), and,

\item[(iv)] Concave sources (V-shaped), that have an upturn in their radio spectrum (the lower - right quadrant of Figure~\ref{alpha_color_spec.fig}).
\end{itemize}

The quadrant description of the radio colour-colour diagram  can be used to classify the broad-spectrum behaviour of a source, depending on where they reside.

When dealing with more than four frequency measurements, this diagnostic is not always able to encompass the true range of properties \citep[see,][]{2018MNRAS.476.2717H}. The radio spectral index measured at the observer-frame frequencies from 144 MHz - 1500 MHz, 2-point spectral index at $\rm{\alpha^{144}_{320}}$ vs $\rm{\alpha^{320}_{656}}$ and $\rm{\alpha^{144}_{320}}$ vs $\rm{\alpha^{656}_{1500}}$
for the stacked sources in the XMM-LSS field is presented in Figure~\ref{alpha_color_spec.fig}. The SFGs shown as open black circles, are predominantly characterised by steepening of the average spectra at low frequencies. 
The correlation between spectral index and redshift for our sample of SFGs is investigated in section~\ref{Linear-steep spectra}.

Overall, the spectral shape of SFGs remain consistently steep from 144 to 656 MHz, and from 656 to
1500 MHz in all redshift bins. The steepness is  more pronounced at low frequencies as depicted by the diagonal line in each panel of Figure~\ref{alpha_color_spec.fig}. The first, second, third and fourth columns of Table~\ref{tab_slope_alpha} summarises the stacked radio spectral indices, $\rm{\alpha_{low}\,-\,\alpha^{144}_{320}}$, $\rm{\alpha_{mid}\,-\,\alpha^{320}_{656}}$, $\rm{\alpha_{high}\,-\,\alpha^{656}_{1500}}$, measured for the selected SFG candidates in each redshift bin. For $\rm{\alpha_{high}}$, this spectral index is thought to characterise the optically thin part of the underlying synchrotron emission. Whereas, $\rm{\alpha_{low}}$ can help distinguish between different kinds of absorption, such as free-free absorption (FFA) or synchrotron self-absorption (SSA) \citep[see,][]{2022ApJ...934...26P}. Since we did not include any higher frequency radio data in our analysis, the terms ‘low frequency’ and ‘high frequency' are purely relative in this case.
We also consider, $\rm{\alpha}$ across all our available data, i.e. $\rm{\alpha_{range}}$ - $\rm{\alpha^{144}_{1500}}$), for comparison given by the fifth column in Table~\ref{tab_slope_alpha}. \citet{2021A&A...648A..14R} investigated the low-frequency radio spectra of a sample of highly SFGs selected at 850 $\mu m$ from the SCUBA-2 Cosmology Legacy Survey (S2CLS). They found evidence of radio spectral flattening at low frequencies. \citet{2021A&A...648A..14R}  proposed that their sources with significant low-frequency spectral flattening have a clumpy distribution of star-forming gas which is attributed to free-free absorption.

In contrast, our stacked SFG  data suggests that the radio-continuum spectrum steepens at low-frequencies, and flattens at higher frequencies.
\citet{2021Galax...9..121S} stated that a substantial difference between the median spectral indices at lower and higher frequencies suggests spectral ageing even in active sources. Although our stacked SFG spectral indices are consistent with that of SFG population, we do not rule out possible of source contamination of AGN from our classification scheme. Also, spectral ageing would cause the spectrum to steepen at high frequency, which is not what we observe in our stacked SFG sample. 
The data we use in this study probe different observer-frame radio frequencies at different redshifts. Consequently, even if all SFGs have the same intrinsic radio spectra, one could still expect to observe the redshift evolution of $\rm{\alpha^{144}_{1500}}$ in the presence of a curved spectrum \citep{2015A&A...573A..45M}.  We discuss the modelling of curved power law described in subsection~\ref{cpl} in subsection~\ref{mod_compare}, where we compare the different models fitted to the stacked SFG  data.

\subsubsection{Linear-steep spectra of SFG\textsc{s} as a function of redshift}\label{Linear-steep spectra}

Figure~\ref{spec_dist.fig} shows a graphical representation of spectral index, $\rm{\alpha}$, as a function of median redshift. 
Here, the spectral indices used are the ones
derived using the equations in subsection~\ref{SED_ppts}, namely: $\rm{\alpha_{low}}$ (green circles), $\rm{\alpha_{mid}}$ (blue circles), $\rm{\alpha_{high}}$ (red circles), $\rm{\alpha_{range}}$ (open black circles).
This plot provides  a visual representation of the observed correlation between the representative radio
spectral index and redshift for the stacked sources. It has been known for several decades that high redshift radio galaxies exhibit ultra steep steep radio spectra \citep[e.g.,][]{1998MNRAS.295..265B,2008A&ARv..15...67M,2016MNRAS.462..917R,2023JApA...44...88S}. 
The stacked SFG sample show an almost negligible variation of spectral indices with redshift.

\citet{2015A&A...573A..45M} reported that SFGs have a radio spectral index consistent with a canonical value of $\mathrm{\alpha^{610}_{1400}\,=\,-0.80\,\pm\,0.29}$ across $0\,<\,z\,<\,2.3$, which suggests that their radio spectra are dominated by non-thermal optically thin synchrotron emission. 
\citet{2019MNRAS.489.5053S} studied a sample of 32 USS sources and found a strong anti-correlation between \textit{z} and $\rm{\alpha}$ among their sample. The exact physical mechanisms behind this phenomenon remain a topic of debate; however, the prevailing view is that the correlation between redshift ($z$) and spectral index $\alpha$ arises from a concave radio spectrum. This includes a flattening of the spectral index at lower radio frequencies, combined with a radio K-correction \citep[see,][]{2019MNRAS.489.5053S}. We refer the reader to \citet{2008A&ARv..15...67M} for a review on the possible causes of the observed $z\,\sim\,\alpha$ relation. 
Previous studies have also shown that there is a strong anti-correlation between the radio spectral index and redshift of the radio sources \cite[see,][]{1999IAUS..183..251A,2008A&ARv..15...67M}. 
We do not find any strong anti-correlation for our stacked radio
SEDs versus redshift. \citet{2023JApA...44...88S} suggested that for SFGs, the nature of the radio SED is mostly dependent on the local parameters within galaxies, such as magnetic fields and properties of the surrounding ISM. Consequently, this is independent of the properties in the large-scale context, where redshift evolution becomes essential.

\begin{figure}
\includegraphics[width=0.48\textwidth]{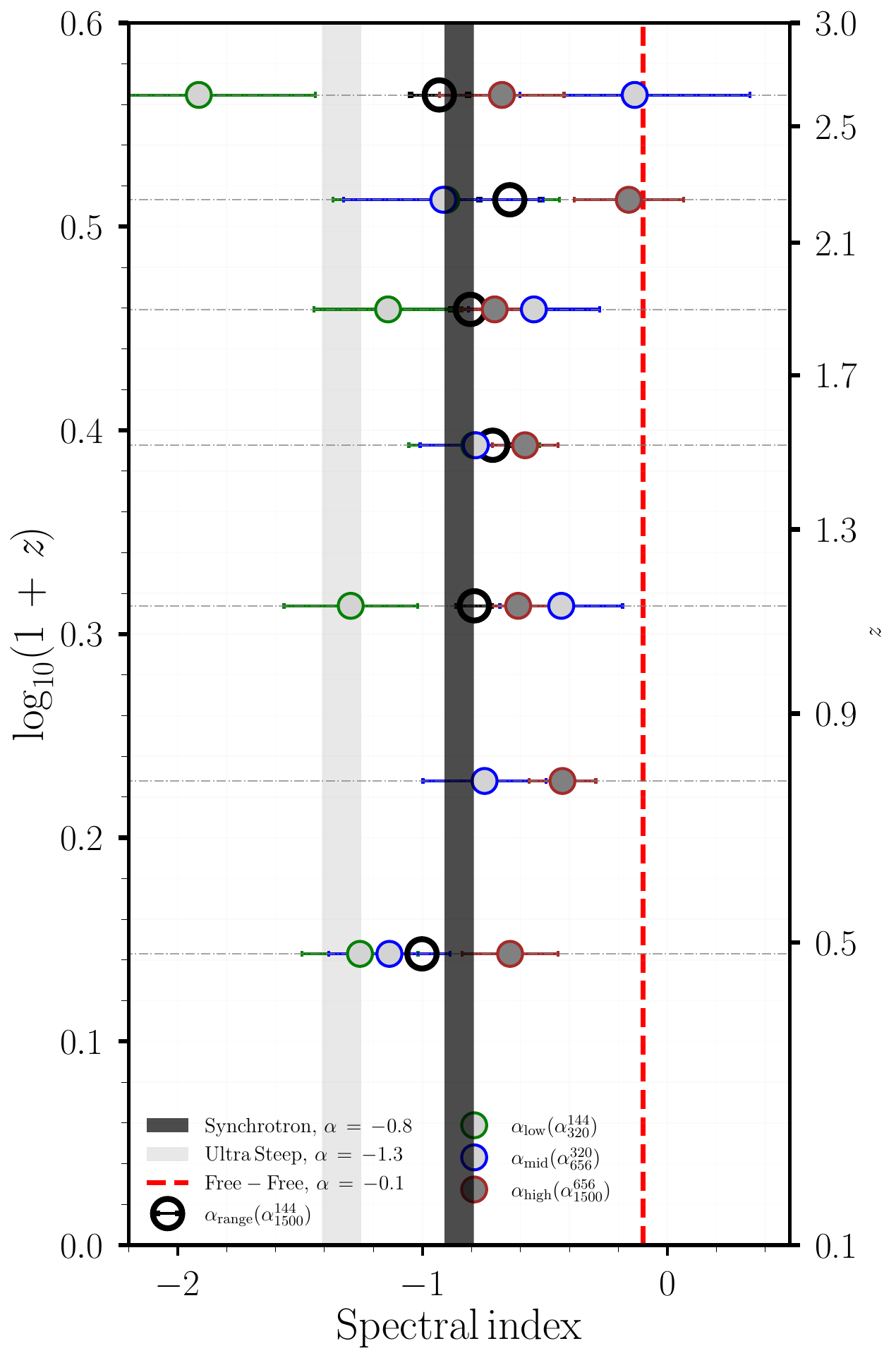}
    \caption{A graphical representation of spectral index, $\rm{\alpha}$, as a function of median redshift. 
   The black and grey shaded areas represents a simple power law with a constant canonical spectral slope
of $\rm{\alpha\,=\,-0.8\pm0.06}$ and a very steep spectral index  with $\rm{\alpha\,=\,-1.3\pm0.06}$, respectively. The red dashed lines show  a free-free emission  with $\rm{\alpha\,=\,-0.1}$.} 
    \label{spec_dist.fig}
\end{figure}

\subsubsection{Comparison of the Stacked SFG Models}\label{mod_compare}
The resulting stacked radio spectra in different redshift bins are shown in Figure~\ref{sed.fig}. 
All spectra are fitted with a power law (Equation~\ref{spec_indx_log.eqn}), curved power law (Equation~\ref{curvedPL.eqn}), and double power law (Equation~\ref{syn_freefree.eqn}). These fits are shown as dashed black lines, solid green lines, and solid red lines in each panel respectively.

In Table~\ref{tab_slope_alpha}, we compile the results obtained from the different fitted models in each redshift bin. We also include the values of the $\chi^{2}_{red}$ for each model in their respective redshift bins. In determining physical properties of galaxies from radio data, the radio SED is often assumed to be described by a simple power law, usually with a spectral index of -0.7 for all sources, and -0.8 for SFGs. Although radio SEDs have been shown to exhibit deviations from this assumption, both in
differing spectral indices and complex spectral shapes, it is often presumed that their individual differences can be canceled out in large samples \citep{2019A&A...621A.139T}.
Upon inspection of Figure~\ref{sed.fig}, it is evident that our stacking experiment shows systematically steep spectral indices (i.e. where steep spectra is defined a $\alpha\,<\,-0.5$). For the power law model, we find a range of spectral indices between -0.86 to -0.62. This model describes a synchrotron component, and is in good agreement with the canonical synchrotron emission spectral index $\rm{\alpha\,\approx\,-0.8}$ for the SF driven population frequently assumed in the literature. 

The curved power law model exhibits a range of spectral indices between -0.75 to -0.50. The \textit{q} value indicates the width of the peak or the degree of spectral curvature. From our fitting, we detect moderate $\textit{q}\,>\,0$ values in all the redshift bins, indicating mild deviations from a simple power law. The \textit{q} values are reported in Table~\ref{tab_slope_alpha}.
Following the solid green lines in Figure~\ref{sed.fig}, we see a convex spectrum in each redshift bin (i.e. the spectrum flattens with frequency.) 
For positive curvature, a more positive $\rm{\alpha}$ implies a  trough at lower frequencies, and a more negative $\alpha$ implies a trough at higher frequencies} (and vice versa for a concave spectrum).
As expected, all the stacked sources with q > 0 lie below the $\alpha(144\,-\,320\,MHz)\,= \,\alpha(320\,-\,656\,MHz)$ and $\alpha(144\,-\,320\, MHz)\,=\,\alpha(656\,-\,1500\,MHz)$ diagonals on the RCC diagrams.

The double power law model is constrained around range of steep spectral indices from -1.96 to -0.80 across all redshift bins. This is consistent with ultra-steep spectral indices, which may be caused by an intrinsically steeper cosmic ray distribution \citep{2018MNRAS.474..779G}. The double power law describes  synchrotron and \textit{FFE} components, which is demonstrated by a slight spectral bend in the fits as seen  in Figure~\ref{sed.fig} (solid red lines). 

\begin{figure*}
\includegraphics[width = 0.42\textwidth]{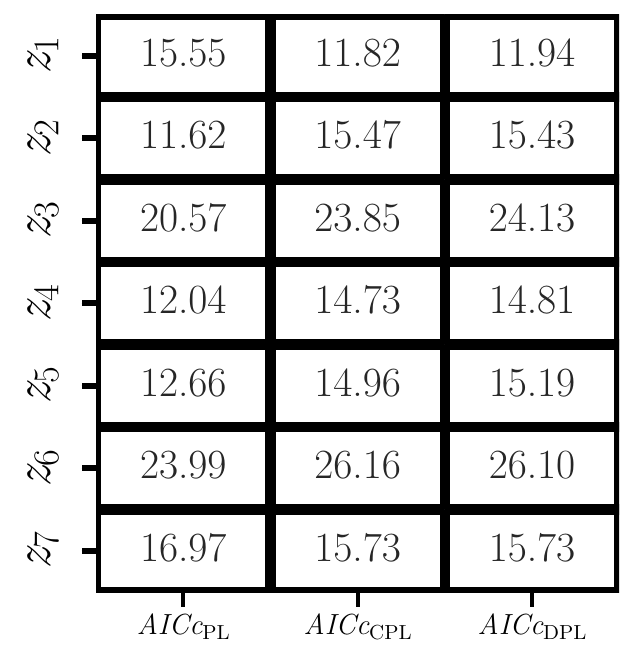}
\includegraphics[width = 0.42\textwidth]{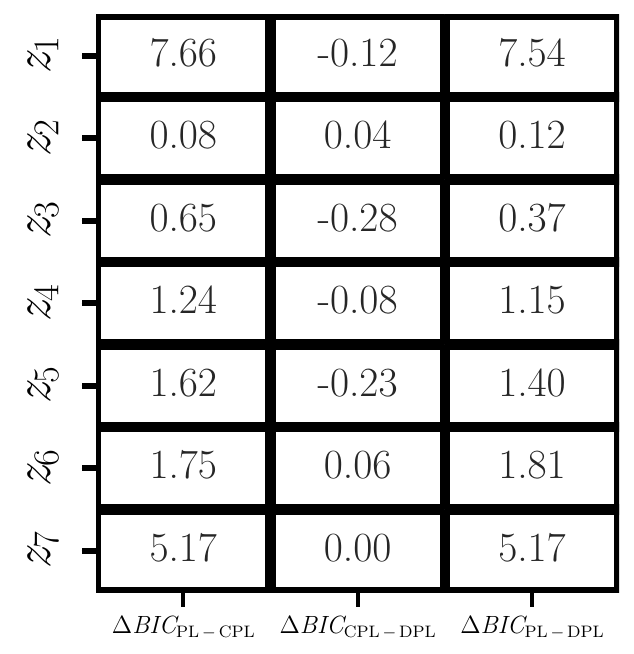}

    \caption{
    Table maps showing the computed  \textit{AICc} (left) and $\mathrm{\Delta \textit{BIC}}$ (right) of the three models of radio SED fits to the various redshift bins of the stacked observer-frame radio SED for SFGs.} 
    \label{mod_compare.fig}
\end{figure*}

The optimal spectral model for characterizing the stacked observer-frame radio SEDs across our sample is determined through a systematic comparison of three candidate models: \textit{PL}, \textit{CPL}, and \textit{DPL}. This selection process relies on two statistical criteria, namely \textit{AICc} and $\rm{\Delta \textit{BIC}}$, presented in tabular format in Figure~\ref{mod_compare.fig}. The figure shows table maps for the computed \textit{AICc} and $\mathrm{\Delta \textit{BIC}}$ of the three models we considered for the stacked observer-frame  radio SED fits. The \textit{AICc} favors all models, with \textit{PL} occasionally better in some cases (i.e., according to the left panel of Figure~\ref{mod_compare.fig}). However, the $\mathrm{\Delta \textit{BIC}}$ weakly favors \textit{CPL/DPL} over \textit{PL} (i.e., since $0\,<\,\mathrm{\Delta \textit{BIC}}\,<\,2$ indicates a weak preference towards the second model). Both criteria differ in detailed comparisons between  the \textit{PL}, \textit{CPL} and \textit{DPL} models. Row seven in Figure~\ref{mod_compare.fig} $\mathrm{\Delta \textit{BIC}}$ = 0.00. No preference between \textit{CPL} and \textit{DPL}, suggests that the data lacks features that distinguish the two models. The fact that the \textit{CPL} model occasionally outperforming the \textit{DPL} model, highlights processes such as synchrotron radiation dominating the physical mechanisms driving the spectral shape.  Hence, the two criteria do not agree on which spectral model is the best for characterising the stacked radio SEDs. In Table~\ref{mod_compare.fig}, the $\rm{\chi^{2}_{red}}$ values look reasonable for all three models (i.e., they are always $<\,\sim2$). Therefore, we conclude that the radio spectra are generally well-fitted by all the models considered.

\begin{table*}
 \centering
 \caption{Summarised stacked radio spectral index properties of the selected SFG candidates. The empty rows in $\rm{\alpha_{low}}$ and $\rm{\alpha_{range}}$ represent redshift range where no median stacked image was produced (i.e $\rm{\textit{z}\in[0.5-0.95]}$ for 144 MHz, see  Figure~\ref{med_stack_fig_append}).}
 \scalebox{0.85}{
 \begin{tabular}{ccccc | cc |ccc |cccc }
 \hline
 \hline
Redshift bin&  $\rm{\alpha_{low}}$ & $\rm{\alpha_{mid}}$&$\rm{\alpha_{high}}$&$\rm{\alpha_{range}}$&\multicolumn{2}{c}{Power Law fit}&\multicolumn{3}{c}{Curved Power Law fit} & \multicolumn{4}{c}{Synchrotron+Free-Free fit}\\
   & & & & &$\rm{\alpha_{PL}}$ & $\rm{\chi^{2}_{red}}$ &$\rm{\alpha_{CPL}}$ &\textit{q}&$\rm{\chi^{2}_{red}}$&$\rm{\alpha_{DPL}}$ & \textit{A}& \textit{B}&$\rm{\chi^{2}_{red}}$\\
 \hline
 \hline
$\rm{\textit{z}_{1}}$& -1.26$\pm$0.24 &-1.13$\pm$0.25&-0.64$\pm$0.20&-1.00$\pm$0.06&-0.86 $\pm$0.05&1.12&-0.75$\pm$0.07&0.22$\pm$0.07&0.30&-1.53$\pm$0.21&1.95$\pm$0.33&1.56$\pm$0.41&0.31\\

$\rm{\textit{z}_{2}}$& -&-0.75$\pm$0.25&-0.43$\pm$0.14&-&-0.62$\pm$0.05&0.68&-0.61$\pm$0.05&0.04$\pm$0.15&0.75&-0.80$\pm$0.498&0.95$\pm$2.17&2.54$\pm$2.31&0.75\\

$\rm{\textit{z}_{3}}$& -1.29$\pm$0.27 &-0.43$\pm$0.25&-0.61$\pm$0.13&-0.79$\pm$0.07&-0.65$\pm$0.04&1.67&-0.63$\pm$0.05&0.06$\pm$0.07&1.80&-0.85$\pm$0.27&1.13$\pm$1.07&2.69$\pm$1.16&1.84\\

$\rm{\textit{z}_{4}}$& -0.79$\pm$0.27&-0.71$\pm$0.23&-0.58$\pm$0.13&-0.71$\pm$0.08&-0.66$\pm$0.04&0.73&-0.63$\pm$0.05&0.08$\pm$0.07&0.66&-0.98 $\pm$0.26&1.63$\pm$0.86&2.49$\pm$0.96&0.67\\

$\rm{\textit{z}_{5}}$& -1.14$\pm$0.30& -0.55$\pm$0.27&-0.78$\pm$0.14&-0.80$\pm$0.08&-0.69$\pm$0.05&0.80&-0.66$\pm$0.06&0.11$\pm$0.08&0.69&-1.09$\pm$0.28&2.20$\pm$0.92&2.84$\pm$1.06&0.72\\

$\rm{\textit{z}_{6}}$& -0.90$\pm$0.41 &-0.92$\pm$0.41&-0.16$\pm$0.18&-0.64$\pm$0.12&-0.67$\pm$0.07&2.05&-0.60$\pm$0.09&0.15$\pm$0.11&2.09&-1.21$\pm$0.39&1.81$\pm$0.68&1.55$\pm$0.80&2.08\\

$\rm{\textit{z}_{7}}$& -1.91$\pm$0.40 &-0.13$\pm$0.22&-0.68$\pm$0.25&-0.93$\pm$0.12&-0.63$\pm$0.07&1.27&-0.50$\pm$0.11&0.35$\pm$0.12&0.79&-1.96$\pm$0.41&2.97$\pm$0.33&0.73$\pm$0.40&0.79\\

 \hline
 \end{tabular}}
  \begin{tablenotes}
\item $\rm{\alpha_{low}\,-\,\alpha^{144}_{320}}$, $\rm{\alpha_{mid}\,-\,\alpha^{320}_{656}}$, $\rm{\alpha_{high}\,-\,\alpha^{656}_{1500}}$, $\rm{\alpha_{range}\,-\,\alpha^{144}_{1500}}$, $\rm{\alpha_{PL}}$ - Power Law, $\rm{\alpha_{CPL}}$ - Curved Power Law, $\rm{\alpha_{DPL}}$ - Double Power Law ( Synchrotron + Free-Free Emission)
\end{tablenotes}
 \label{tab_slope_alpha} 
 \end{table*}

\begin{figure*}
\includegraphics[width=0.83\textwidth]{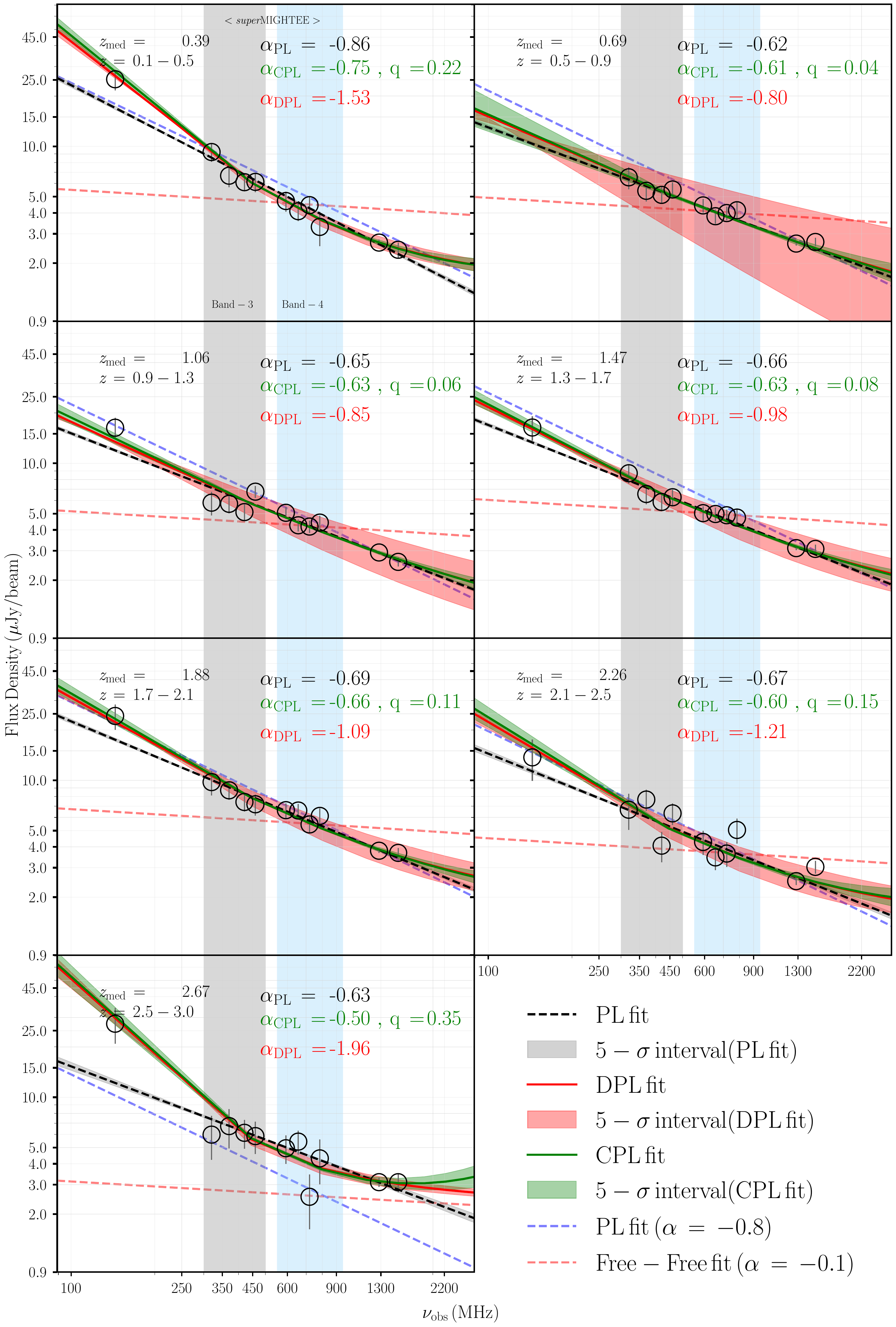}
    \caption{A compilation of the stacked observed-frame radio SEDs (flux density vs. frequency) for the 144 - 1500 MHz median stacked SFGs  up to $\rm{\textit{z}\simeq3}$. The error bars represent the measured background noise in the median stacked images. The blue and red dashed lines show a simple power law  and a free-free emission with a constant canonical spectral slope
$\rm{\alpha\,=\,-0.8}$ and $\rm{\alpha\,=\,-0.1}$, respectively, normalised to the 656 MHz point for the SFGs as a reference. 
The fit we obtain by following Equations~\ref{spec_indx_log.eqn}, ~\ref{curvedPL.eqn} and ~\ref{syn_freefree.eqn}, are shown as dashed black lines, solid green lines, and solid red lines in each panel respectively. Also shown are as grey, green and red shaded areas corresponding 5$\sigma$ dispersion in each panel, respectively. The grey and lightblue vertical shaded areas represent the \textit{super}MIGHTEE Band-3 and Band-4 frequency range respectively.}
    \label{sed.fig}
\end{figure*}

\subsubsection{Estimates for Thermal Fraction at 1500 MHz}
In previous sections, we have shown that the stacked SFGs  display complex radio SEDs.  
Using the spectral index information over this wide frequency range (144-1500 MHz), we can estimate the
thermal emission fraction at 1500 MHz by assuming a fixed non-thermal spectral index. The thermal fraction ($f_{T}$) measures the contribution of thermal free-free emission to the observed total radio continuum. It is an excellent indicator of current star formation in galaxies as it probes the $\rm{H_{II}}$ region.

Following \citet{1990ApJ...357...97C,1992ARA&A..30..575C} in the literature, and assuming that the free–free emission is optically thin, we estimate the thermal fraction  at the reference
frequency 1500 MHz, which is given by:

\begin{equation}
 \frac{1}{f_{T}}\,\equiv\,\bigg\langle\frac{S}{S_{T}}\bigg\rangle\,\sim\,1\,+\,10\left(\frac{\nu}{GHz}\right)^{0.1\,+\,\alpha}
\label{eqn:therm_frac}
\end{equation}
where,  $S_T$ is the flux density of the thermal free-free component and $S$ the total flux density. For our specific case, $\rm{\nu\,= 1500\,MHz}$ and $\alpha$ is the typical non-thermal spectral index we measure in this work. The non-thermal flux typically exceeds the thermal flux for all frequencies below $\rm{\nu\,\sim\,30\,GHz}$ \citep{1992ARA&A..30..575C}.

\begin{figure}
\centerline{\includegraphics[width=0.42\textwidth]{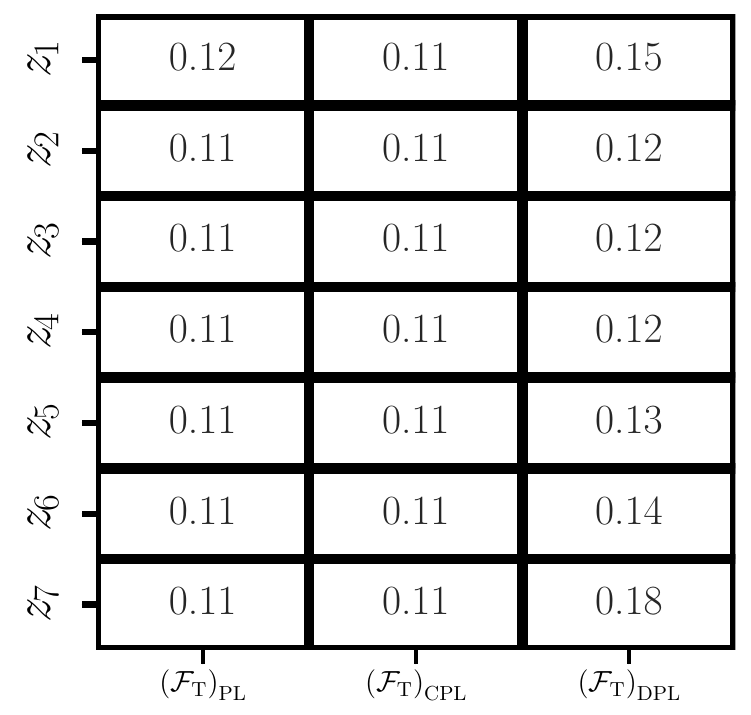}}
    \caption{Table showing the  estimates for the thermal fraction at 1500 MHz of the three models considered for the radio SED fits to the various redshift bins.
   }
    \label{therm_frac.fig}
\end{figure}

At increasing frequencies, due to the steep spectral index of non-thermal synchrotron emission, the thermal free-free emission begins to dominate the total observed continuum. For each redshift bin, using the fitted models, we follow Equation~\ref{eqn:therm_frac} in order to derive appropriate nominal thermal fraction values.
Figure~\ref{therm_frac.fig} presents a table showing the estimates for the thermal fraction, $S_{T}/S$, at 1500 MHz for the various redshift bins. The simple power law model shows a thermal fraction of about 12$\%$ at redshift $\textit{z}_{1}$ and approximately 11$\%$ for other bins. The curved power law model consistently indicates a thermal fraction of 11$\%$ across all bins. In contrast, the double power law model presents varying fractions: 15$\%$ for $\textit{z}_{1}$, 12$\%$ for $\textit{z}_{2}$, $\textit{z}_{3}$, and $\textit{z}_{4}$, around 13$\%$ for $\textit{z}_{5}$, 14$\%$ $\textit{z}_{6}$, and 18$\%$ for $\textit{z}_{7}$.

We find that at 1500 MHz frequencies, the thermal fraction makes up only a small fraction of the total radio continuum emission. This is consistent to earlier studies \citet{1990ApJ...357...97C,1992ARA&A..30..575C,1992ApJ...401...81P}, where at 1.4 GHz the typical thermal fraction was estimated to be $\rm{\sim10\,per\,cent}$.  \citet{1992ARA&A..30..575C} observed that the exact shape of the radio SED of SFGs  is
usually assumed to be a superposition of the steep synchrotron
spectrum, described by a power law, and a $\sim10$ per cent contribution at 1400 GHz of a flat free-free spectrum.
 
\citet{2021A&ARv..29....2P} presented an
overview of the modelling of the SEDs of starburst-dominated Luminous Infrared Galaxies (\textit{LIRGs}), using the value of optically thin spectral index of $\alpha\,\approx\,-0.8$. At 1.4 GHz, \citet{2021A&ARv..29....2P} reported that Equation~\ref{eqn:therm_frac} predicted the fraction of thermal free-free emission to be \textbf{$\sim\,0.11$.} They conclude  that the low-frequency radio emission in starburst-dominated \textit{LIRGs} is predominantly of non-thermal, synchrotron origin.  This is consistent with our results, however, \citet{2021A&ARv..29....2P} indicated that multi-frequency observations above and below $\sim$30 GHz are necessary to estimate the true thermal component.
At $\nu\,\approx\,30\,GHz$, Equation~\ref{eqn:therm_frac} predicts $f_{T}\,\sim\,0.52$, while at $\nu\,\approx\,200\,GHz$, $f_{T}\,\sim\,0.80$, when the radio continuum peaks at sub-mm wavelengths and is powered by dust emission. 
Studies by \citet{2017ApJ...836..185T} showed that low-frequency radio observations are particularly useful to
study synchrotron emission from galaxies, since they measured a mean thermal fraction of  10$\%$ at $\sim$1.4 GHz over their entire sample of nearby galaxies. 

\section{Conclusions and outlook}\label{conclusion.sec}

In this paper we have presented a data-driven stacked radio spectral properties of SFGs at the observer-frame frequencies using radio observations from 144 - 1500 MHz of the XMM-LSS field. 

We applied various criteria in order to minimise the impact of contaminating the radio flux density from potentially elongated sources. By using the rest-frame  \textit{u - V} vs \textit{V - J}  colors we separate SFGs from quiescent systems and AGN in seven redshift bins up to $\textit{z}\sim3$ in order to study the average radio spectral index. 
 
We derive the photometric redshift scatter ($\sigma_{\text{NMAD}}$), outlier fraction (O$_{f}$), and  the  continuous ranked probability score ($\overline{\rm{CRPS}}$) for the separate populations. These statistical tests validate the overall redshift accuracy and performance of SFGs is better than that of the quiescent galaxies, and AGN populations with respect to their photo-\textit{z} estimates as a function of $\textit{K}_{S}$ magnitude.  
We focus on radio SEDs below the survey threshold using median image stacking techniques that is best applied in the radio regime where the angular resolution is high. A correction is applied to the median stacked flux densities to account for median flux boosting.

By analysing the shape and characteristics of the SEDs, it becomes possible to gain insights into the underlying physical processes driving the observed emission.

Our principal findings can be summarised as follows:
\begin{itemize}

\item We subdivided our sample into  seven redshift bins spanning $\rm{0.1\leq\,\textit{z}\,\leq\,3.0}$, and stack into radio images ranging from 144 - 1500 MHz using observations from LOFAR, \textit{u}GMRT, MeerKAT and VLA  at the positions of $\textit{K}_{S}$ selected sources in
the XMM-LSS field. The median stacking technique provides invaluable compendium of useful information about the broadband radio SED of SFGs.

\item The multi-frequency nature of this study allows the estimation of the stacked radio spectral properties at observer-frame 144 – 1500 MHz. We populate our stacked radio spectral indices on the radio colour–colour plots, $\rm{\alpha^{144}_{320}}$ versus $\rm{\alpha^{320}_{656}}$, and $\rm{\alpha^{144}_{320}}$ versus $\rm{\alpha^{656}_{1500}}$. The representative radio colour-colour diagram reveals a tentative steepening at low frequency. 
\item It is evident that the majority of our stacked representative sources lie in the steep quadrant in both panels of the radio colour-colour plot for the SFG population. 

\item We observe a lack of strong correlation among the radio
sources below the survey threshold in the XMM-LSS field for their radio spectral index with redshifts. For our stacked SFGs,
this may mean that the nature of the radio SED is
mostly dependent on the local parameters within the
galaxies, like magnetic fields, properties of the surounding ISM etc. As such, this suggests the radio spectral index is independent of the properties
in the large-scale context for which redshift evolution
becomes crucial.

\item Due to the relatively moderate frequency range we probe, i.e.  144 – 1500 MHz, we model the stacked radio spectra to study the radio emission mechanism of the SFGs in the different redshift bins. We apply  single, curved and double  power law  models  to the observer frame SFGs stacked spectra. We fix the thermal free-free emission with $\rm{\alpha_{FF}\,=\,-0.1}$. Our statistical modelling suggests a typical non-thermal synchrotron spectral steepening at low frequency, for all models, comparable to those measured for star-forming sources above the survey thresholds in the literature. We observe a steepening over the range of several low-frequencies, and may flatten at higher frequencies due to the thermal radio emission. 
mechanisms that drive the star-formation processes in SFGs  below the survey threshold.

\item The asymptotic low-frequency synchrotron spectra derived in this study cannot be only modelled by simple power laws as believed hitherto. The curved shape of the synchrotron spectra due to the increasing contribution from free-free components at high frequency is mildly visible across the entire redshift range. Therefore, the \textit{CPL} and \textit{DPL} are reasonably good fits in each redshift bin.

\item Model comparisons using the \textit{AICc} and $\rm{\Delta \textit{BIC}}$ indicate that the radio
spectra are generally well-fitted by all the models considered (i.e., \textit{PL}, \textit{CPL} and \textit{DPL}), with the \textit{CPL} model being slightly preferred. The occasional superiority of the \textit{CPL} over the \textit{DPL} in specific redshift bins is attributed to synchrotron radiation dominating the physical processes of the stacked radio SEDs. 
\item From our fitted models, we find at low frequencies the thermal fraction makes up only a small fraction of the total radio continuum emission. At 1500 MHz the estimated thermal fraction is fairly constant at around $\sim11$ to $\rm{\sim18\,per\,cent}$ across $\rm{0.1\leq\,\textit{z}\,\leq\,3.0}$, with the average thermal fraction being $\rm{\sim11\,per\,cent}$. This reaffirms what has long been known that SFGs comprise both a steep synchrotron component powered by shocks of supernovae, and a flat component from free-free emission from $\rm{H_{II}}$ regions.
 
\end{itemize}

This work demonstrates the potential of median stacking analyses to extract meaningful results in flux density regimes comparable to those that will be achieved with next-generation radio surveys.
In the future, we plan to  probe frequency ranges from 1 - 10 GHz in order to probe the dominant emission mechanism. This is a key science project in the exploration of the radio universe that will be made by the SKA Phase 1 \citep[which will operate from 50 MHz to at least 15 GHz, see,][]{2009IEEEP..97.1482D,SKA2015,SKA2019} and the next-generation Very Large Array \citep[ngVLA, will cover a frequency range from 1.2 – 116 GHz,][]{2018ASPC..517...15S}. The radio sky at high-frequency  has historically remained largely unexplored due to the typical faintness of sources in this regime, and the slow survey speed at high frequency due to the scaling of interferometer primary-beam areas with frequency. High-frequency radio surveys offer an invaluable tracer of high-redshift star formation, as they directly target the faint radio free–free emission. 
A radio spectral decomposition of the non-thermal and thermal radio emission at increasing frequencies will help to confirm the assertion that SFGs  are dominated by synchrotron radiation. Hence our results have important implications since new facilities will provide broad-band capabilities allowing for long-integration mosaic imaging observations to create ultra-deep continuum and full-polarisation images of the sky over wide frequency ranges. The low and mid frequency surveys that will be undertaken with these facilities will provide the needed combination of spatial resolutions and spectral coverage for spectra-morphological SFG/AGN studies.

\section*{Acknowledgements}

This work was supported by the Korea Astronomy and Space Science Institute under the R\&D program (Project No. 2025-1-831-02), supervised by the Korea AeroSpace Administration.
EFO would like to acknowledge the hospitality of the Inter-University Institute for Data Intensive Astronomy (IDIA) which is a partnership of the University of Cape Town, the University of Pretoria and the University of the Western Cape.
JMS acknowledges the support of the Natural Sciences and Engineering Research Council of Canada (NSERC), 2019-04848.
CHIC acknowledges the support of the Department of Atomic Energy, Government of India, under the project 12-R\&D-TFR-5.02-0700.
MV acknowledges financial support from the Inter-University Institute for Data Intensive Astronomy (IDIA), a partnership of the University of Cape Town, the University of Pretoria and the University of the Western Cape, and from the South African Department of Science and Innovation's National Research Foundation under the ISARP RADIOMAP Joint Research Scheme (DSI-NRF Grant Number 150551) and the CPRR HIPPO Project (DSI-NRF Grant Number SRUG22031677).
We acknowledge the use of the ilifu cloud computing facility - \url{https://www.ilifu.ac.za}, a partnership between the University of Cape Town, the University of the Western Cape, the University of Stellenbosch, Sol Plaatje University, the Cape Peninsula University of Technology and the South African Radio Astronomy Observatory. The ilifu facility is supported by contributions from the Inter-University Institute for Data Intensive Astronomy (IDIA - a partnership between the University of Cape Town, the University of Pretoria, the University of the Western Cape and the South African Radio Astronomy Observatory), the Computational Biology division at UCT and the Data Intensive Research Initiative of South Africa (DIRISA). We thank Ben Keller for sharing his PASTA stacking code with us before its public release and for his helpful advice on the installation of the code on the ilifu cloud computing facility at the Inter-University Institute for Data Intensive Astronomy (IDIA).
\section*{Data Availability}
The data underlying this article will be shared on reasonable request to the corresponding author.
\section*{ORCID IDS}
E.F. Ocran: \url{https://orcid.org/0000-0002-3749-533X}\\
A.R. Taylor: \url{https://orcid.org/0000-0001-9885-0676}\\
J.M. Stil: \url{https://orcid.org/0000-0003-2623-2064}\\
M. Vaccari: \url{https://orcid.org/0000-0002-6748-0577}\\
S. Sekhar: \url{https://orcid.org/0000-0002-8418-9001}\\
C.H. Ishwara-Chandra: \url{https://orcid.org/0000-0001-5356-1221}\\
Jae-Woo Kim: \url{https://orcid.org/0000-0002-1710-4442}
\section*{Software}
This work relies on the \textsc{Python} programming language (\url{https://www.python.org/}).
The Python Astronomical Stacking Tool Array (\texttt{PASTA}) programme developed at the University of Calgary by Ben Keller and Jeroen Stil\, is available at \url{https://github.com/bwkeller/PASTA}. We used astropy ( \url{https://www.astropy.org/}; \citet{2013A&A...558A..33A, 2018AJ....156..123A}), numpy (\url{https://numpy.org/}), matplotlib (\url{https://matplotlib.org/}). \textsc{seaborn}\citep{2021JOSS....6.3021W}
\bibliographystyle{mnras}
\bibliography{mnras} 

\appendix
\section{Definitions of statistical metrics used to evaluate photometric redshift accuracy and quality}\label{AppendA}

Metrics for photometric redshift precision and accuracy are defined as in \citet[][and references therein]{2018MNRAS.473.2655D}.

$\sigma_{\text{NMAD}}$ | Normalised median absolute deviation | 
\begin{equation}
\rm{1.48 \times \text{median} ( \left | \Delta z \right | / (1+z_{\text{spec}}))}
\label{eq:sigma_nmad}
\end{equation}
\\
 O$_{f}$ | Outlier fraction | Outliers defined as
 \begin{equation}
 \rm{\left | \Delta z \right | / (1+z_{\text{spec}}) > 0.15}
 \label{eq:out_frac}
\end{equation}
\\
$\overline{\rm{CRPS}}$ | Mean continuous ranked probability score | 
\begin{equation}
\rm{\overline{\rm{CRPS}} = \frac{1}{N} \sum_{i=1}^{N} \int_{-\infty}^{+\infty} [ \rm{CDF}_{i}(z) -  \rm{CDF}_{z_{s},i}(z)]^{2} dz}
 \label{eq:crps}
\end{equation}

\begin{table*}
\scalebox{0.95}{
    \begin{subtable}{0.98\textwidth}
        \centering
        \begin{tabular}{ll}
       Data & Source \\ \midrule
The VIMOS Public Extragalactic Redshift Survey (VIPERS)              &  \text{\citep{2014A&A...562A..23G,2014A&A...566A.108G,2018A&A...609A..84S}}            \\
The UKIDSS Ultra-Deep Survey (UDSz)&\text{\citep{2013MNRAS.433..194B,2013MNRAS.428.1088M}}\\
The Complete Calibration of the Color-Redshift relation
survey (C3R2) &\text{\citep{2017ApJ...841..111M}} \\
The \textit{Subaru} compilation of objects in the SXDF&\text{\citet{2005ApJ...620L...1O,2008ApJS..176..301O}}, \citet{2016ApJ...828...26M},\citet{2016ApJ...833..195W}\\
The X-UDS compilation
consolidated spectroscopic redshifts & \text{\citet{2005ApJ...634..861Y}}, \citet{2006MNRAS.372..741S}, \citet{2015PASJ...67...82A}  \\
       \end{tabular}
       \caption{Spectroscopic data}
       \label{tab:week1}
    \end{subtable}
    }
    \hfill
    \scalebox{0.95}{
    \begin{subtable}{0.98\textwidth}
        \centering
        \begin{tabular}{ll}
        Data & Source \\ \midrule
The VISTA Deep Extragalactic Observations (VIDEO) \textit{YJHK}$_{s}$         	& \citet{2013MNRAS.428.1281J}          \\
The CFHT Legacy Survey (CFHTLS) \textit{ugri}& \citet{2012yCat.2317....0H} \\
Suprime-Cam (SupCam) \textit{BVR}$_{c}$& \citet{2008ApJS..176....1F}
        \end{tabular}
        \caption{Photometric data}
        \label{tab:week2}
     \end{subtable}
     }
     \caption{Spectroscopic (top) and Photometric (bottom) Redshift Samples for Objects in the SPLASH-SXDF Catalogue.}
     \label{phot_spec.tab}
\end{table*}

A multitude of spectroscopic and photometric surveys have been undertaken in the XMM-LSS field.
Table~\ref{phot_spec.tab} presents the photometric and spectroscopic redshift samples for objects in the SPLASH-SXDF catalogue.

\section{STACKING ANALYSIS: DETAILS AND TESTS}\label{stack_analysis}
\subsection{Classifying the stacked sources as resolved or unresolved}

\begin{figure}
\centering
\centerline{\includegraphics[width=0.5\textwidth]{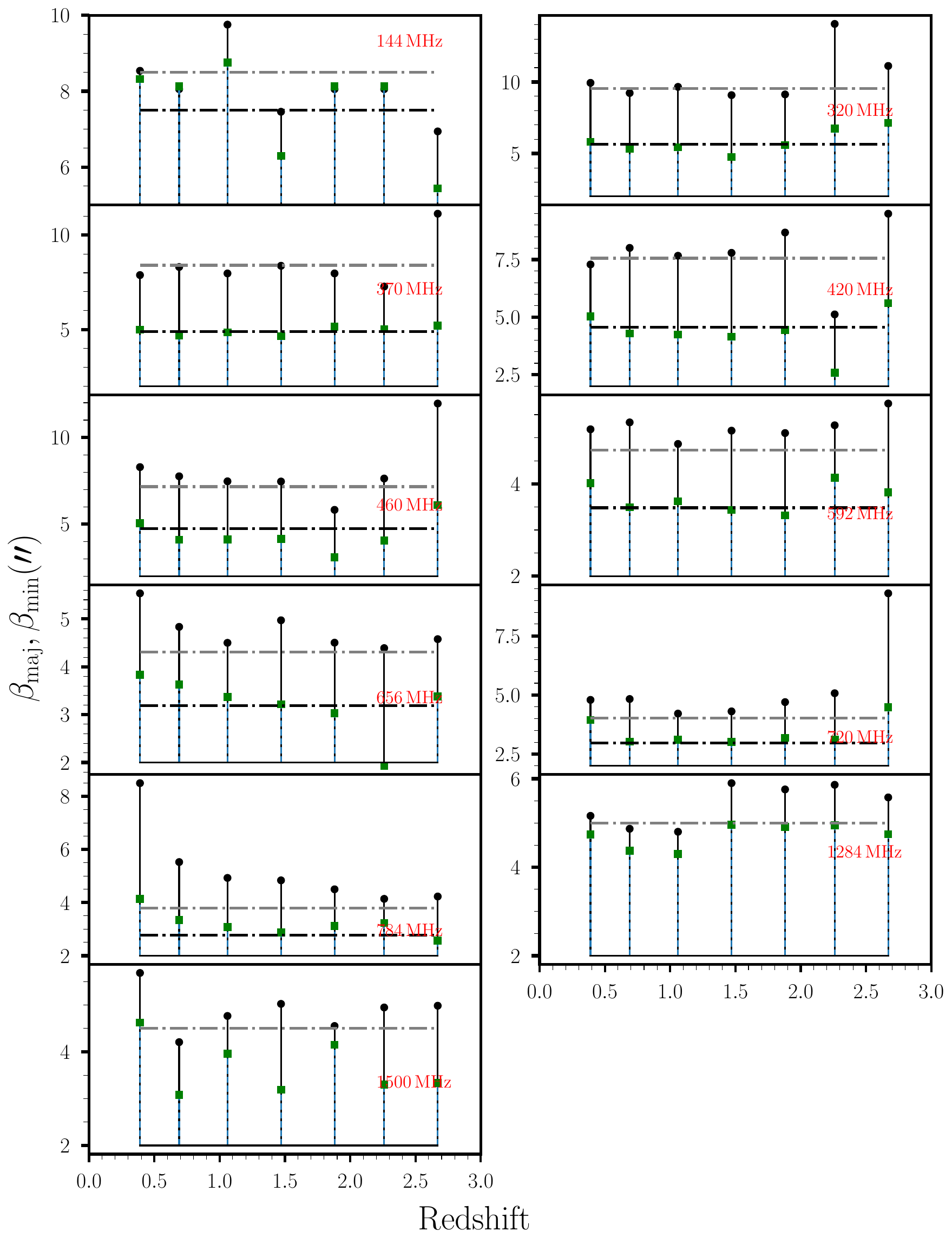}}
    \caption{Stacked median source fitted axes, $\rm{\textit{B}_{maj}}$ and $\rm{\textit{B}_{min}}$, as a function of median redshift across $\rm{144\leq\,\nu\,\leq 1500\,MHz}$ up to $\rm{\textit{z}\simeq3}$. The green spines represent the measure semi-minor axes $\rm{\textit{B}_{min}}$, whereas the solid black spines represent the semi-major axes $\rm{\textit{B}_{maj}}$ retrieved from each median stacked image. The horizontal black and grey  lines represent the original major and minor axes of the synthesised beam, respectively.}
\label{theta.fig}
\end{figure}

An extended galaxy (the most massive, and hence largest galaxies) will obviously have a larger solid angle than a compact galaxy at the same distance. This will affect the flux observed from that galaxy. Figure~\ref{theta.fig} shows the stacked median source fitted axes, $\rm{\textit{B}_{maj}}$ and $\rm{\textit{B}_{min}}$ as a function of median redshift. The median axis are the deconvolved FWHM of the  semi major and minor axis of the source as determined by \textsc{PyBDSF}.

\begin{figure}
\centering
\centerline{\includegraphics[width=0.5\textwidth]{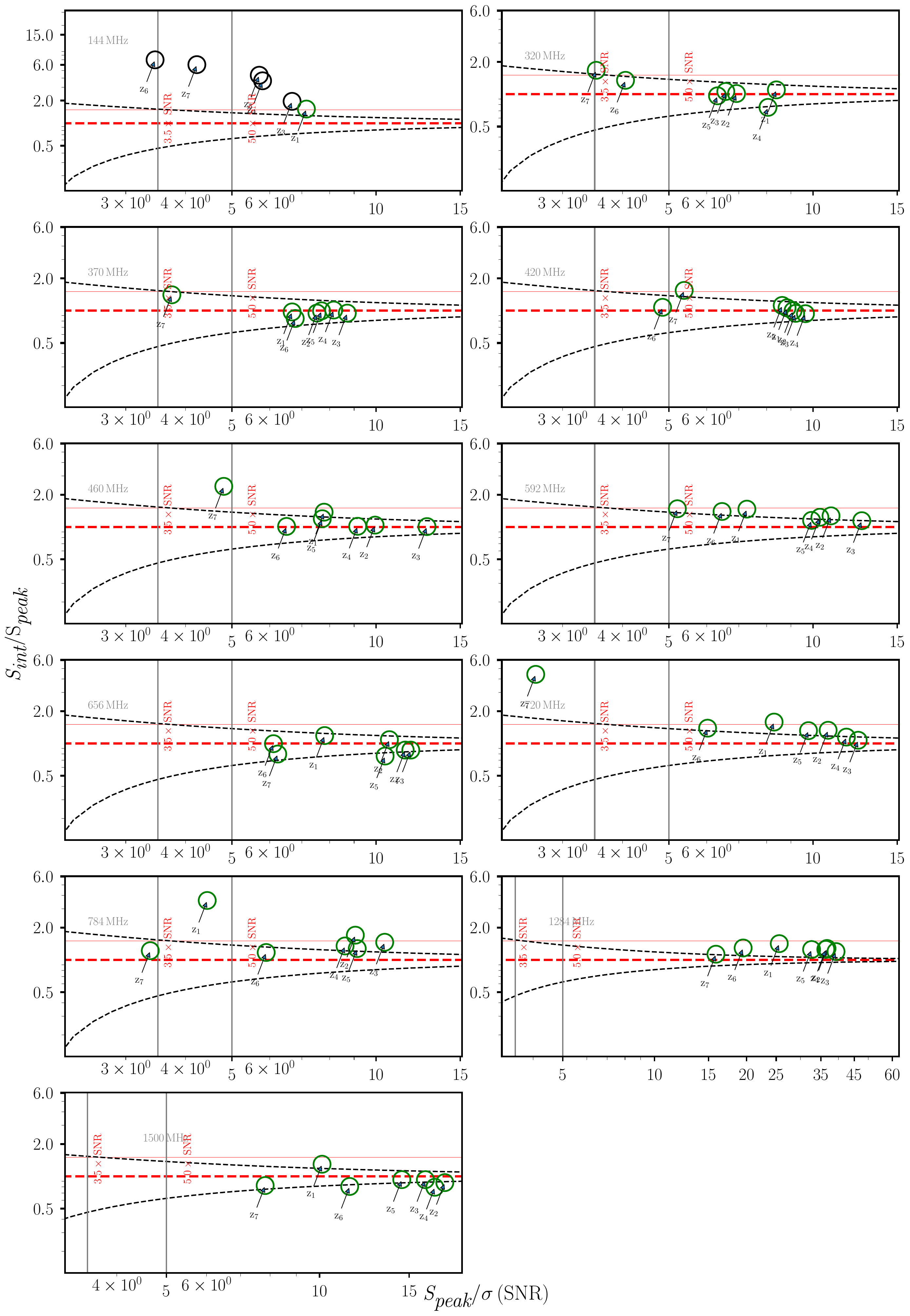}}
    \caption{Ratio of the integrated flux density to peak flux density as a function of signal-to-noise ratio ($\mathrm{S_{p}/\sigma}$). In our analyses, following equation~\ref{rms}, unresolved stacked sources at each frequency we study are depicted by the open \textbf{green} circles in each panel of Figure~\ref{sp_rms.fig} whereas resolved stacked sources are depicted by open black circles. The black dashed lines show the envelope defined by Equation~\ref{src_size}. Sources lying above the upper envelope are usually classified as resolved and those lying below the upper envelope are classified as unresolved, 
during the source-fitting procedure.  The solid red lines in each panel show $\mathrm{S_{i}/ S_{p}\,=\,1.5}$, whereas the dashed red lines show $\mathrm{S_{i}/ S_{p}\,=\,1}$, illustrating what a perfect point source would have.}
\label{sp_rms.fig}
\end{figure}

In order to assess whether our sources are clearly resolved, we follow the same criterion applied to GMRT 610 MHz detections in \cite{2020MNRAS.491.1127O} to identify resolved sources:

\begin{equation}\label{src_size}
\centering
\rm{\frac{\textit{S}_i}{\textit{S}_p}\,=\,1.0\,\pm\, \frac{3}{\textit{SNR}}}
\end{equation}

where SNR is simply the peak flux density divided by the rms of the image. This expression was obtained empirically to define an envelope containing 95\% of unresolved sources, below such threshold. Sources lying above the upper envelope are classified as resolved and those lying below the upper envelope are classified as unresolved. We compare our selection of resolved and unresolved stack sources in subsection~\ref{stack.sec} to the envelope defined by equation~\ref{src_size}. Figure~\ref{sp_rms.fig} illustrates the ratio of the integrated flux density to peak flux density as a function of signal-to-noise ratio ($\mathrm{\textit{S}_{p}/\sigma}$). Sources which are classified as resolved/unresolved using the method by \citet{2015MNRAS.453.4020F} usually reside above/below the upper envelope. The solid red lines in each panel show $\mathrm{\textit{S}_{i}/ \textit{S}_{p}\,=\,1.5}$. Whereas the dashed red lines show $\mathrm{\textit{S}_{i}/ \textit{S}_{p}\,=\,1}$, what a perfect point source would have. 

\subsection{Verification of flux boosting}{\label{boosting.sec}}
There is one caveat that applies when using the median stack images and it is usually not addressed is the issue of flux boosting. This represents the estimated ratio between recovered and intrinsic flux density of the same source. \citet{2007ApJ...654...99W} demonstrated that, given a set of point-source flux densities ${\textit{S}_{i}}$ in an image with local RMS-noise ${\sigma_{i}}$,the stacked median $\tilde{\textit{S}}_{stack}$ does not necessarily represent the true sample median $\tilde{\textit{S}}_{true}$\citep[e.g. see,][]{Algera_2020,2022ApJ...924...76A}. 

\begin{figure}
\centering
\centerline{\includegraphics[width=0.45\textwidth]{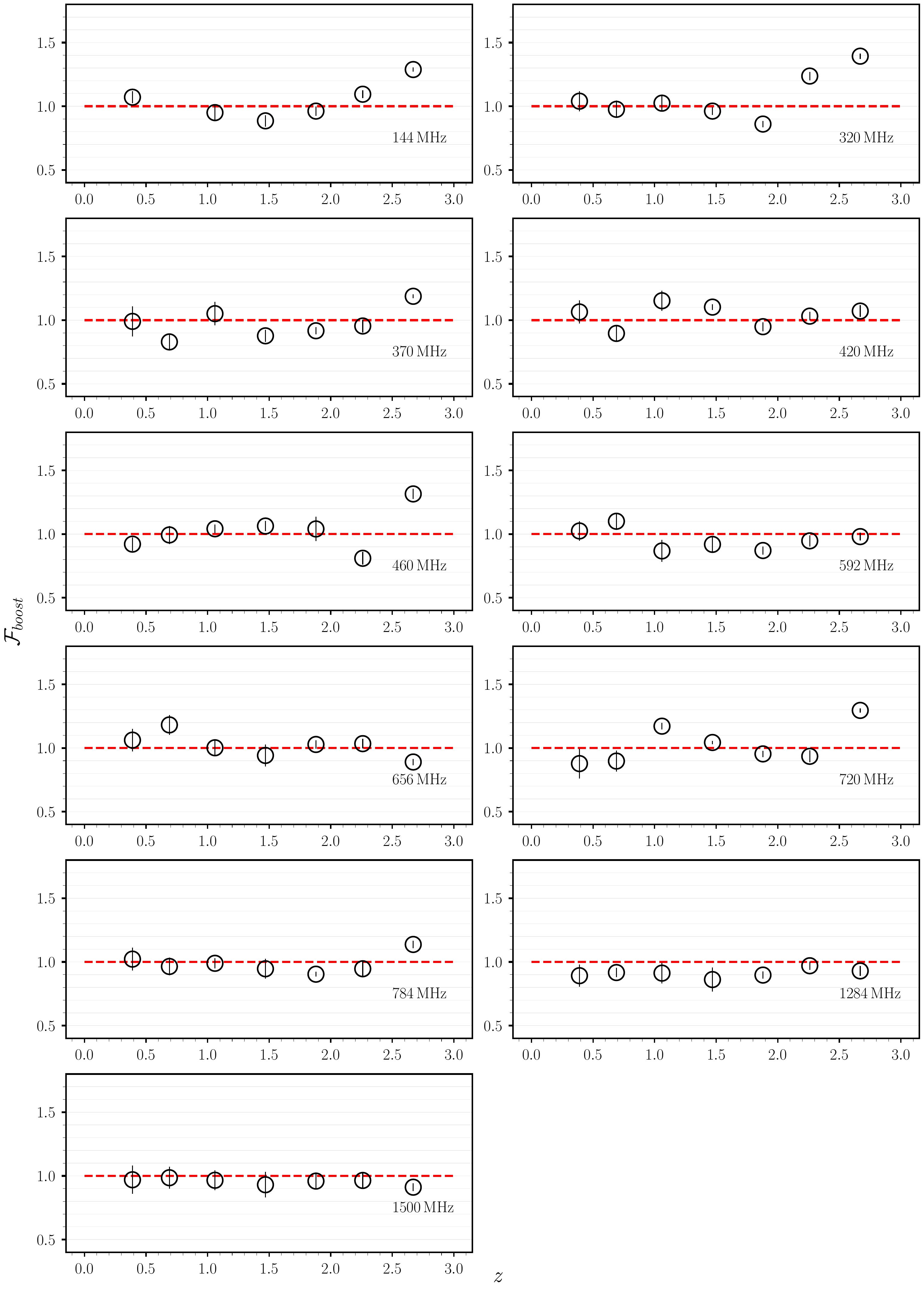}}
    \caption{Median boosting as a function of median redshift for stacks of simulated sources measured across $144\leq\,z\,\leq\,1500$ MHz up to $z\,\simeq\,3$.}
\label{fboost.fig}
\end{figure}
To quantify flux boosting effects, we conduct a stacking analysis on simulated sources with known intrinsic flux densities. Redshifts ($0.1\,\leq\,z\,\leq\,3.0$) are randomly sampled from a normal distribution, and mock sources are generated by assigning radio flux densities and redshifts to synthetic galaxies. Simulated sources are binned into seven redshift intervals matching those applied in the stacking analysis. The number of sources in each simulated redshift bin follows the total number of SFGs in Table~\ref{tab_ppts}. Following \citet{2022ApJ...924...76A}, we inject these sources into residual maps to preserve realistic noise properties while avoiding artificial proximity to the confusion limit. Following \citet{2022ApJ...924...76A}, we measure the local noise is measured directly from the residual images, with scatter of $0^{{\prime}{\prime}}30$ added in R.A. and DEC., to account for small offsets, thereby  capturing realistic spatial offsets between sources. The residual maps for  LOFAR, uGMRT, MeerKAT, and VLA are generated by running  \textsc{PyBDSF} source finder \citep{2015ascl.soft02007M} on their original total intensity maps.
A total of 5,000 simulated sources, distributed according to the flux density distribution of the original total intensity images, are inserted at random positions. For consistency with the radio maps (144 – 1500 MHz) we use in this study, the World Coordinate System (WCS) information from FITS headers is used to maintain native image dimensions and astrometry. Unresolved mock sources are assigned positions in the residual maps using a Gaussian beam model that replicates the resolution and position angle of the LOFAR, \textit{u}GMRT, MeerKAT, and VLA maps. Finally, we stack the mock sources and compare their input and output median flux densities. The ratio of these two quantities then defines the boosting factor, which is used to correct the flux densities probed in the real stacks. Figure~\ref{fboost.fig} presents the median boosting as a function of median redshift for stacks of simulated sources. The derived flux density boosting factor, $\mathcal{F}_{boost}\,=\,S_{recovered}/S_{input}$, remains near unity in each stack with high signal-to-noise ratios (S/N) but deviates slightly below unity for some stacks. The uncertainties were estimated by repeating the simulations for 100 times with different realisations, i.e. different input values for random variables like the position and flux density, and computing the standard deviation. The error on the derived correction factor is negligible compared with the Gaussian fitting error and the stacked background noise. Hence, the error is not added to the total error estimate. This procedure mirrors the stacking methodology applied to observational data, differing only in the substitution of simulated source catalogues and residual maps. All median stacked flux densities reported in the main text incorporate corrections for median boosting effects.
%
\bsp	
\label{lastpage}
\end{document}